\documentclass[fleqn,usenatbib]{mnras}

\usepackage{newtxtext,newtxmath}
\usepackage[T1]{fontenc}
\usepackage{graphicx}   % Including figure files
\usepackage{amsmath}    % Advanced maths commands
\usepackage{comment}
\usepackage{epsfig}
\usepackage{xcolor}
\usepackage{tabularx}
\usepackage{threeparttable}
\usepackage{hyperref}
\hypersetup{colorlinks=true, citecolor=blue, linkcolor=blue}
\usepackage{xspace}
\let\oldAA\AA
\renewcommand{\AA}{\text{\oldAA}\xspace}

\newcommand{\redtxt}[1]{\textcolor{black}{#1}}
\newcommand{\realredtxt}[1]{\textcolor{black}{#1}}

\newcommand{\hii}{H\,{\sc ii}}
\newcommand{\hei}{He\,{\sc i}}
\newcommand{\heii}{He\,{\sc ii}}

\newcommand{\oiii}{[O\,{\sc iii}]}
\newcommand{\oiiip}{O\,{\sc iii}}
\newcommand{\oii}{[O\,{\sc ii}]}
\newcommand{\ciii}{C\,{\sc iii}]}
\newcommand{\civ}{C\,{\sc iv}}

\newcommand{\feiii}{[Fe\,{\sc iii}]}
\newcommand{\niv}{N\,{\sc iv}]}
\newcommand{\niii}{N\,{\sc iii}]}
\newcommand{\nii}{[N\,{\sc ii}]}

\newcommand{\ha}{H$\alpha$}
\newcommand{\hb}{H$\beta$}

\newcommand{\jwst}{\textit{JWST}\xspace}
\newcommand{\pyneb}{\textsc{PyNeb}\xspace}
\newcommand{\cloudy}{\textsc{Cloudy}\xspace}
\newcommand{\somename}{NOEGs\xspace}

\newcommand{\kms}{$\rm km~s^{-1}$\xspace}

\definecolor{vasilycolor}{HTML}{228B22}

\title[Globular cluster progenitors]{\centering Connecting \jwst\ discovered N/O-enhanced galaxies to globular clusters:\\ Evidence from chemical imprints}

\author[Ji et al.]{Xihan Ji,$^{1,2}$\thanks{E-mail: \href{mailto:xj274@cam.ac.uk}{xj274@cam.ac.uk}}
Vasily Belokurov,$^{3}$
Roberto Maiolino,$^{1,2,4}$
Stephanie Monty,$^{3}$
Yuki Isobe,$^{1,2,5}$
\newauthor
Andrey Kravtsov,$^{6,7,8}$
William McClymont,$^{1,2}$
Hannah Übler$^9$
\\
$^{1}$Kavli Institute for Cosmology, University of Cambridge, Madingley Road, Cambridge, CB3 0HA, UK\\
$^{2}$Cavendish Laboratory, University of Cambridge, 19 JJ Thomson Avenue, Cambridge, CB3 0HE, UK\\
$^{3}$Institute of Astronomy, University of Cambridge, Madingley Road, Cambridge CB3 0HA, UK\\
$^{4}$Department of Physics and Astronomy, University College London, Gower Street, London WC1E 6BT, UK\\
$^{5}$Waseda Research Institute for Science and Engineering, Faculty of Science and Engineering, Waseda University, 3-4-1, Okubo, Shinjuku, Tokyo 169-8555, Japan\\
$^{6}$Department of Astronomy and Astrophysics, The University of Chicago, Chicago, IL 60637 USA\\
$^{7}$Kavli Institute for Cosmological Physics, The University of Chicago, Chicago, IL 60637 USA\\
$^{8}$Enrico Fermi Institute, The University of Chicago, Chicago, IL 60637 USA\\
$^{9}$Max-Planck-Institut für extraterrestrische Physik, Gie{\ss}enbachstra{\ss}e 1, 85748 Garching\\
}

\begin{document} 
\label{firstpage}
\pagerange{\pageref{firstpage}--\pageref{lastpage}}
\maketitle
 
\begin{abstract}
  Recent \jwst observations have revealed a growing population of galaxies at $z>4$ with elevated nitrogen-to-oxygen ratios. These ``N/O-enhanced'' galaxies (NOEGs) exhibit near- to super-solar N/O at sub-solar O/H, clearly deviating from the well-established scaling relation between N/O and O/H observed in local galaxies. The origin of this abundance anomaly is unclear. Interestingly, local globular clusters also exhibit anomalous light-element abundances, whose origin remains debated. 
  In this work, we compare the chemical abundance patterns of 22 known NOEGs at $0\lesssim z\lesssim 12$ -- primarily discovered with \jwst -- to those observed in local globular clusters. We find similarities in the abundances of C, N, O, Fe, and He between the two populations.
  The similar abundance patterns support the scenario in which globular cluster stars formed within proto-cluster environments -- similar to those traced by NOEGs -- that were self-enriched.
  Indeed, the enhancement in N/O in early galaxies appears to be only found in dense stellar environments with $\Sigma _{\star}\gtrsim 10^{2.5}~M_\odot~{\rm pc^{-2}}$, as expected for the progenitors of globular clusters in the Milky Way, and similar to those of star clusters identified in strongly lensed high-redshift galaxies. Furthermore, we find a tentative positive correlation between N/O ratios and stellar mass among NOEGs. The apparent high occurrence rate of NOEGs at high redshift is consistent with the picture of cluster-dominated star formation during the early stages of galaxy evolution.
  Measuring chemical abundances across diverse stellar environments in high-redshift galaxies will be crucial for elucidating the connection between NOEGs and globular clusters.
\end{abstract}

%
%-------------------------------------------------------------------

\section{Introduction}

%\purpletxt{The mysterious N emitters at high $z$ might be connected to globular clusters or nuclear star clusters, as already been discussed qualitatively by individual works. Here we make a more systematic comparison between the chemical abundance patterns of GC stars and those of high-$z$ N emitters.}

Chemical enrichment is an essential part of galaxy evolution, which is closely related to the baryonic cycle inside and outside of galaxies \citep{maiolino2019,Kobayashi_2020}.
With observations of local galaxies and stars, abundances of elements such as oxygen, nitrogen, carbon, iron, helium, and so forth have been routinely measured on galactic and sub-galactic scales in statistical samples.
Notably, observational studies of galaxies in the local Universe show there exist tight correlations between elemental abundance ratios such as N/O, C/O, and O/H in the interstellar medium (ISM) as well as stellar atmospheres, spanning more than one order of magnitude in these abundances \citep[e.g.,][]{cve_1993,Andrews_2013,Belfiore_no_2017,Nicholls_2017}.

The observed correlations imply a standard path for chemical evolution, and
current chemical evolution models can well reproduce the observed abundances \citep[e.g.,][]{Vincenzo_2016}, {although there is still debate on the details of the enrichment processes \citep{Matteucci_2021}.}
Based on current chemical evolution models, the relation between N/O and O/H can be interpreted as a combination of two enrichment channels. 
More specifically, nitrogen synthesis in stars proceeds through the CNO cycle, which relies on trace carbon and oxygen nuclei as catalysts; it is the origin of those catalysts, as well as the stellar layer where they are recycled, that divides nitrogen production into two distinct regimes, \redtxt{which are primary and secondary productions}. 
Low-metallicity massive (and/or fast-rotating) stars \redtxt{produce primary nitrogen at the early time of galaxy evolution,} when freshly synthesized C and O from the He-burning core are mixed down into the adjacent H-burning shell; there the CNO cycle converts them to $^{14}$N, and core-collapse supernovae expel this metal-independent yield.
%, producing
\redtxt{By adjusting the yield of primary N from massive stars, chemical evolution models can reproduce} 
the well-known constant N/O ``plateau'' at low O/H \citep[see][]{Matteucci_1986,Henry_2000,Prantzos_2018,Kobayashi_2020}. 
Intermediate-mass asymptotic giant branch (AGB) stars, \redtxt{on the other hand, produce both primary and secondary nitrogen.} 
%\redtxt{Here primary N is made from C made by the triple-$\alpha$ process from hot-bottom burning. Secondary N is made when }preexisting C and O inherited from the birth cloud are dredged into the deep convective envelope and at its $\gtrsim$50 MK base, where hot-bottom burning allows convection and CNO cycling operate in the same region, these two elements are repeatedly transformed into $^{14}$N.
\realredtxt{Here secondary N is made through CNO cycling of the preexisting C and O inherited from the birth cloud, at the base of the convective envelope reaching temperatures of $\gtrsim$50 MK (hot-bottom burning). Primary N is made from C made in the interior by the triple-$\alpha$ process and ``dredged up'' to the hot-bottom burning region.}
%\redtxt{While AGB stars also produce primary N from C made by the triple-$\alpha$ process from hot-bottom burning, the yield of primary N is typically much smaller than that of secondary N \citep{Ventura_2013}.}
Since the seed C and O now reflect the star's initial metallicity, the secondary yield scales with O/H, \redtxt{which, combined with the potential differential outflows at high O/H,} driving the observed rise in N/O at higher metallicity \redtxt{with a typical delay time of 0.1\,-\,1 Gyr} \citep[e.g.,][]{Meynet2002,Vincenzo_2016,maiolino2019}.
%At low metallicities (i.e., low O/H), which are characteristic for the gasous environments in the early phase of star formation, the production of nitrogen is mainly through the primary reaction chain in massive stars and it is released into the ISM by core collapse supernovae (CCSNe), which are explosions of massive stars ($M_\star > 8~M_\odot$). At high metallicities, with 0.1\,-\,1 Gyr after the initial star formation, the production of nitrogen is more efficient through the CNO cycle, and its enrichment in the ISM is mainly through the winds of intermediate mass stars ($2~M_\odot < M_\star < 8~M_\odot$) during their asymptotic giant branch (AGB) phase.
%The above processes lead to a nearly constant N/O at low O/H and a nearly linear dependence of N/O on O/H at high O/H.
These primary and secondary channels successfully reproduce the N/O–O/H trend in nearby galaxies.
%and form the basis of the standard framework for nitrogen enrichment. 

Expanding on this established picture requires probing elemental abundances across cosmic time, particularly in the early Universe, where the physical conditions and enrichment pathways may diverge significantly from those in the local galaxy population.
%To have a complete picture of chemical enrichment, it is vital to monitor elemental abundances in galaxies throughout cosmic time.
However, the observational constraints on elemental abundances in the early Universe are more difficult due to the redshift of spectral features typically used to make abundance measurements (e.g, \oiii$\lambda 5007$ goes to $>2~{\rm \mu m}$ at $z>3$).
In recent years, thanks to the sensitivity and wavelength coverage of \jwst and its instruments, especially the Near Infrared Spectrograph \citep[NIRSpec,][]{jakobsen2022,boker2023}, elemental abundances have been estimated for galaxies even close to cosmic dawn, i.e. up to $z>10$ \citep[e.g.,][]{bunker2023,cameron2023,castellano2024,napolitano_ghz9agn_2024,Carniani_gsz14_2024,Schouws_gsz14_2024}.
These \jwst observations have revealed a puzzling population of ``chemically abnormal'' galaxies at high redshift,
%However, one of the puzzles that come with the fruitful \jwst\ observations is the increasing number of ``chemically abnormal'' galaxies at high redshift.
which show unexpectedly bright nitrogen emission lines indicative of super-solar abundance ratios of N/O at sub-solar O/H,
in contrast to the significantly sub-solar N/O ratios seen in metal-poor galaxies in the local Universe.
%making them significantly above the N/O versus O/H relation observed in local galaxies.
A number of these galaxies are found at $z\gtrsim 6$, meaning that 
\redtxt{if the measured abundances are interpreted as a result of global chemical evolution,}
their cosmic age is even less than the typical delay time of \redtxt{significant} nitrogen enrichment by intermediate-mass AGB stars ($\sim 1$ Gyr) \redtxt{according to standard chemical evolution models.
Still, while galaxy-integrated N/O upturns often appear on Gyr scales, the onset of AGB hot-bottom-burning N production is much earlier: massive/intermediate-mass AGBs ($\sim$ 3.5\,–\,8 $M_\odot$) eject N-rich winds from $\sim$ 30\,–\,40 Myr after birth and largely cease by $\sim$ 300 Myr. Thus, in principle, AGB-wind N pollution is plausible even at $z\gtrsim6$ \realredtxt{\citep{dantona_2023}}.
}

While such ``N/O enhanced''
%\footnote{\color{red} This might not be an exact term as the enhanced N/O can be made by depletion of oxygen as we discuss later. Therefore, here the nitrogen loudness should be interpreted as a relative loudness, or ``N/O'' loudness.} 
galaxies (hereafter NOEGs) were identified in the pre-\jwst\ era -- mainly among quasars ($\sim$1\% of QSOs at $1.6 \lesssim z \lesssim 4$; \citealp{Bentz_osmer_nlqso_2004,Bentz_nlqso_2004,Jiang_nlqso_2008}), where they were speculated to reflect super-solar metallicities in broad-line regions \citep{Hamannferland_1999} -- they were very rare in local star-forming (SF) galaxies \citep{james2009}. Only now, with \jwst, has strong N/O enhancement been confirmed in statistical samples of metal-poor galaxies at high redshift.

Thus far, there are more than 20 galaxies with established N/O enhancement at $z>2$, and there are potentially more \somename missing due to the wavelength coverage and depths of current \jwst observations \citep{Hayes_stack_2025,isobe2023,Isobe_2025}.
Standard chemical evolution models fail to explain the nitrogen enhancement in \somename, and alternative enrichment channels such as fast-rotating massive stars,
very massive stars (VMS; $M_\star>100~M_\odot$), extremely massive stars (EMSs; $10^3~M_\odot<M_\star<10^4~M_\odot$), super massive stars (SMSs; $M_\star>10^4~M_\odot$), Wolf-Rayet (WR) stars with direct collapses, tidal disruption events (TDEs), early enrichment of AGB stars (combined with inflows and/or differential outflows), and so forth, have been proposed \redtxt{\citep{Tsiatsiou_rrs_2024,Vink_2023,Gieles_gcmp_2025,Nagele_2023,Charbonnel_2023,Nandal_ems_2024,Nandal_frms_2024,Nandal_2025,watanabe2024,cameron2023,dantona_2023,Rizzuti_no_2024,will_nomore}}
Still, the exact enrichment mechanism remains poorly constrained.
%\footnote{As an alternative explanation for the high N/O, \citet{Flury_shock_2024} recently suggested prevalent shock ionization in the early Universe that biases the abundance derivation, which, however, relies on various model assumptions due to the intrinsically high dimensionality of shock parameter space.}.

The abundance of \somename found at high redshift raises a key question of whether we can see the relics of these systems in the local Universe.
If the gas showing abnormal abundances at high redshift later formed stars, one might expect to observe such abundance patterns in certain populations of old stars.
Interestingly, it has been known for a long time that stars in globular clusters (GCs) of the Milky Way (MW) exhibit peculiar abundances in light elements, including He, C, N, O, Na, Mg, Al, and so forth \citep{Gratton_2004}.
In addition, many of these GCs are estimated to have formed at $z>4$ \citep[e.g.,][]{VandenBerg_2013}, which coincides with the redshift of \somename that are still actively forming stars at the time of \jwst observations.
It has been suggested that the deep gravitational potentials of proto-GCs had allowed them to imprint the anomalous chemical abundances in their stars.
Observations of local GCs have revealed a spread of chemical abundances related to the multiple
%, possibly discrete, 
populations of stars \citep[e.g.,][and references therein]{Smith_gcmetrev_1987,Kraft_gcchemrev_1994,Gratton_nao_2001}. {
%SM: I suggest rephrasing the following sentences slightly, just to clear up confusion around the origin of light element anti-correlations in GCs. Here are my suggestions: 
Traditionally, stars in GCs have been split into two generations, those with light element abundances similar to MW field stars at the same iron abundances (1G), and those with enhanced N, Na, and depleted O (2G) \citep[see][for recent reviews on the subject]{bastian2018, gratton2019, milone2022}. 
%Note that there are stars occupying the transition zone between these two generations, suggesting that they may not be born from exactly discrete star-forming events. 
%\redtxt{We note that stars in GCs generally have smooth abundance distributions spanning typical halo abundances to extreme enhancements in some elements, meaning there are no strictly discrete star-forming events and the two-generation classification should be considered as a rough cut in the formation time of stars.}
Depending on the mass of the cluster, light-element anomalies may extend to an enhancement in Al and depletion in Mg, and, in extreme cases, an enhancement in K \citep[e.g., in the massive GCs, NGC~2419, NGC~2808, NGC~5139, and NGC~4833;][]{cohen2012, mucciarelli2012, mucciarelli2015, meszaros2020, carretta2021}. Importantly, the fraction of chemically peculiar stars in a GC strongly correlates with the cluster mass \citep[see e.g.,][]{Milone2017,milone_hegc_2018,bastian2018, gratton2019}. This correlation is noticeably tighter if the cluster's initial (i.e., before evaporation and tidal shredding) mass is considered \citep[see e.g., Figures 6 and 4 in][respectively]{gratton2019, Belokurov2023}. While the nucleosynthetic site is debated, these anomalies are clear evidence of the CNO cycle, NeNa and MgAl chains operating at ever-increasing temperatures between the formation of 1G and 2G stars.
Suggested sites have included WR stars, AGB stars \redtxt{\citep{Ventura2001,dantona_2023,will_nomore}}, fast-rotating massive stars \citep[see e.g.,][]{Maeder2006, Prantzos2006}, interactive massive binaries \citep[][]{deMink2009}, very massive stars \citep[VMSs, see e.g.,][]{Vink2018}, super-massive stars \citep[SMSs, see e.g.,][]{Denissenkov2014,Prantzos2017}. \redtxt{Also}, extremely massive \redtxt{stars \citep[EMSs,][]{Gieles_gcmp_2025} and AGB stars \citep{Ventura_2013,Ventura_2018,AlvarezGaray_2024}} have been shown to explain the exceptionally high N abundances
%found at high-$z$ 
\textit{and} to achieve temperatures high enough to reach the MgAl chain in massive GCs.}

Given the close resemblance of the two unsolved problems in the chemical enrichment of \somename and GCs, some recent works have suggested that \somename discovered by \jwst host proto-GCs and thus the two problems are interconnected \citep[e.g.,][]{Belokurov2023,isobe2023,Senchyna_2024,Marques-Chaves_2024,Gieles_gcmp_2025,Renzini_baryoncon_2025}.
%\redtxt{Indeed, [obs of bright clusters tba]}
Indeed, recent \jwst\ observations have hinted that \somename are likely more compact than normal star-forming galaxies at similar redshift, implying that star formation in these systems occurs in dense, spatially concentrated regions \citep[e.g.,][]{topping2024,schaerer_gnz9_2024,Harikane_2025}.
For example, in the gravitationally lensed galaxy, Sunburst Arc at $z=2.26$, where several star clusters are spatially separated, it has been shown that the nitrogen enhancement occurs in a Lyman-continuum leaking ``super star cluster'' \citep[LyC,][]{pascale_sbarc_2023}, but not seen in the integrated spectrum of the galaxy \citep{welch_sba_2024}. Such young, compact and extremely massive (age~$\lesssim4$ Myr, $R_{\rm e}\lesssim20$ pc, $M_\star\sim10^{7} ~M_\odot$) super star clusters are very rare in the local Universe but their lower-mass analogs have been spotted in extreme starbursts and dramatic, gas-rich mergers \citep[][]{Portegies2010}.

More recently, \citet{Isobe_2025} show that nitrogen enhancement in \jwst-observed galaxies is correlated with high gas densities and presence of active galactic nuclei (AGN), which indicate a connection between proto-GC formation and black hole seeding through VMSs/EMSs/SMSs and/or runaway merging.
In the early Universe, it is expected that star formation is bursty \citep[][]{Caplar2019,Tacchela2020,Sun2023,Kravtsov2024,McClymont_burstysf_2025} and mostly contributed by clumpy components manifested as massive stellar clusters, whereas more smooth, disk-dominated star formation only becomes prevalent at a later epoch \cite[e.g.,][]{Yu_mwdiskfire2_2021,yu_mwdiskfire2_2023,Hafen_discfire_2022,Dillamore_mwdisk_2024,Semenov_mwdisk_2024}.
In this picture, if the ``chemical anomaly'' is a natural outcome of proto-cluster evolution, nitrogen enhancement is expected to be more common in the early Universe, when star formation is bursty and before the emergence of dynamically stable stellar disks. Recent simulations point to a characteristic mass scale around $ 10^{11}~M_\odot$ in halo mass or a few $\times 10^{9}~M_\odot$ in stellar mass, above which long-lived, kinematically cold disks emerge \citep[e.g.,][]{Dillamore_mwdisk_2024,Semenov_mwdisk_2024}. If a galaxy’s dynamical state is linked to its ability to form massive star clusters, then anomalously high N/O should be found preferentially in systems below this threshold, and it is more likely to happen in the early Universe.

Thus far, however, there has yet to be a systematic comparison between the chemical imprints of stars in GCs and ISM in \somename to assess this connection. In this work, we compiled a sample of 21 \somename at $z > 2$ together with a local analog of high-$z$ \somename, Mrk 996. Our aim is to compare the measured chemical abundance patterns of these systems with MW stars, especially those identified in GCs, and investigate whether \somename can host progenitors of GCs observed today, if we assume MW GCs are representative.

The structure of this manuscript is as follows.
In Section~\ref{sec:data}, we introduce our sample galaxies and stars as well as the relevant observations.
In Section~\ref{sec:method}, we describe derivations of the physical properties of these sources.
In Section~\ref{sec:chem_com}, we compare the derived abundance patterns of \somename and MW GCs.
We discuss the connection between \somename and MW GCs as well as the validity of NOEG identifications in Section~\ref{sec:discuss}, and we draw our conclusions in Section~\ref{sec:conclude}.
Throughout this work, we assume a flat $\rm \Lambda CDM$ cosmology with $h=0.674$ and $\Omega _{\rm m} = 0.315$ \citep{planck2020}. We adopt the solar abundance set from \citet{Lodders_2021}, with $\rm 12+\log(He/H)_\odot=10.924$, $\rm 12+\log(C/H)_\odot=8.47$, $\rm 12+\log(N/H)_\odot=7.85$, $\rm 12+\log(O/H)_\odot=8.71$, and $\rm 12+\log(Fe/H)_\odot=7.48$.
We follow the convention of the nebular community by referring to the relative oxygen abundance, O/H, as the metallicity, unless otherwise specified.

\section{Observational data}
\label{sec:data}

In this section, we describe the observational data we used, including those of galaxies and stars.
All sources are drawn from existing publications.

\subsection{N/O-enhanced galaxies}

\begin{figure*}
    \centering
    \includegraphics[width=0.44\textwidth]{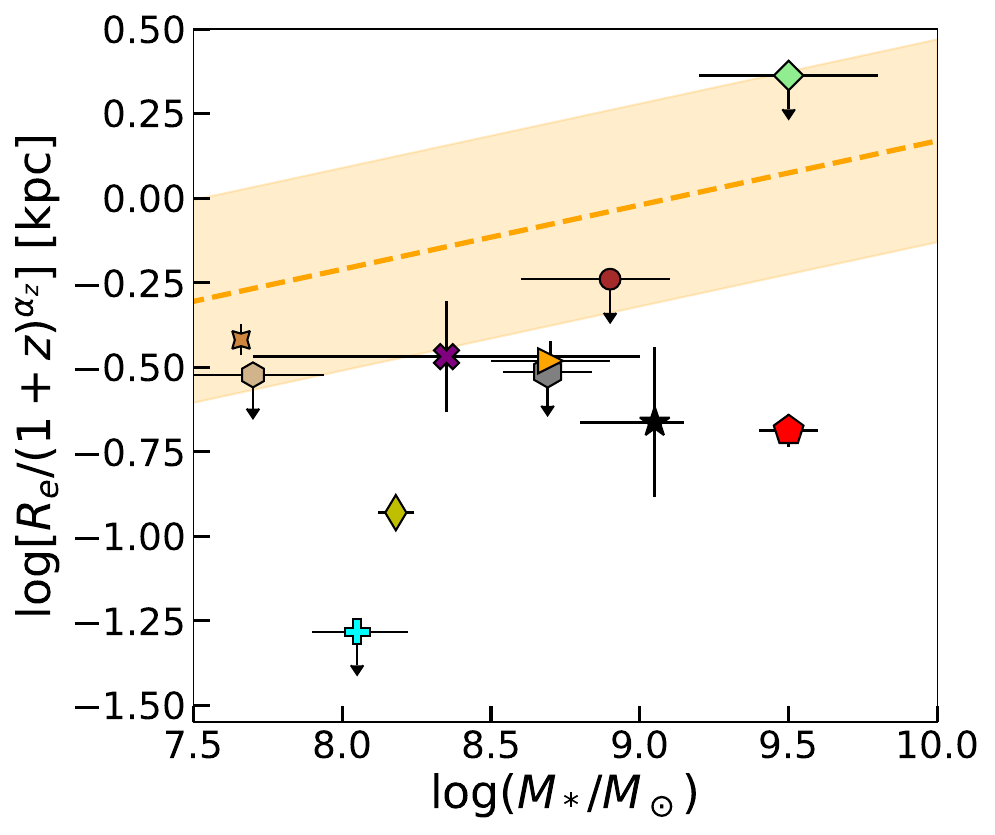}
    \includegraphics[width=0.55\textwidth]{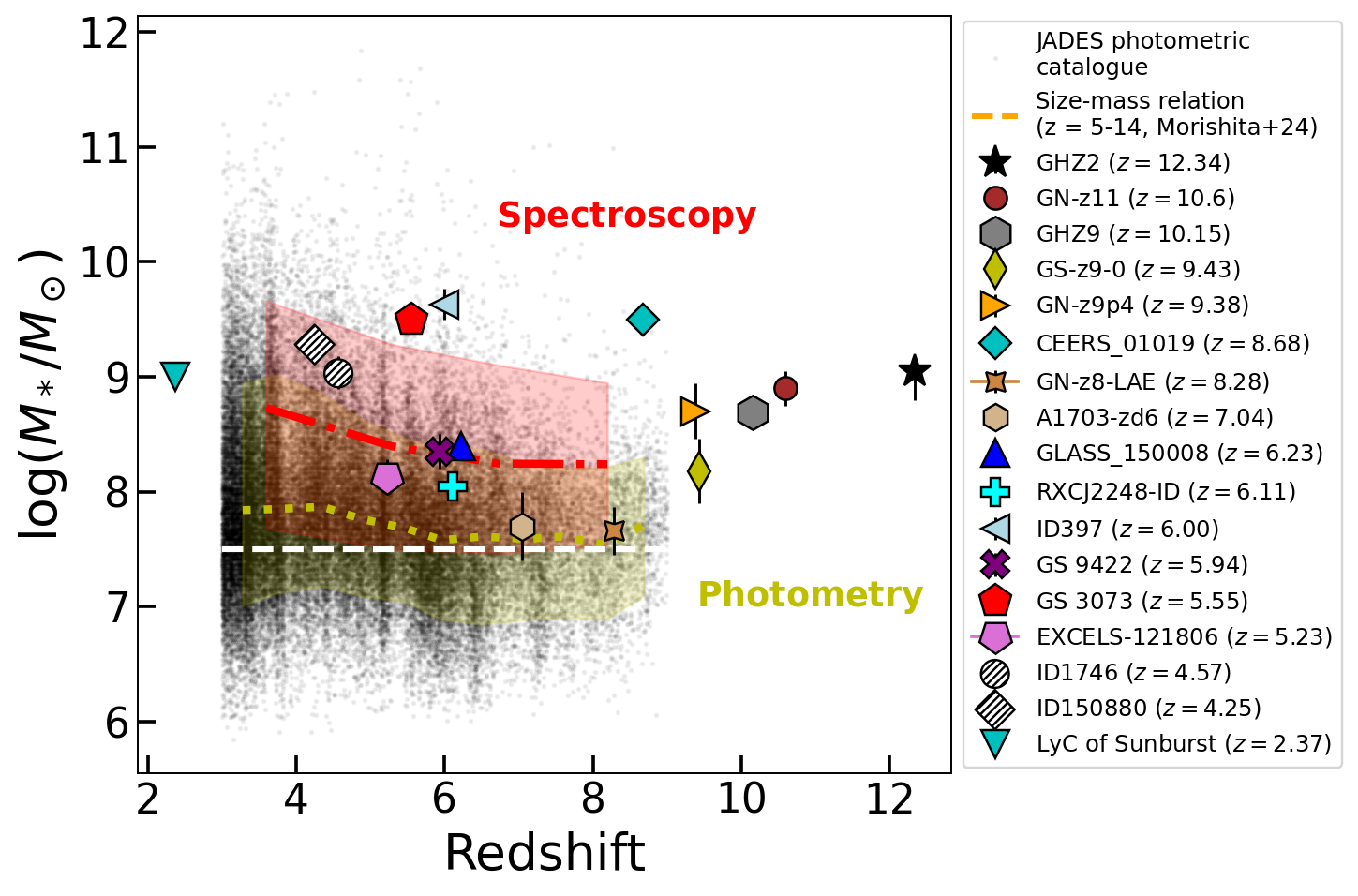}
    \caption{Global properties of high-$z$ \somename in our sample.
    \textit{Left:} size-stellar mass relation for \somename and a compilation of \jwst galaxies at photometric redshift of $5<z_{\rm photo}<14$ studied by \citet{Morishita_2024}. For the \jwst sample, \citet{Morishita_2024} fitted the redshift evolution of the relation with the parameter $\alpha_z=-0.44$, which we use to normalize $R_e$. 
    We plot the best-fit linear model of \citet{Morishita_2024} as the dashed orange line together with the $1\sigma$ dispersion in $\log R_e$ indicated by the shaded region.
    The \somename are generally more compact compared to galaxies at similar stellar masses.
    \textit{Right:} stellar mass-redshift relation for \somename and JADES photometric sample at $3<z_{\rm photo}<9$.
    The JADES sample has a $90\%$ completeness at $M_\star \gtrsim 10^{7.5}~M_\odot$ as indicated by the dashed white line \citep[see][]{Simmonds_2024}.
    The dotted yellow line represents the median trend of the JADES photometric sample and the shaded region is bounded by the 10 percentile and 90 percentile lines.
    The dash-dotted red line represents the median trend of the JADES subsample with spectroscopic observations and the shaded region is bounded by the 10 percentile and 90 percentile lines.
    The \somename are relatively massive, and their selection is highly inhomogeneous and probably biased by their UV brightness, which allows for measurements of faint nitrogen lines.
    }
    \label{fig:jades_mass}
\end{figure*}

We compiled a sample of 21 \somename at $2\lesssim z\lesssim12$ in the literature, corresponding to a range of cosmic ages of 350 Myr - 2.8 Gyr after the Big Bang \citep{pascale_sbarc_2023,labbe_monster_2024,ji2024,cameron2023,cameron_gs9422_2024,topping2024,isobe2023,schaerer_gnz9_2024,curti_gsz9_2024,napolitano_ghz9agn_2024,bunker2023,castellano2024,Navarro-Carrera_nloud_2024,Arellano-cordova_nloud_2024,Stiavelli_nloud_2024}.
These sources are all reported to have $\rm (N/O)\gtrsim (N/O)_\odot$ at $\rm (O/H)< (O/H)_\odot$.
We emphasize that the \somename are selected from parent samples with different depths and complex selection functions.
Thus, our aim is not to quantify, for example, the actual fraction of \somename in early galaxies and the relevant statistical interpretations, but rather understand whether the \somename, as a potentially distinct population of early galaxies, exhibit any connections to the local relics.
%We list these sources in Table~\ref{tab:nemitters_measurements}.
Spectroscopic observations that confirm nitrogen enhancement in these sources were primarily made via \jwst/NIRSpec except for the Lyman continuum (LyC) leaking super star cluster in the gravitationally lensed Sunburst Arc at $z=2.37$, whose chemical abundances were derived from observations via the ground-based Very Large Telescope (VLT) with MUSE integral field unit (IFU) and X-shooter spectroscopy in addition to  \jwst/NIRSpec \citep{pascale_sbarc_2023,welch_sba_2024}.
In addition to these high-$z$ \somename, we include a blue compact dwarf (BCD) galaxy, Mrk 996, at $z\sim0$, which shows nitrogen enhancement with respect to the local relation of N/O-O/H via VLT VIMOS observations and is a rare local analog of \somename at high redshift \citep{james2009}.
We refer the readers to the sources we listed for details of data reduction and emission-line measurements.
We summarize some derived properties for these sources in Table~\ref{tab:global}, including redshift, stellar mass, and effective radius.
{Figure~\ref{fig:jades_mass} shows comparisons between the stellar masses, sizes, and redshifts of \somename and those derived photometrically for \jwst-observed galaxies at $5<z_{\rm photo}<14$ by \citet{Morishita_2024} as well as at $3<z_{\rm photo}<9$ in the \jwst Advanced Deep Extragalactic Survey \citep[JADES,][]{bunker2023a,eisenstein2023,eisenstein2023b,deugenio_dr3_2024}.
Overall, the high-$z$ \somename show relatively compact morphologies and well as high stellar masses, which we discuss in further details in Section~\ref{sec:discuss}.
}

\subsection{Gas-phase abundances in low-redshift galaxies}
\label{subsec:local_galaxymet}

For comparison, we included a sample of local galaxies and nebulae with abundance measurements from \citet{izotov2006}, \citet{pilyugin2012}, \citet{berg_2016,berg2019}, \citet{Annibali_2019}, \citet{mendezdelgado_feo_2024}, and \citet{Grossi_2025} with $z\lesssim 0.1$. 
Specifically, the sample of metal-poor galaxies of \citet{izotov2006} come from the Data Release 3 (DR3) of the Sloan Digital Sky Survey \citep[SDSS,][]{york2000}.
The abundances were derived through the so-called $T_{\rm e}$ method, where electron temperatures are estimated through flux ratios of optical auroral and strong lines of metals (e.g., \oiii$\lambda 4363$/\oiii$\lambda 5007$), and the majority of the sample have $\rm 12+\log(O/H) \lesssim 8.5$.
We took the abundances of O/H and N/O in the gas phase derived by \citet{izotov2006} for these galaxies.
There are 5 galaxies with $\rm \log(N/O)>-0.5$, all of which, however, have signal-to-noise (S/N) in \oiii$\lambda 4363$ of $\rm S/N\approx 1$ and thus do not have reliable temperature estimates.
Therefore, we did not include these galaxies.
We also used N/O and O/H derived by \citet{pilyugin2012} for a sample of galaxies in the DR6 of SDSS using both the $T_e$ method with auroral lines and the strong-line method.
This sample covers $\rm 12+\log(O/H)$ above 8.5, although most of the galaxies still have lower metallicities.
%These galaxies are observed to form a tight sequence in the abundance space spanned by N/O and O/H, implying a ubiquitous pathway for chemical enrichment.
We further include N/O and O/H of 3 extremely metal-poor \hii\ regions with $\rm 12+\log(O/H)\sim 7$ in the galaxy DDO 68 derived by \citet{Annibali_2019}, as well as 6 metal-poor dwarf galaxies with $\rm 12+\log(O/H)\approx 7.4-7.9$ from \citet{Grossi_2025}.
We also included a sample of dwarf galaxies from \citet{berg_2016,berg2019} with abundances of O/H and C/O. Since the measurement of C/O usually relies on the UV carbon lines, there are not many determinations of the gas-phase C/O in local galaxies.
{Finally, the sample of \citet{mendezdelgado_feo_2024} includes 451 galaxies and
\hii\ regions with Galactic or extra-galactic origins, where the local galaxy sample of \citet{izotov2006} is also included.
From this sample, we took the measurements of O/H and Fe/O made by \citet{mendezdelgado_feo_2024} with corrections for temperature inhomogeneities.
We note that whether or not to include the temperature corrections do not have any significant impact on our conclusions.
}

\subsection{Stellar abundances in the Milky Way}
\label{subsec:derive_starmet}

We included a sample of MW stars observed by the SDSS Apache Point Observatory Galaxy Evolution Experiment (APOGEE) in the data release 17 \citep[DR17,][]{sdss_dr17}. APOGEE is a medium resolution\footnote{By the standard of the stellar community.} ($R\sim$ 22,500), near infrared spectroscopic survey (15,140\,-\,16,940\AA) covering $\sim650,000$ MW stars across both hemispheres.  To derive chemical abundances for up to 20 different species, the APOGEE
Stellar Parameter and Chemical Abundance Pipeline \citep[ASPCAP,][]{aspcap} is used to fit individual absorption lines (either atomic or molecular) in the stellar spectra using the full-spectrum-fitting code, \textsc{ferre} \citep{ferre}. Specifically, we used the \texttt{allStarLite} catalogue, applying the same quality cuts as in \citet{belokurov2022} to exclude stars with problematic flags (e.g., \verb|STAR_BAD, LOW_SNR, PERSIST|-related issues, \verb|EXTRATARG| duplicates). We only consider red giant stars, i.e. those with $1.5<\log(g)<3$ and $T_{\rm eff}<5300$~K excluding the most evolved red giant branch (RGB) stars as they show spurious N/O abundances. To be included in our sample, stars have to exhibit moderately small uncertainties (i.e., $<0.2$ dex) in measurements of abundances of the following chemical elements: [Fe/H], [Mg/Fe], [C/Fe], [N/Fe], [O/Fe], [Si/Fe], [Al/Fe].  We also explicitly exclude stars that are part of the \texttt{magcloud} observing program. Distances are taken from the \texttt{AstroNN} catalogue \citep[][]{Mackereth2018,Leung2019}, and proper motions from Gaia EDR3 \citep[][]{gaia_edr3, Lindegren2021}.

Where possible, we have compared chemical abundances in our APOGEE DR17 sample with those in GALAH DR4 \citep[][]{galah_dr4} and have detected an unphysical trend in the behaviour of [O/Fe] as a function of [Fe/H] as reported by APOGEE DR17. We correct for this trend using median residuals between APOGEE DR17 and GALAH DR4 in bins of metallicity. The resulting corrected [O/Fe] values are validated using abundances of a sample of stars in the old and the intermediate-age disks of the Galaxy (sometimes referred to as the thick and the thin disks) published by \citet{Amarsi2019}. Our corrected [O/Fe] abundances show a perfect agreement with those in the literature. Note that for this calibration procedure and for the analysis presented in this paper, we select old (thick) and intermediate-age (thin) disks using values of  orbital eccentricity and $z_{\rm max}$ (amplitude of the vertical excursion about the Galactic plane) as published in the \texttt{AstroNN} catalogue.

To select GC stars from the clean APOGEE catalogue we cross-match with the GC membership catalogues of \citet{vasiliev2021}. Using Gaia EDR3 \citep{gaiadr3} astrometry and photometry, \citet{vasiliev2021} assign a membership probability between 0 and 1 for stars in the vicinity of 170 MW GCs. Adopting a membership probability of $>50\%$ we recovered $\sim2500$ stars across 27 clusters. Because we are interested in differentiating between generation one and two stars in GCs, we keep only the GCs with five or more member stars in our APOGEE catalogue. 

Following \redtxt{the above} initial cross-match, we then split members within each cluster into likely first- and second-generation stars.
\redtxt{We used chemical abundance patterns to define the two generations, where the stars formed later are enhanced in N, Na, and Al but depleted in O and Mg compared to stars formed earlier \citep{bastian2018}, and the demarcation is set at the average abundance pattern in each cluster.}
%using the following definitions. 
\redtxt{Starting with N and O, whose abundances are best measured in our sample,} first-generation (1G) stars are defined as having $\mathrm{[N/O]}\leq\mathrm{[O/Fe]_{average}}$ and $\mathrm{[O/Fe]}\geq\mathrm{[O/Fe]_{average}}$. For the second-generation (2G) stars we apply the opposite cuts with the modification that $\mathrm{[O/Fe]}\leq\mathrm{[O/Fe]_{average}} - \sigma(\mathrm{[O/Fe]})$, where $\sigma$ represents the standard deviation in [O/Fe]. This modification is necessary because the [N/O] versus [O/Fe] distribution in GCs is, in reality, a continuum, not two distinct populations and we would like to sample the endpoint of the distribution to capture the 2G stars (this is especially important in massive clusters like $\omega$-Centauri which represents $\sim30\%$ of our sample).
\redtxt{We note that while the abundance distributions in GCs are usually continuous from the 1G to 2G stars, this does not necessarily imply continuous star formation in the past, but could be a result of bursts of star formation with stellar ejecta mixed with the ambient gas in different proportions \citep{dercole_2008}.}
%We emphasize that this division between first- and second-generation stars is a simplified approximation of the true, likely continuous formation history within globular clusters. For our purposes, however, it provides a useful framework for comparing abundance patterns between GCs and \somename, especially given the relatively large uncertainties in abundance measurements at high redshift.
\redtxt{The above definition with N/O automatically separates two generations in the N/O-O/H space.
We also checked an alternative selection not involving N/O but leveraging the well-known Al-Mg anticorrelations among GC stars \citep[e.g.,][]{Gratton_nao_2001}.
We did not use the Na-O anticorrelation due to the large uncertainties of Na in APOGEE \citep{sdss_dr16} and the large spread in Na differences between APOGEE and GALAH \citep{galah_dr4}.
We selected 1G stars by requiring $\rm [Al/Mg] \leq [Al/Mg]_{average}-0.5\sigma([Al/Mg])$ and 2G stars by requiring $\rm [Al/Mg] \geq [Al/Mg]_{average}+0.5\sigma([Al/Mg])$.
Our conclusions remain unchanged with the alternative selection.
We used the cut involving N/O as our fiducial selection since 1) N measurements have lower uncertainties compared to Al measurements [$\rm S/N(N)_{median}=9>S/N(Al)_{median}=6$]; 2) the Al-Mg cut is physically not independent of the N-O cut as the Mg-Al cycle occurs at higher temperatures compared to the CNO cycle.
}

{Within our catalogue of 
%globular clusters (GCs), 
GCs, we identified known nuclear star clusters (NSCs) following classifications by \citet{Pfeffer_nsc_2021} and \citet{McKenzie_nsc_2022}. These include Omega Centauri (NGC 5139), NGC 6273, NGC 6656, and NGC 1851, which are thought to be the remnant nuclei of accreted dwarf galaxies. Among these, NGC 5139 alone contributes approximately 80\% of the stars in the NSC subsample and around 40\% of all stars in our combined GC/NSC sample. It is therefore essential to distinguish this cluster from the broader GC population. Beyond this numerical dominance, we also flag NSCs because they are believed to have hosted extended star formation episodes and to harbour more complex stellar populations than typical GCs \citep{Neumayer_nsc_2020}. As a result, NSCs may exhibit distinct chemical abundance patterns, a point we return to in the analysis that follows.
}

\section{Derivation of physical properties of \somename}
\label{sec:method}

\begin{table}
    \centering
    \caption{Global properties of a compilation of \somename from the literature at $0\lesssim z \lesssim 12$ investigated in this work.}
    \begin{tabular}{l |c c c}
    \hline
    ID & Redshift & $\log M_\star/M_\odot$ & $R_e~{\rm (pc)}$ \\
    \hline
    Mrk 996$^{\rm a}$ & 0.00544 & $8.7\pm 0.1$ & 710 (220\,-\,430$^\dagger$) \\
    Sunburst Arc (LyC)$^{\rm b}$ & 2.37 & 9 ($7^\dagger$) & 31 ($<20^\dagger$) \\
    ID150880$^{\rm c}$ & 4.25 & $9.28^{+0.02}_{-0.04}$ & - \\
    UNCOVER-45924$^{\rm d}$ & 4.46 & - & $\lesssim 70$ \\
    ID1665$^{\rm c}$ & 4.48 & - & - \\
    ID1746$^{\rm c}$ & 4.57 & $9.03$ & - \\
    ID1477$^{\rm c}$ & 4.63 & - & - \\
    ID60001$^{\rm e}$ & 4.69 & $9.34^{+0.08}_{-0.10}$ & 600, 290 \\
    EXCELS-121806$^{\rm f}$ & 5.23 & $8.13^{+0.09}_{-0.06}$ & - \\
    %EXCELS-70684$^{\rm hh}$ & 5.26 & $8.52\pm 0.13$ & - \\
    GS\_3073$^{\rm g}$ & 5.55 & $9.5\pm 0.1$ & 80\,-\,180 \\
    GS\_9422 (tentative)$^{\rm h}$ & 5.94 & 7.7\,-\,9.0 & 90\,-\,200 \\
    ID397$^{\rm c}$ & 6.00 & $9.63$ & - \\
    RXJ2248-ID$^{\rm i}$ & 6.11 & $8.05^{+0.17}_{-0.15}$ & $< 22$ \\
    GLASS\_150008$^{\rm j}$ & 6.23 & $8.4^{+0.4}_{-0.2}$ & - \\
    A1703-zd6$^{\rm k}$ & 7.04 & $7.7\pm 0.24$ & $<120$ \\
    GN-z8-LAE$^{\rm l}$ & 8.28 & $7.66\pm 0.02$ & $143\pm 15$ \\
    CEERS\_01019$^{\rm m}$ & 8.68 & $9.5\pm 0.3$ & 790\,-\,910$^*$ \\
    GN-z9p4$^{\rm n}$ & 9.38 & $8.7\pm 0.2$ & $118\pm 16~(<190)$ \\
    GS-z9-0$^{\rm o}$ & 9.43 & $8.18\pm 0.06$ & $110\pm 9$ \\
    GHZ9$^{\rm p}$ & 10.15 & 8.52\,-\,8.86 & $<106$ \\
    GN-z11$^{\rm q}$ & 10.60 & $8.9^{+0.2}_{-0.3}$ & $196\pm 12^*$ \\
    GHZ2$^{\rm r}$ & 12.34 & $9.05^{+0.10}_{-0.25}$ & 34\,-\,105 \\
    \hline
    \end{tabular}
    \begin{tablenotes}
%	    \centering
        \small
        \item $\bf Notes.$
        \item References: $\rm ^{a}$ \citet{berg_2016,james2009,Thuan_mrk996_1996}; $\rm ^{b}$ \citet{vanzella_sbarc_2022,Sharon_2022,pascale_sbarc_2023,welch_sba_2024};
        $\rm ^{c}$ \citet{Stiavelli_nloud_2024};
        $\rm ^{d}$ \citet{labbe_monster_2024}; 
        $\rm ^{e}$ \citet{Zhangyechi_2025};
        $\rm ^{f}$ \citet{Arellano-cordova_nloud_2024};
        $\rm ^{g}$ \citet{Vanzella_2010,ubler2023a,ji2024};
        $\rm ^{h}$ \citet{cameron_gs9422_2024,tacchella2024};
        $\rm ^{i}$ \citet{topping2024};
        $\rm ^{j}$ \citet{isobe2023};
        $\rm ^{k}$ \citet{Topping_2025};
        $\rm ^{l}$ \citet{Navarro-Carrera_nloud_2024};
        $\rm ^{m}$ \citet{isobe2023,larson_ceersagn_2023};
        $\rm ^{n}$ \citet{schaerer_gnz9_2024};
        $\rm ^{o}$ \citet{curti_gsz9_2024};
        $\rm ^{p}$ \citet{napolitano_ghz9agn_2024};
        $\rm ^{q}$ \citet{cameron2023,tacchella2023,maiolino_nature_gnz11_2024,alvarez-marquez_gnz11_2024};
        $\rm ^{r}$ \citet{castellano2024,calabro_ghz2_2024}.
        \item $\rm ^{\dagger}$ Values corresponding to a star cluster or an \hii\ region/association rather than the whole galaxy.
        \item $\rm ^{*}$ With an additional point-source like component.
    \end{tablenotes}
    \label{tab:global}
\end{table}

\begin{table*}
        \centering
        \caption{Derived nebular properties for a compilation of \somename from the literature at $0\lesssim z \lesssim 12$ investigated in this work.}
        \label{tab:nemitters_measurements}
        \begin{tabular}{l | c c c c c c c}
                \hline
                ID & $T_{\rm e}~({\rm 10^4~K})$ & $n_{\rm e}~{\rm (cm^{-3})}$ & $\rm 12+log(O/H)$ & $\rm \log(N/O)$ & $\rm \log(C/O)$ & $\rm \log(Fe/O)$ & He/H \\
                \hline
                Mrk 996 (high density)$^{\rm a}$ & 1 & $10^7$  & $8.38^{+0.07}_{-0.08}$  & $-0.13\pm 0.28$
                & - & - & $0.072\pm 0.018$ \\
                Mrk 996 (low density)$^{\rm a}$ & & $170 \pm 40$ & $8.36^{+0.17}_{-0.28}$ & $-1.43\pm 0.14$
                & $-0.22\pm 0.25$ & $\bf -1.60^{+0.22~\dagger}_{-0.33}$ & $0.091\pm 0.017$ \\
                \hline
                LyC (high density)$^{\rm b}$ & - & $10^5$ & $8.03\pm 0.06$ & $-0.21^{+0.10}_{-0.11}$
                & $-0.51\pm 0.05$ & - & - \\
                LyC (low density)$^{\rm b}$ & & $10^{3-4}$ & $8.03\pm 0.06$ & $-1.31\pm 0.12$
                & - & $-1.7\pm 0.2$ & $0.08\pm 0.02$ \\
                \hline
                ID150880$^{\rm c}$ & 1.49 & $203^{+500}_{-203}$ & $8.00^{+0.10}_{-0.08}$ & $-0.88^{+0.15}_{-0.16}$
                & - & - & - \\
                \hline
                UNCOVER-45924$^{\rm d}$ & $\bf 2.6^{+0.5~\dagger}_{-0.4}$ & $10^{2-4~*}$ & $\bf 7.15$\,-\,$\bf 7.67^{\dagger}$ & $\bf 0.20^{+0.06~\dagger}_{-0.05}$ & {$\bf -0.06^{+0.06~\dagger}_{-0.05}$} & - & - \\
                \hline
                ID1665$^{\rm c}$ & 1.2 & $204^{+390}_{-204}$ & $8.26^{+0.15}_{-0.11}$ & $-0.93\pm 0.15$
                & - & - & - \\
                \hline
                ID1746$^{\rm c}$ & 1.59 & $790^{+2300}_{-590}$ & $7.92^{+0.13}_{-0.10}$ & $-0.85^{+0.15}_{-0.14}$
                & - & - & - \\
                \hline
                ID1477$^{\rm c}$ & 1.84 & $300^*$ & $7.72^{+0.09}_{-0.07}$ & $-0.62^{+0.11}_{-0.10}$
                & - & - & - \\
                \hline
                ID60001$^{\rm e}$ & $1.61\pm 0.04$ & $350\pm 180$ & $7.75\pm 0.03$ & $-0.76\pm 0.03$ & - & - & $\bf 0.12\pm 0.02^{\dagger}$ \\
                \hline
                EXCELS-121806$^{\rm f}$ & $1.49\pm 0.11$ & $870\pm 450$ & $7.97^{+0.05}_{-0.04}$ & $-0.86^{+0.15}_{-0.11}$
                & $-1.02\pm 0.22$ & - & - \\
                \hline
                GS\_3073 (high density)$^{\rm g}$ & $1.45\pm 0.13$ & $5\pm 1\times 10^5$ & $8.00^{+0.12}_{-0.09}$ & $0.42^{+0.13}_{-0.10}$ & $-0.38^{+0.13}_{-0.11}$ & $>-1.84$\,-\,$-1.24$ & $\bf 0.07\pm 0.01^{\dagger}$ \\
                GS\_3073 (low density)$^{\rm g}$ & & $10^{2-4}$ & $8.00^{+0.12}_{-0.09}$ & $-1.10^{+0.18}_{-0.20}$ 
                & - & $>-1.84$\,-\,$-1.24$  & $\bf 0.10\pm 0.02^{\dagger}$ \\
                \hline
                GS\_9422 (tentative)$^{\rm h}$ & $1.83\pm 0.15$ & $\lesssim 10^3$ & $7.59\pm 0.01$ & $<-0.85$ & $-0.73\pm 0.03$ & $-1.03^{+0.15}_{-0.20}$ & $0.11\pm 0.01$ \\
                \hline
                ID397$^{\rm c}$ & 1.43 & $300^*$ & $7.96^{+0.10}_{-0.08}$ & $-0.67^{+0.14}_{-0.13}$
                & - & - & - \\
                \hline
                RXJ2248-ID$^{\rm i}$ & $2.46\pm 0.26$ & 0.64\,-\,$3.1\times 10^5$ & $7.43_{-0.09}^{+0.17}$ & $-0.39_{-0.08}^{+0.10}$
                & $-0.83_{-0.10}^{+0.11}$ & - & $0.166^{+0.018}_{-0.014}$ \\
                \hline
                GLASS\_150008$^{\rm j}$ & $1.93\pm 0.12$ & $\lesssim 10^4$ & $7.65_{-0.08}^{+0.14}$ & $-0.40_{-0.07}^{+0.05}$
                & $-1.08_{-0.14}^{+0.06}$ & - & $0.142\pm 0.054$ \\
                \hline
                A1703-zd6$^{\rm k}$ & $2.30\pm 0.32$ & $9.4\pm 4.2\times 10^4$ & $7.47\pm 0.19$ & $-0.6\pm 0.3$
                & $-0.74\pm 0.18$ & - & $\bf 0.082\pm0.017^\dagger$ \\
                \hline
                GN-z8-LAE$^{\rm l}$ & $1.73\pm 0.24$ & $\lesssim 10^4$ & $7.85\pm 0.17$ & $-0.44\pm 0.36$
                & $-0.69\pm 0.21$ & - & - \\
                \hline
                CEERS\_01019 (AGN)$\rm ^{m}$ & $1.5\pm 0.5^{*}$ &   $1.1^{+3.3}_{-1.0}\times 10^3$ & $8.37_{-0.14}^{+0.13}$ & $>-0.01$
                & $<-0.36$ & - & - \\
                CEERS\_01019 (SF)$\rm ^{m}$ & & $1.1^{+3.3}_{-1.0}\times 10^3$ & $7.94^{+0.46}_{-0.31}$ & $>0.28$
                & $<-1.04$  & -  & - \\
                \hline
                GN-z9p4$^{\rm n}$ & $2.30\pm 0.37$ & $100^*$ & $7.38\pm 0.15$ & $-0.59\pm 0.24$
                & $<-1.18$ & - & - \\
                \hline
                GS-z9-0$^{\rm o,~s}$ & $2.2\pm 0.2$ & $650\pm 430$ & $7.49_{-0.15}^{+0.11}$ & $-0.93\pm 0.37$ & $-0.90\pm 0.26$ & $-1.33^{+0.37}_{-0.85}$ & - \\
                \hline
                GHZ9$^{\rm p}$ & - & $100^*$ & 6.69\,-\,7.69  & $-0.08$\,-\,$0.12$
                & $-0.96$\,-\,$-0.45$ & - & - \\
                \hline
                GN-z11$^{\rm q}$ & $1.36\pm 0.13$ & $100^*$ & $7.91\pm 0.07$ & $>-0.25$
                & $>-0.78$ & $-0.86^{+0.22}_{-0.43}$ & $\bf 0.073^{+0.046~\dagger}_{-0.029}$ \\
                \hline
                GHZ2$^{\rm r,~s}$ & $2.12^{+0.27}_{-0.24}$ & $10^{3-5}$ & $7.44_{-0.24}^{+0.26}$ & $-0.25\pm 0.05$
                & $-0.74\pm 0.20$ & $-1.50^{+0.83}_{-1.00}$ & - \\
                \hline
        \end{tabular}
        \begin{tablenotes}
%	    \centering
        \small
        \item $\bf Notes.$
        \item References: $\rm ^{a}$ \citet{berg_2016,james2009,Marques-Chaves_2024}; $\rm ^{b}$ \citet{vanzella_sbarc_2022,pascale_sbarc_2023,welch_sba_2024}; 
        $\rm ^{c}$ \citet{Stiavelli_nloud_2024}; 
        $\rm ^{d}$ \citet{labbe_monster_2024}; 
        $\rm ^{e}$ \citet{Zhangyechi_2025}; 
        $\rm ^{f}$ \citet{Arellano-cordova_nloud_2024}; 
        $\rm ^{g}$ \citet{ubler2023a,ji2024};
        $\rm ^{h}$ \citet{cameron_gs9422_2024,tacchella2024};
        $\rm ^{i}$ \citet{topping2024};
        $\rm ^{j}$ \citet{isobe2023};
        $\rm ^{k}$ \citet{Topping_2025};
        $\rm ^{l}$ \citet{Navarro-Carrera_nloud_2024}; 
        $\rm ^{m}$ \citet{isobe2023,larson_ceersagn_2023};
        $\rm ^{n}$ \citet{schaerer_gnz9_2024};
        $\rm ^{o}$ \citet{curti_gsz9_2024};
        $\rm ^{p}$ \citet{napolitano_ghz9agn_2024};
        $\rm ^{q}$ \citet{cameron2023,tacchella2023,maiolino_nature_gnz11_2024,alvarez-marquez_gnz11_2024,ji_gnz11_2024,Nakane_gnz11_2024};
        $\rm ^{r}$ \citet{castellano2024,calabro_ghz2_2024,Zavala_ghz2_2024,Zavala_ghz2_2025};
        $\rm ^{s}$ \citet{Nakane_2025}.
        \item $\rm ^{\dagger}$ These values are newly derived in this work based on the reported fluxes in the literature. See Section~\ref{sec:method} for details.
        \item $\rm ^{*}$ These are assumed values as no observational constraints from narrow emission lines are available.
        \item All temperatures listed are for the high-ionization zone traced by $\rm O^{2+}$.
    \end{tablenotes}
\end{table*}

In this section, we describe the derivation of physical properties, mainly the chemical abundances, for the compiled sample.
The majority of the physical properties we used for our sample sources are directly from the literature listed in Tables~\ref{tab:global} and~\ref{tab:nemitters_measurements}.
We refer readers to the original sources for details of the derivation and summarize several key points relevant to our analysis below.

In Table~\ref{tab:global}, we summarize the global properties of our sample \somename including their redshift, stellar masses, and effective radii.
The stellar mass is reported for the whole galaxy.
For LyC of the Sunburst Arc, the stellar mass of this super star cluster alone is also reported.
For the effective radius, it is measured in different photometric bands for different systems.
For ID60001, the two effective radii correspond to two different components identified in F090W band of \jwst/NIRCam by \citet{Zhangyechi_2025}.
For CEERS\_01019 and GN-z11, there are both a point-source like component and an extended component and the effective radii correspond to the extended component.
For Mrk 996, \citet{Thuan_mrk996_1996} measured a scale length of 0.42 kpc for the exponential disk component excluding the central starburst region, corresponding to an effective radius of 0.71 kpc for the disk.
We also report the size of the central \hii\ region in Mrk 996 according to \citet{james2009}; for LyC, we report the upper limit on the size of the super star cluster described in \citet{pascale_sbarc_2023}.

In Table~\ref{tab:nemitters_measurements}, we report the derived chemical abundances including O/H, N/O, C/O, Fe/O, and He/H.
Not all sources have all these abundances measured or constrained due to the lack of required nebular emission lines in observations.
We emphasize that, similar to the work by \citet{schaerer_gnz9_2024}, we do not attempt to re-derive abundances in published works.
%However, we do comment on the methods to derive abundances adopted by different authors below.
We discuss the validity of previous abundance derivations in Section~\ref{sec:discuss}.
For some sources, emission line fluxes required for abundance derivations were measured but the abundances were not reported in the corresponding works.
We supplement derivations of these chemical abundances previously not reported for the following sources, Mrk 996, UNCOVER-45924, ID60001, GS\_3073, A1703-zd6, and GN-z11.
In what follows, we give a brief overview of the previous and new abundance derivations.
We further discuss details of specific abundance measurements in Sections~\ref{sec:chem_com} and \ref{sec:discuss}.

First of all, for 19 out of 22 sources in our sample \somename, the oxygen abundances are constrained by the so-called direct method, where the electron temperature, $T_{\rm e}$, is derived using the flux ratio of \oiii$\lambda 4363$/\oiii$\lambda 5007$ or \oiiip]$\lambda 1666$/\oiii$\lambda 5007$.
For LyC, while \citet{pascale_sbarc_2023} applied photoionization models instead of the direct $T_{\rm e}$ method to fit all abundances, they included measured fluxes of \oiiip]$\lambda 1666$ and \oiii$\lambda 5007$. Thus, we consider that the temperature is also well constrained in this case.
For CEERS\_01019, \citet{isobe2023} assumed a typical temperature of $(1.5\pm 0.5)\times 10^4$ K as no observational constraint is available.
For GHZ9, \citet{napolitano_ghz9agn_2024} estimated O/H indirectly based on the equivalent width of \ciii$\lambda \lambda 1906,1908$ as well as the strong emission line ratio of [Ne\,{\sc iii}]$\lambda 3869$/\oii$\lambda \lambda 3726,3729$, obtaining a range of 1 dex in O/H.
In principle, since \citet{napolitano_ghz9agn_2024} measured the fluxes of \oiiip]$\lambda 1666$ and \oiii$\lambda 4363$, the electron temperature can be estimated assuming no significant density variation is present (e.g., presence of a broad-line region).
However, since GHZ9 is an X-ray detected AGN, it is unclear whether the UV \oiiip]$\lambda 1666$ has contamination from the broad-line region (BLR).
Thus, we adopt the conservative range of metallicity estimated by \citet{napolitano_ghz9agn_2024}.
We note that for GN-z11 and GHZ2, although \oiii$\lambda 5007$ is redshifted out of the wavelength range of NIRSpec, this line is detected in the follow-up MIRI observations and is combined with \oiii$\lambda 4363$ or \oiiip]$\lambda 1666$ to provide temperature estimates \citep{alvarez-marquez_gnz11_2024,calabro_ghz2_2024}. 
With the derived temperatures, abundance ratios of N/O, C/O, Fe/O, and He/H are subsequently derived using flux ratios of relevant strong emission lines from the rest-frame UV to the optical and the calculated emissivities.

Next, we comment on the abundance derivations for specific sources.
{For Mrk 996, LyC, and GS\_3073, we report two abundance patterns for each source, since at least two nebular components with different gas densities have been found in observations \citep{james2009,pascale_sbarc_2023,ubler2023a,ji2024}. 
Notably, the derived N/O is always higher for the high-density component compared to the low-density component in these sources.
The existence of high-density regions is also reported for RXJ2248-ID, A1703-zd6, and GN-z11 \citep{topping2024,Topping_2025,maiolino_nature_gnz11_2024}.
For GN-z11, there is a component with a extremely high density reaching $n_{\rm e}\sim 10^{10}~{\rm cm^{-3}}$ and potentially tracking the BLR of an AGN (\citealp{maiolino_nature_gnz11_2024}; Maiolino et al. in prep.). The fiducial abundances reported by \citet{cameron2023} and \citet{alvarez-marquez_gnz11_2024} were derived by assuming observed metal lines come from a low-density region with $n_{\rm e} = 100~{\rm cm^{-3}}$. Regardless, the contribution from a high density region would likely only strengthen the upper limit on the derived N/O based on UV emission lines \citep[see e.g.,][]{ji2024}.
The density effect is further discussed in Section~\ref{sec:discuss}.
%Finally, for some sources, emission line fluxes required for abundance derivations were measured but the abundances were not reported in the corresponding works.
}

For Mrk 996, \citet{james2009} did not calculate the abundance of Fe, although \feiii$\lambda 4658$ is significantly detected in the VLT/VIMOS spectrum.
Following the approach adopted by \citet{Thuan_fe_1995} and \citet{izotov2006}, we calculated $\rm Fe^{2+}/H^+$ using
\begin{equation}
    \rm 12+\log(\frac{Fe^{2+}}{H^+})=\log(\frac{[FeIII]\lambda 4658}{H\beta})+6.498+\frac{1.298}{t}-0.498\log(t),
\end{equation}
where $t\equiv T_{\rm e}/10^4~{\rm K}$ and we assumed $t\approx 1$ to be consistent with the analysis of \citet{james2009}.
We obtained $\rm 12 +\log (Fe^{2+}/H^+) = 6.31\pm 0.02$.
The ionization correction factor (ICF) for $\rm Fe^{2+}$ to account for ionization species not observed is given by
\begin{equation}
    \rm ICF(Fe^{2+})=-1.377\nu + 1.606+1.045/\nu,
\end{equation}
where $\rm \nu \equiv O^+/(O^++O^{2+})$.
The above equation is applicable when $\rm 12+\log(O/H)>8.2$, which is the case for Mrk 996 based on the measurements of \citet{james2009}.
According to \citet{james2009}, $\nu=0.54\pm 0.19$, and thus $\rm ICF(Fe^{2+})= 2.8^{+1.3}_{-0.8}$.
As a result, $\rm 12+\log(Fe/H)=6.76_{-0.14}^{+0.17}$ and $\rm \log(Fe/O)=-1.60^{+0.22}_{-0.33}$.
We note that this calculation only applies to the low-density component identified by \citet{james2009}, as \feiii$\lambda 4658$ is only observed in this component in Mrk 996.

For UNCOVER-45924, \citet{labbe_monster_2024} provide flux measurements for emission lines from the \jwst micro shutter assembly \citep[MSA,][]{ferruit2022} PRISM spectrum (with a nominal spectral resolution of $R\sim 100$) but do not derive the chemical abundances.
We derived the abundances of O, N, C, and He using fluxes reported in Tables 2, 4, and 5 of \citet{labbe_monster_2024}.
For the oxygen abundance, we first estimated the electron temperature for the high-ionization zone from \oiii$\lambda 4363$/\oiii$\lambda 5007$.
We used the \textsc{Python} package \pyneb \citep{luridiana2015} with the atomic database of CHIANTI \citep[v10,][]{chianti0,chianti1} and \citet{oiii_as_tz17}.
We also corrected the line ratio using the Balmer decrement assuming an average SMC bar extinction curve of \citet{gordon2003} and an intrinsic \ha/\hb\ of 3.1 typical for the narrow-line region (NLR) of AGN \citep{agn3_2006}.
At the low density limit of \oiii$\lambda 5007$ ($n_{\rm e}<10^4~{\rm cm^{-3}}$), we derived $T_{\rm e}({\rm O^{2+}})=2.6^{+0.5}_{-0.4}\times 10^4$ K.
Using \oiii$\lambda 5007$/\hb, we obtained $\rm 12+\log(O^{2+}/H^+)=7.07\pm 0.20$.
For the temperature of the low ionization zone, we used \oii$\lambda \lambda 3726,3729$/\oii$\lambda \lambda 7320,7331$ also measured by \citet{labbe_monster_2024} and corrected for dust attenuation.
However, we found $T_{\rm e}({\rm O^{+}}) > 3\times10^4$ K, even higher than $T_{\rm e}({\rm O^{2+}})$, which is not expected for NLRs of AGN \citep{dors2020}.
A potential explanation is the low-density approximation no longer holds in this case, as \oii$\lambda \lambda 3726,3729$/\oii$\lambda \lambda 7320,7331$ is more sensitive to the density variation compared to \oiii$\lambda 4363$/\oiii$\lambda 5007$ due to the low critical density of \oii$\lambda \lambda 3726,3729$ ($\sim 10^4~{\rm cm^{-3}}$).
This would result in overestimation of $T_{\rm e}({\rm O^{+}})$.
Unfortunately, no density constraints are available for the NLR of UNCOVER-45924.
If we considered the low density limit and kept $T_{\rm e}({\rm O^{+}}) > 3\times10^4$ K, $\rm O^+$ would be negligible compared to $\rm O^{2+}$.
The ICF for $\rm O^{2+}+O^+$ is usually derived from $\rm \frac{He^{2+}+He^+}{He^+}$ \citep{Torres-Peimbert1977}.
However, \cite{labbe_monster_2024} argue the measured fluxes of \hei\ and \heii\ lines are dominated by the BLR.
As an alternative solution, we took an average ICF(O) of 1.2 from a sample of AGN NLRs analyzed by \citet{dors2020}, yielding $\rm 12+\log(O/H)=7.15\pm 0.20$.
If we assumed $n_{\rm e}=10^4~{\rm cm^{-3}}$, $T_{\rm e}({\rm O^{2+}})$ remains largely unchanged but $T_{\rm e}({\rm O^{+}})=1.14\pm 0.04 \times 10^4$ K, leading to $\rm 12+\log(O/H)=7.67\pm 0.10$.
We list both values in Table~\ref{tab:nemitters_measurements} and consider the range as a systematic uncertainty.
%We obtained $T_{\rm e}({\rm O^{+}})=1.6\pm 0.1\times 10^4$ K and derived $\rm 12+\log(O^{+}/H^+)=7.07\pm 0.20$ based on the dereddened \oii$\lambda \lambda 3726,3729$/\hb.
%Such a large temperature difference between high and low ionization zones are expected for NLRs of AGN \citep{dors2020}.

We also derived N/O and C/O for UNCOVER-45924. In the rest-frame optical, \nii$\lambda 6583$ is significantly blended with \ha\ in the PRISM spectrum and is a subdominant component within the blended line.
Thus, we did not use the \nii\ flux fitted by \citet{labbe_monster_2024}.
Instead, we used the UV lines of \niii$\lambda 1746$-$1754$ and \oiiip]$\lambda \lambda 1661,1666$.
While these lines can have both narrow and broad components and they cannot be separated in the PRISM spectrum, \citet{labbe_monster_2024} argue all UV lines are broad based on their spectral energy distribution (SED) fitting.
If we follow the interpretation of \citet{labbe_monster_2024}, the derivation of N/O would be subject to the complex physical conditions in the BLR \citep[e.g.,][]{temple2021}.
Due to the similar ionization potentials and critical densities, \niii/\oiiip] is a good tracer of N/O when the ionization condition and density are not extreme.
At densities of $n_{\rm e}\lesssim 10^9~{\rm cm^{-3}}$, the emissivity ratio of \niii/\oiiip] is insensitive to density.
At $n_{\rm e}> 10^{9}~{\rm cm^{-3}}$, the emissivity ratio of \niii/\oiiip] starts to drop with increasing density until $n_{\rm e}\sim 10^{11}~{\rm cm^{-3}}$.
Therefore, deriving N/O from \niii/\oiiip] at $n_{\rm e}< 10^{9}~{\rm cm^{-3}}$ gives a \textit{lower limit} (see, e.g., photoionizaiton models of \citealp{ji2024}).
Another difficulty is the temperature of the gas.
To solve these problems, we compared the measured \niii/\oiiip] with a set of \cloudy \citep[c17.03,][]{ferland2017} photonionization models similar to those computed by \citet{ji2024}, with $n_{\rm H}=10^9~{\rm cm^{-3}}$ and a range of ionization parameters ($U$)\footnote{The ionization parameter is defined as the ratio between the ionizing flux and the product of the hydrogen density and the speed of light.}, O/H, and N/O. 
Details of the models can be found in Appendix~\ref{appendix:monster}.
We fixed $\rm 12+\log(O/H)=7.41$, the average of the two values listed in Table~\ref{tab:nemitters_measurements}.
Thus, the only free parameters are $U$ and N/O, which are jointly constrained by \niv$\lambda 1486$/\niii$\lambda 1746$-1754 and \niii$\lambda 1746$-1754/\oiiip]$\lambda \lambda 1661,1666$.
We show the comparison between observed ratios and model predictions in Appendix~\ref{appendix:monster}. The best-fit value we obtained is $\rm \log(N/O)=0.20^{+0.06}_{-0.05}$.
Similarly, we derived the relative abundance of carbon for UNCOVER-45924 by using the flux ratio of \ciii$\lambda \lambda 1906,1908$/\oiiip]$\lambda \lambda 1661,1666$.
To constrain the ionization parameter, we tried both \niv$\lambda 1486$/\niii$\lambda 1746$-1754 and \civ$\lambda \lambda 1548,1551$/\ciii$\lambda \lambda 1906,1908$.
The ionization parameter constrained by \civ$\lambda \lambda 1548,1551$/\ciii$\lambda \lambda 1906,1908$ is roughly 0.5 dex lower than that constrained by \niv$\lambda 1486$/\niii$\lambda 1746$-1754 (see Appendix~\ref{appendix:monster}).
This could be due to either a highly stratified ionization structure (i.e., \niv\ probes a more highly ionized region compared to \civ) or resonant scattering of \civ$\lambda \lambda 1548,1551$.
Regardless, the two ionization parameters would only result in 0.1 dex difference in the derived C/O, with the lower-ionization case we obtained from \civ/\ciii\ predicting a lower C/O.
%\redtxt{The two ionization parameters result in 0.1 dex difference in the derived C/O, and thus we took the average C/O and added a 0.05 dex systematic uncertainty quadratically to the measurement uncertainty of C/O.}

Finally, for ID60001, GS\_3073, A1703-zd6, and GN-z11, there are significant detections of \hei$\lambda 5876$ and \hei$\lambda 3889$ in their corresponding \jwst/NIRSpec spectra \citep{ubler2023a,ji2024,ji_gnz11_2024,bunker2023,maiolino_nature_gnz11_2024,Zhangyechi_2025}, but no derivation of He/H has been carried out thus far.
Among all optical transitions of \hei, \hei$\lambda 5876$ is usually the strongest one and is relatively insensitive to density variations and radiative transfer effects \citep[e.g.,][]{Benjamin_he1_2002,Olive_he1_2004,Aver_he1_2015}.
This makes \hei$\lambda 5876$ a good tracer of $\rm He^+/H^+$ once combined with a close-by hydrogen recombination line such as \hb\footnote{Here we do not consider stellar absorptions underlying these lines \citep[e.g.,][]{Olive_he1_2004,Hsyu_he_2020,yanagisawa_he_2024}, which should be modest for the high-$z$ SF galaxies and have a much smaller impact compared to measurement uncertainties and systematic uncertainties we discussed.}.
For ID60001, we use the measured ratio of \hei$\lambda 5876$/\hb\ and \heii$\lambda 4686$/\hb\ to calculate $\rm He^+/H^+$ and $\rm He^{2+}/H^+$ using \pyneb.
We assumed $T_{\rm e}= (1.61\pm 0.04)\times 10^4$ K and $n_{\rm e}=345^{+184}_{-172}~{\rm cm^{-3}}$ as derived by \citet{Zhangyechi_2025}.
While the $T_{\rm e}$ was derived for $\rm O^{2+}$, which lies in an ionization zone in between $\rm He^{+}$ and $\rm He^{2+}$, the contribution from $\rm He^{2+}/H^+$ to the He/H is very small ($\sim 2\%$ of $\rm He^{+}/H^+$) and $\rm T_{\rm e}(He^+)/T_{\rm e}(O^{2+})$ is usually close to 1 (in this case $\sim 0.99$ based on the average relation computed with photoionization models for \hii\ regions by \citealp{Izotov_he_2013}).
Since $\rm He^0$ might also coexist with $\rm H^+$, one also needs to estimate $\rm He^0/H^+$. However, the abundance ratio of $\rm He^0/(He^++He^{2+})$ is usually less than $2\%$ especially for ionized gas with high ionization parameters \citep{Izotov_he_2014}.
Therefore, we assume the contribution from $\rm He^0$ is negligible.
The above analysis leads to a final helium abundance of $\rm He/H=0.12\pm 0.02$, higher than the solar reference of 0.084 but with a relatively large uncertainty.
Such an enhancement of helium has also been found in a few other \somename as derived by \citet{yanagisawa_he_2024}, although density variation could in principle bias the result, which we further discuss in Section~\ref{subsec:helium}.
Similarly, we derived He/H for A1703-zd6 using the measurements of \hei$\lambda 5876$/\hb, $T_{\rm e}({\rm O^{2+}})$, and $n_{\rm e}$ reported by \citet{Topping_2025}.

For GS\_3073, we used the narrow component of \hei$\lambda 5876$, \heii$\lambda 4686$, and \hb\ measured from the NIRSpec/IFU high-resolution grating spectrum by \citet{ubler2023a} and used $T_{\rm e}=(1.45\pm 0.13)\times 10^4$ K derived by \citet{ji2024}.
Since all lines involved are permitted, we used both densities from the high-density component and low-density component in GS\_3073 to obtain a range of He/H, which is $\rm He/H\sim 0.07-0.10$.
For GN-z11, we used \hei$\lambda 3889$, \heii$\lambda 1640$, and $\rm H\delta$ and used $T_{\rm e}=(1.36\pm 0.13)\times 10^4$ K \citep{alvarez-marquez_gnz11_2024,ji_gnz11_2024}.
We assumed $n_{\rm e}=100~{\rm cm^{-3}}$ and obtained $\rm He/H=0.073^{+0.046}_{-0.029}$. Raising the density to $n_{\rm e}=1000~{\rm cm^{-3}}$ would lower He/H by 0.003.
One caveat of this approach is the radiative transfer of \hei$\lambda 3889$.
This is because the lower level of \hei$\lambda 3889$ is the ground level of the triplet state, which is metastable.
As a result, even low-density gas can become optically thick to \hei$\lambda 3889$.
The value of He/H we derived for GN-z11 at the optically thin limit can thus be considered as a lower limit.

%\subsection{Measurements of other properties}

\section{Comparisons of chemical abundance patterns}
\label{sec:chem_com}

In this section, we compare chemical abundance patterns of \somename to those of local stars and galaxies.
It is worth noting that the chemical abundances for galaxies including \somename are gas-phase abundances traced by emission lines from ionized gas.
In contrast, the chemical abundances of local stars are obtained through observations of absorption lines in the stellar atmosphere (see Section~\ref{subsec:derive_starmet}).
{We emphasize that we make no assertion on what population of stars is illuminating the gas in the \somename (as possible proto-GCs).
Rather, we are inspecting whether the chemical composition of stars in the MW GCs matches that in the ISM of the \somename -- linking the chemical composition of stellar atmospheres to the gas-phase abundances measured in our sample of high-$z$ galaxies.}
Since carbon, oxygen, and iron are observed to be depleted onto dust grains in the ISM \citep[see e.g.,][]{jenkins2009}, one expects the gas-phase abundances of these metals to differ from the abundances in stars born later from the same ISM by certain depletion factors \citep{Gunasekera_2023}.
Among the elements we investigated, nitrogen and helium do not deplete.
For oxygen and carbon, the relative depletions are likely modest, with a typical value of 0.1-0.2 dex in the MW \citep[][although carbon depletion has a large uncertainty]{jenkins2009}.
The element that is most affected by dust depletion is Fe, where the depletion is usually larger than 1 dex along MW sightlines \citep[][and references therein]{jenkins2009}.
We discuss this effect in more detail in Section~\ref{subsec:iron}.
%We discuss the impact of dust depletion in the corresponding results in this section. \redtxt{[I could either add a depletion vector or move this to the discussion]}
%Still, we emphasize that the depletion factors, if assumed to be the typical values, do not impact our analyses as we describe later in this section.

\begin{figure*}
    \centering
    \includegraphics[width=0.85\textwidth]{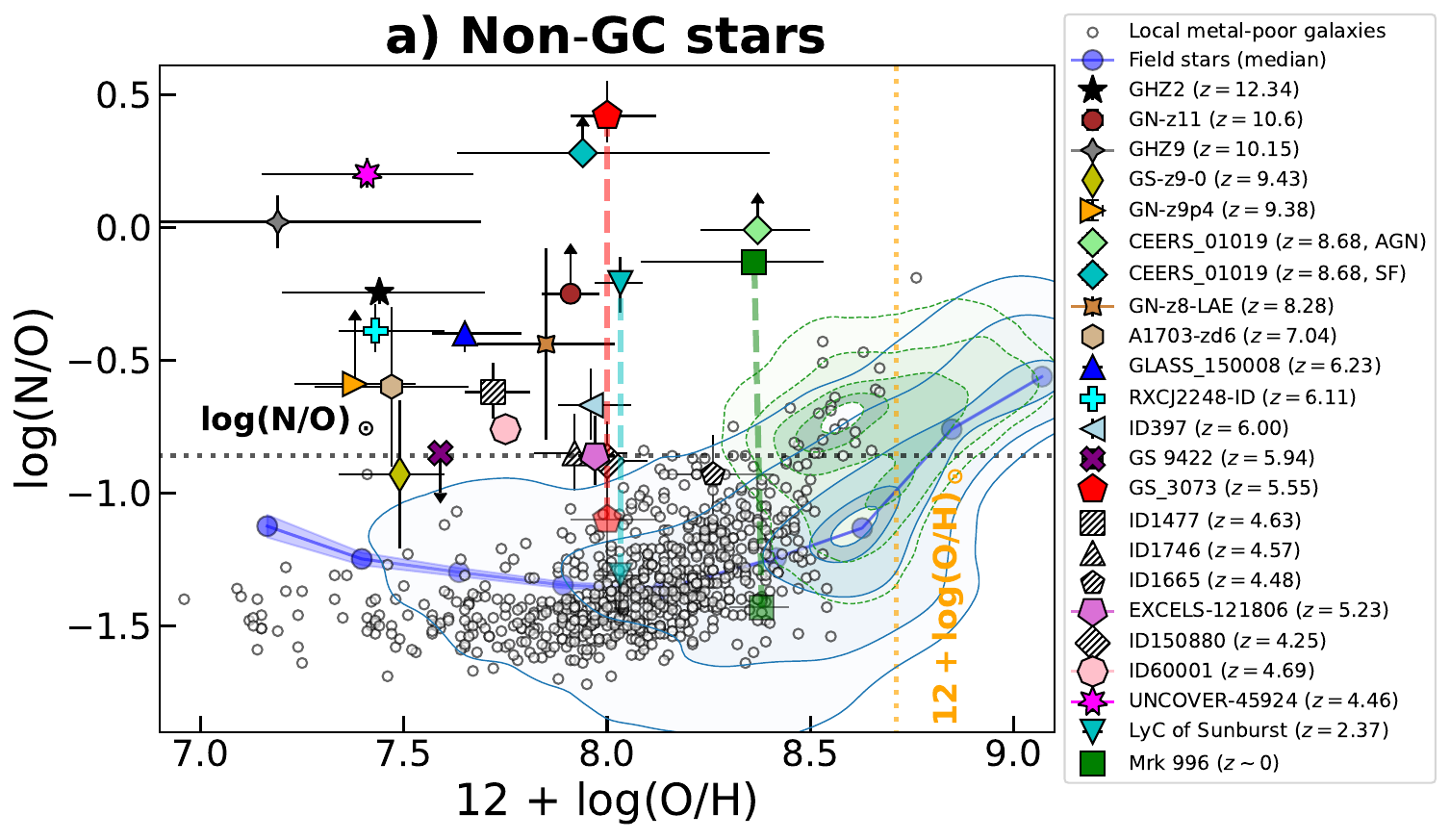}
    \includegraphics[width=\textwidth]{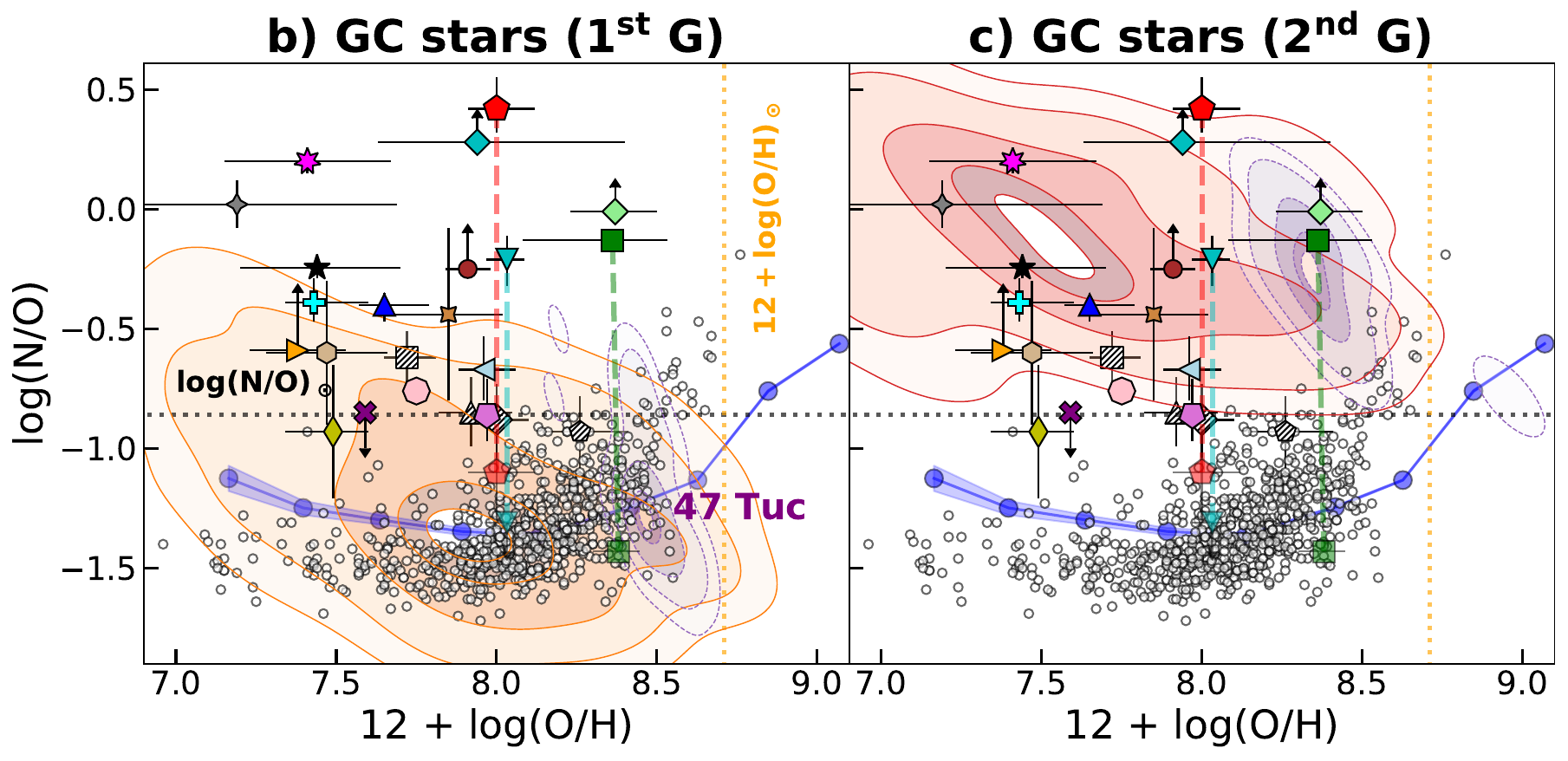}
    \caption{
    Distributions of \somename in the N/O versus O/H space in comparison with local galaxies and stars.
    Derived abundances for \somename come from \citet{castellano2024,cameron2023,cameron_gs9422_2024,curti_gsz9_2024,schaerer_gnz9_2024,isobe2023,larson_ceersagn_2023,topping2024,Topping_2025,ji2024,ubler2023a,pascale_sbarc_2023,james2009,Arellano-cordova_nloud_2024,napolitano_ghz9agn_2024,Navarro-Carrera_nloud_2024,Stiavelli_nloud_2024,Zhangyechi_2025} and also this work.
    Systems showing two abundances connected by dashed lines have their higher abundances derived for high-density components ($n_{\rm e}\gtrsim 10^5~{\rm cm^{-3}}$) decomposed from the observed spectra, and their lower abundances derived for low-density components ($n_{\rm e}< 10^4~{\rm cm^{-3}}$).
    Derived abundances for local galaxies come from \citet{izotov2006,pilyugin2012,Annibali_2019,Grossi_2025}.
    The contours correspond to probability distributions of abundances of MW stars compiled from \citet{sdss_dr17} computed by the kernel density estimation function \textsc{kdeplot} from the \textsc{python} package \textsc{seaborn}. Five contour levels are plotted corresponding to 5, 16, 50, 84, and 95 percentiles of the distributions, respectively.
    %\redtxt{The colored density distributions correspond to MW stars with non-GC origins and GC origins. In the top panel, the density distribution of non-GC stars are normalized by the number of stars in individual $\rm 12+\log(O/H)$ bins to show the tail of the distribution towards low metallicities. In the bottom panel, the distributions of GC stars are relatively uniform at low metallicities and thus are not normalized in $\rm 12+\log(O/H)$ bins.}
    \textit{Top:} abundance patterns of non-GC stars in the MW, where stars with $\log g<1.5$, $\log g>3$, or $T_{\rm eff}>5300$ K are excluded.
    Stars with azimuthal velocities of $v_{\rm t} \geq 150~{\rm km~s^{-1}}$ trace the thin disk and have an overall higher N/O, which are plotted as green contours with dashed boundaries.
    {The cyan shaded region represents the median trend of MW field stars \redtxt{with no error cut in abundances (selected with $\rm 1.5< \log g < 3$, $T_{\rm eff}<5300$ K, and $v_{\rm t}<150$ \kms)} with $1\sigma$ median uncertainties.}
    \textit{Bottom:} abundance patterns of GC stars {(excluding NSC stars)} in the MW divided into $\rm 1^{st}$ and $\rm 2^{nd}$ generations.
    The abundance patterns of stars in the GC 47 Tuc are plotted as purple contours with dashed boundaries.
    Overall, the abundances of \somename are more consistent with those of GC stars, among which the $\rm 2^{nd}$-generation GC stars better overlap with \somename at $\rm \log(N/O)>-0.5$.
    }
    \label{fig:no_com}
\end{figure*}

\begin{figure*}
    \centering
    \includegraphics[width=0.85\textwidth]{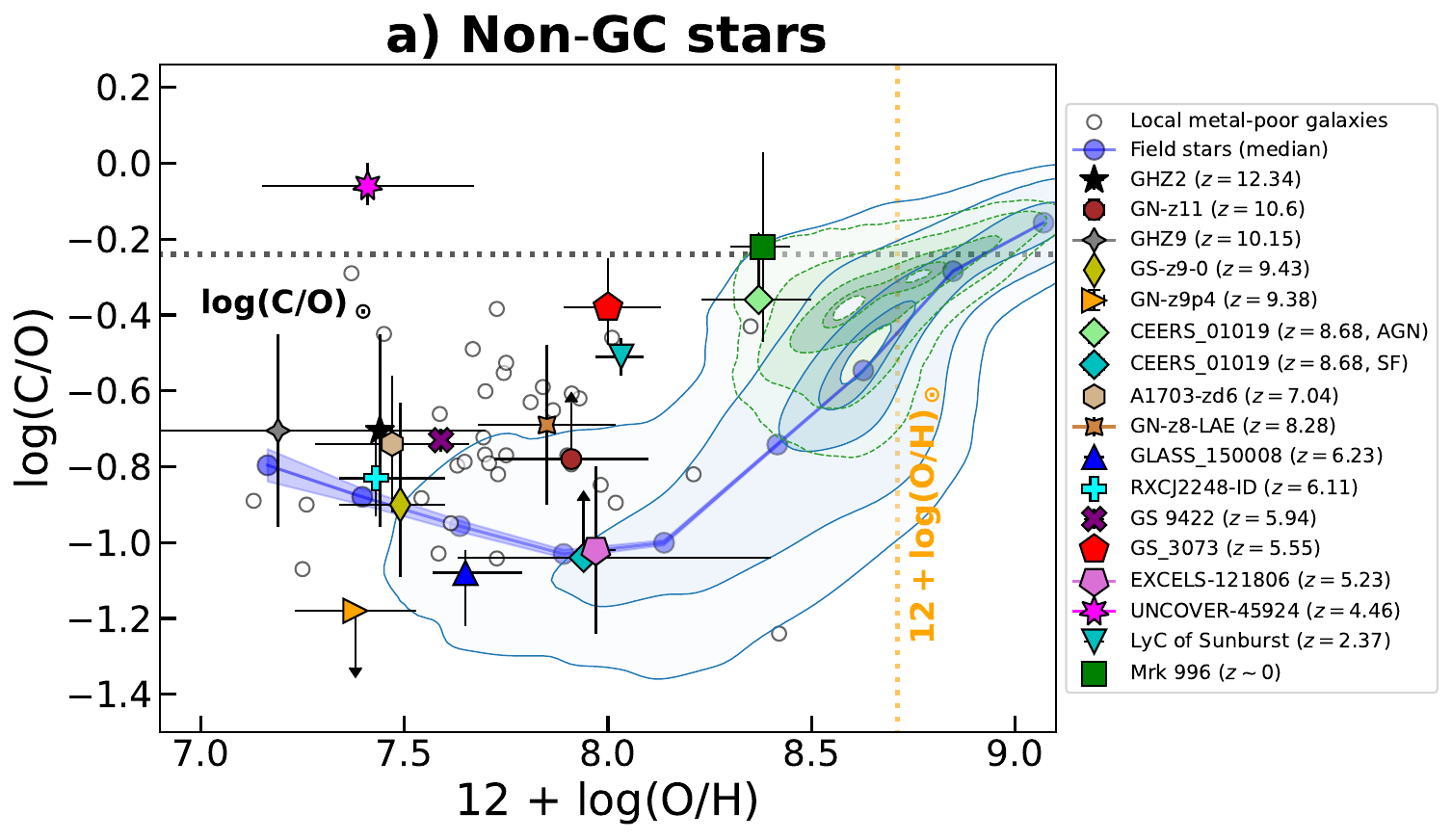}
    \includegraphics[width=\textwidth]{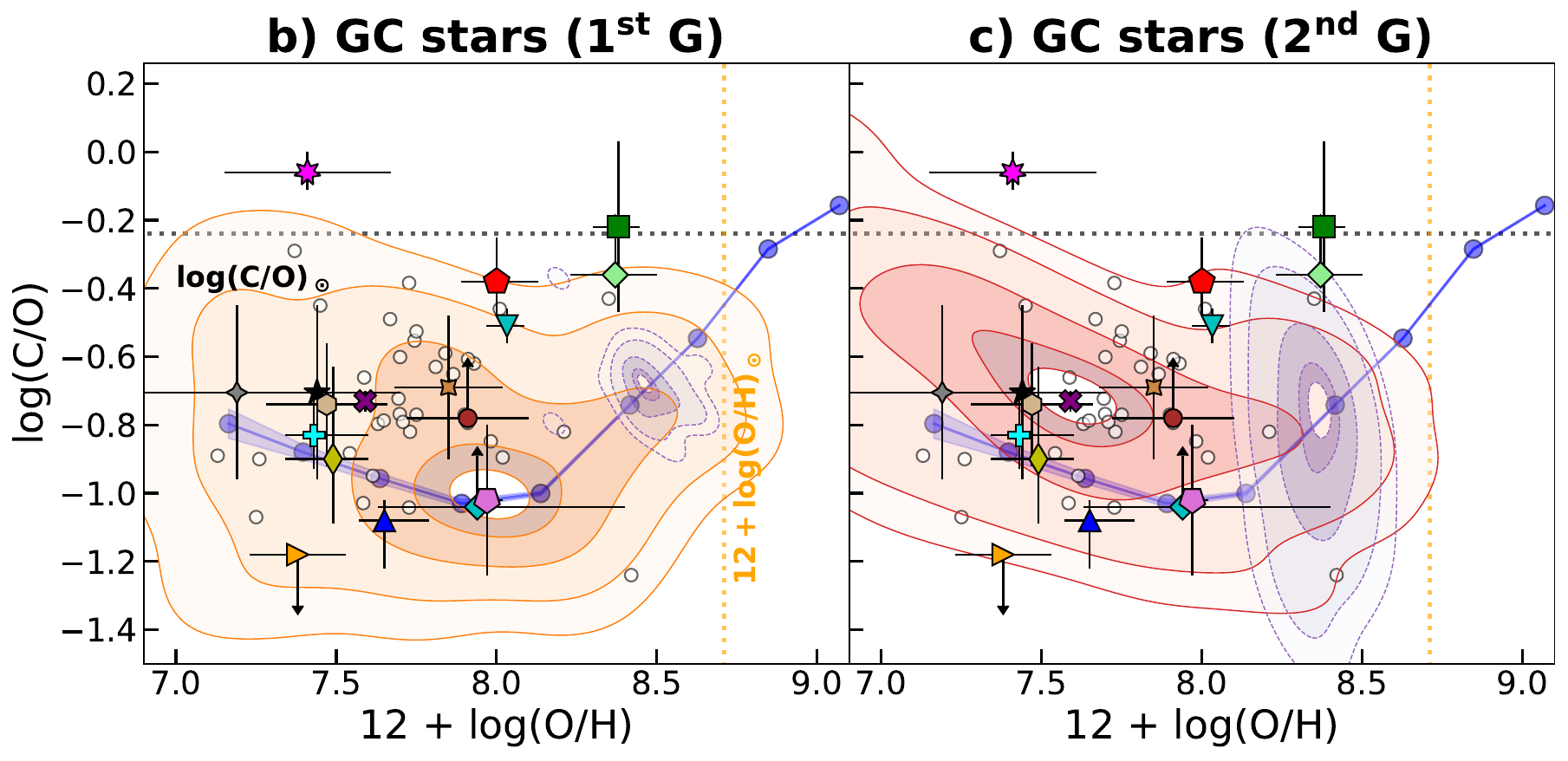}
    \caption{
    Distributions of \somename in the C/O versus O/H space in comparison with local galaxies and stars.
    Derived abundances for \somename come from \citet{castellano2024,cameron2023,cameron_gs9422_2024,curti_gsz9_2024,schaerer_gnz9_2024,isobe2023,larson_ceersagn_2023,topping2024,Topping_2025,ji2024,ubler2023a,pascale_sbarc_2023,Marques-Chaves_2024,Arellano-cordova_nloud_2024,Navarro-Carrera_nloud_2024,napolitano_ghz9agn_2024} and also this work.
    Derived abundances for local galaxies come from \citet{berg_2016,berg2019}.
    The contours correspond to probability distributions of abundances of MW stars compiled from \citet{sdss_dr17} computed by the kernel density estimation function \textsc{kdeplot} from the \textsc{python} package \textsc{seaborn}. 
    Five contour levels are plotted corresponding to 5, 16, 50, 84, and 95 percentiles of the distributions, respectively.
    \textit{Top:} abundance patterns of non-GC stars in the MW, where stars lying on the thin disk and stars with $\log g<1.5$, $\log g>3$, or $T_{\rm eff}>5300$ K are excluded.
    Stars with azimuthal velocities of $v_{\rm t} \geq 150~{\rm km~s^{-1}}$ trace the thin disk and have an overall higher C/O, which are plotted as green contours with dashed boundaries.
    {The cyan shaded region represents the median trend of MW field stars with $1\sigma$ median uncertainties.}
    \textit{Bottom:} abundance patterns of GC stars {(excluding NSC stars)} in the MW divided into $\rm 1^{st}$ and $\rm 2^{nd}$ generations.
    While \somename appear to overlap with the low-metallicity tail of the non-GC stars, they better overlap with GC stars.}
    \label{fig:co_com}
\end{figure*}

\begin{figure*}
    \centering
    \includegraphics[width=0.85\textwidth]{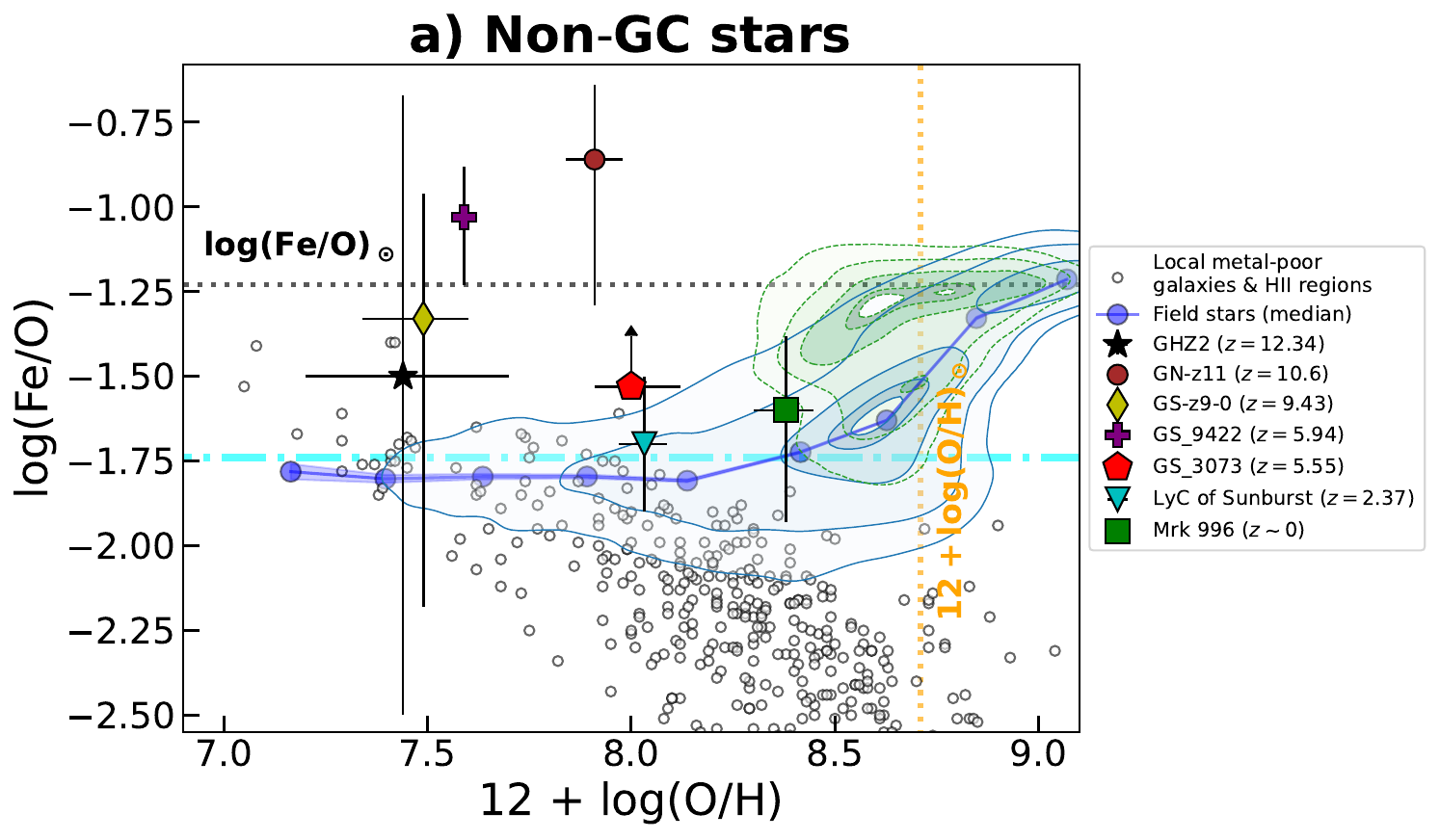}
    \includegraphics[width=\textwidth]{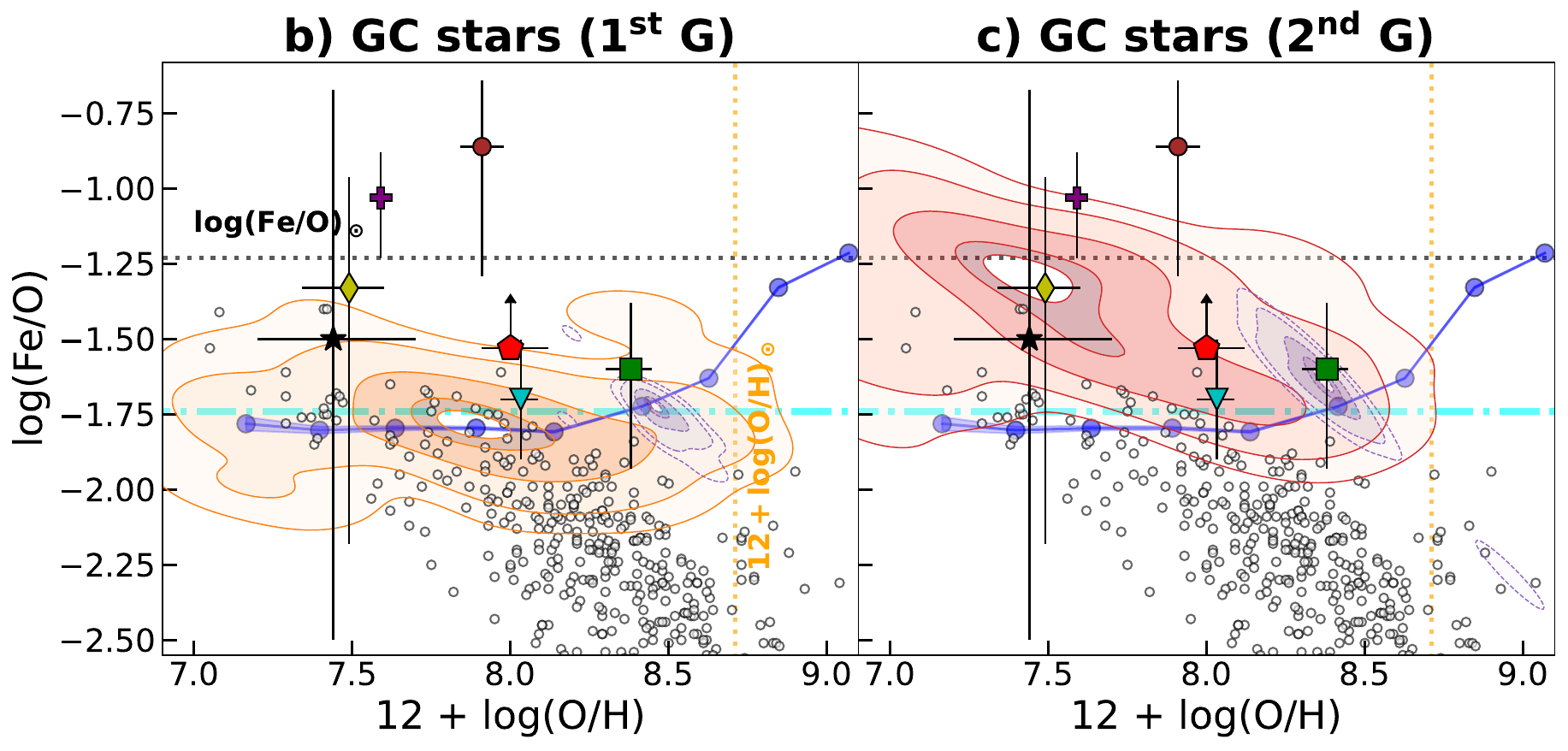}
    \caption{
    Distributions of \somename in the Fe/O versus O/H space in comparison with local galaxies and stars.
    Derived abundances for \somename come from \citet{cameron2023,cameron_gs9422_2024,Nakane_gnz11_2024,Nakane_2025,ji_gnz11_2024,ji2024,ubler2023a,welch_sba_2024} and also this work.
    Derived abundances for local galaxies and \hii\ regions from \citet{mendezdelgado_feo_2024} are plotted as open circles, which reach $\rm log(Fe/O)\sim -3$ outside the range shown in the current figure.
    The horizontal dash-dotted cyan line is the asymptotic line of Fe/O for local galaxies and \hii\ regions derived by \citet{mendezdelgado_feo_2024}.
    %The local \hii\ regions of \citet{mendezdelgado_feo_2024} has Fe/O systematically lower than the galaxies of \citet{izotov2006} due to different ways of calculating the ICF for $\rm Fe^{2+}$.
    The contours correspond to probability distributions of abundances of GC stars compiled from \citet{sdss_dr17} computed by the kernel density estimation function \textsc{kdeplot} from the \textsc{Python} package \textsc{seaborn}. Five contour levels are plotted corresponding to 5, 16, 50, 84, and 95 percentiles of the distributions, respectively.
    \textit{Top:} abundance patterns of non-GC stars in the MW, where stars lying on the thin disk and stars with $\log g<1.5$, $\log g>3$, or $T_{\rm eff}>5300$ K are excluded.
    Stars with azimuthal velocities of $v_{\rm t} \geq 150~{\rm km~s^{-1}}$ trace the thin disk and have an overall higher Fe/O, which are plotted as green contours with dashed boundaries.
    {The cyan shaded region represents the median trend of MW field stars with $1\sigma$ median uncertainties.}
    \textit{Bottom:} abundance patterns of GC stars {(excluding NSC stars)} in the MW divided into $\rm 1^{st}$ and $\rm 2^{nd}$ generations.
    While some \somename appear to overlap with the low-metallicity tail of the non-GC stars, they also overlap with GC stars.
    %\redtxt{SM: after doing some digging, the secondary clump in present both 1G and 2G stars is associated with 47 Tuc.}
    }
    \label{fig:feo_com}
\end{figure*}

\begin{figure*}
    \centering
    \includegraphics[width=\textwidth]{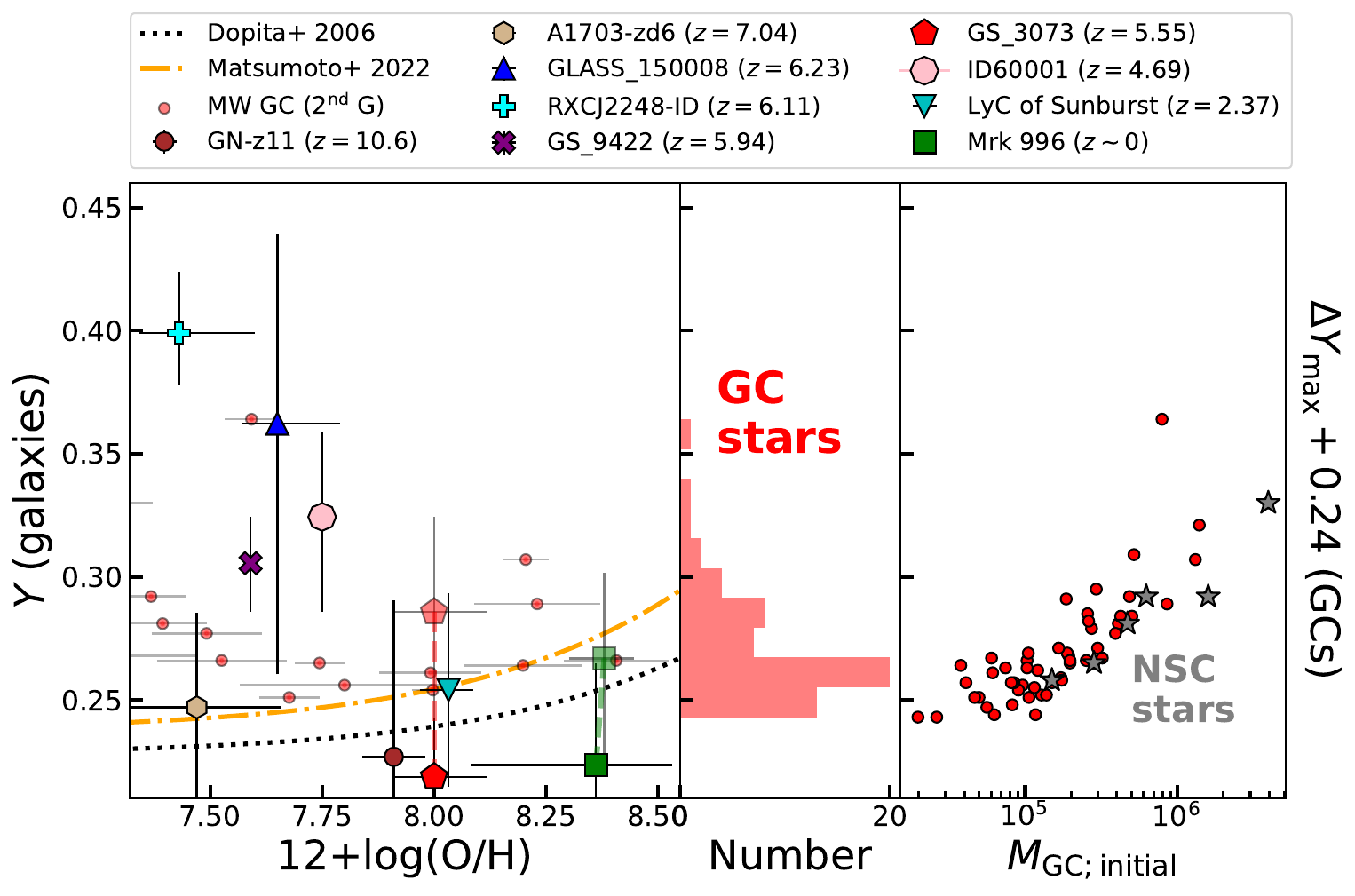}
    \caption{\textit{Left:} Distributions of \somename in the Y (i.e., mass fraction of He) versus O/H space in comparison with local galaxies and stars.
    Derived abundances for the LyC of the Sunburst Arc come from \citet{welch_sba_2024}.
    Derived abundances for GS\_9422, RXCJ2248-ID, and GLASS\_150008 come from \citet{yanagisawa_he_2024}.
    For GN-z11, A1703-zd6, GS\_3073, ID60001, and Mrk 996, we derived their He abundances based on the measurements of \hei\ and \heii\ emission lines.
    In comparison, we also plot 1) Y as a function of O/H for local galaxies fitted by \citet{dopita2006} and \citet{Matsumoto2022}, and 2) maximum Y enhancement in individual MW GCs as a function of O/H averaged over the 2G stars.
    \textit{Middle:} distributions of Y for a sample of local GC stars measured by \citet{milone_hegc_2018}.
    \textit{Right:} maximum enhancement in Y (see text) for a sample of local GC stars versus the inferred initial masses of the GCs measured by \citet{milone_hegc_2018}, where accreted NSCs of previous dwarf galaxies are marked as star symbols.
    The \somename show a tentative trend of decreasing Y with increasing metallicity, which might be connected to the masses of clusters formed in these systems or caused by the bias in deriving Y \redtxt{(see text for details)}.
    }
    \label{fig:heh_com}
\end{figure*}

\subsection{Nitrogen and oxygen abundance}
\label{subsec:nitrogen}

We start by inspecting the abundance patterns of nitrogen and oxygen.
While oxygen is mainly enriched by massive stars ($M_\star > 8~M_\odot$), there are different proposed channels for nitrogen enrichment.
At low metallicities [$\rm 12+\log(O/H)< 8$],
nitrogen can be enriched either by massive stars, leading to a nearly constant N/O with small scatters \citep[e.g.,][]{Matteucci_1986}, or by intermediate-mass fast-rotating stars ($4~M_\odot <M_\star < 7~M_\odot$) through rotational mixing and released by stellar winds, which would lead to a relatively large scatter in N/O \citep{Henry_2000,Meynet2002,Kobayashi_2020}.
%\redtxt{there are two proposed channels for }
%both nitrogen and oxygen are believed to be mainly produced by massive stars ($M_\star > 8~M_\odot$) and are released by CCSNe into the ISM (\citealp{Matteucci_1986}; although see \citealp{Kobayashi_2020} for a different picture of $4~M_\odot <M_\star < 7~M_\odot$ AGB stars that dominate the enrichment).
At high metallicities [$\rm 12+\log(O/H)> 8.2$], the enrichment of nitrogen in galaxies is mainly ascribed to winds from AGB stars with $2~M_\odot <M_\star < 8~M_\odot$ and is typically delayed by $\sim 0.1-1$ Gyr compared to the enrichment of oxygen \redtxt{for the global chemical evolution of galaxies} \citep[e.g.,][]{Henry_2000}. The above physical picture fits a plateau of subsolar N/O at low O/H and a nearly linear scaling between N/O and O/H at high O/H in observations of local galaxies, although the transitional O/H likely depends on the star formation efficiency \citep[SFE,][]{Vincenzo_2016}.
\redtxt{While the above picture applies to the chemical evolution of the whole galaxies, the chemical imprints in star clusters can be drastically different with enhanced N from massive AGB stars with a delay time as short as 40 Myr.}

In Figure~\ref{fig:no_com} we compare N/O and O/H between \somename (large colored symbols), MW stars (density contours), and local metal-poor galaxies (small open circles). 
Additionally, we distinguish MW stars currently in the field -- most of which originated in the disk (blue and green contours) -- from those observed in surviving Galactic globular clusters (orange and red contours). 
When interpreting these distributions, it is important to consider the different selection biases that affect the reported chemical measurements. 
In particular, the metallicity distributions (traced by O/H) for ionized gas in local SF galaxies and stars in the MW are shaped by distinct observational and physical factors. 
Two main challenges complicate the detection and chemical characterization of local low-metallicity galaxies. 
First, such systems are intrinsically rare at $z\sim0$ \citep{izotov2006,pilyugin2012}. Second, they are also faint, as lower-mass galaxies tend to be less chemically evolved \citep[i.e., the mass-metallicity relation,][]{Tremonti_2004,Andrews_2013}. 
Together, these effects lead to an under-representation of both low and high O/H values in the metallicity distribution of the local metal-poor galaxy sample (small open circles). 
These can be compared to stars in the Galactic disk, which trace a different environment: the MW -- a more massive ($\sim 2$ orders of magnitude higher than the metal-poor galaxies) and chemically evolved system, comparable in stellar mass to an $L_*$ galaxy. As shown in Figure~\ref{fig:no_com}, most MW stars occupy the high-O/H region of the diagram. In contrast, the metallicity distributions of Galactic GCs in panels b) and c) reflect the varying star formation histories of their progenitor systems, such as accreted dwarf galaxies and the early MW. Since the contribution of globular clusters to overall star formation declines sharply with increasing metallicity \citep[see][]{Belokurov2023}, their stars exhibit lower average O/H than both local galaxies and MW field stars.

Local SF galaxies from SDSS form a relatively tight sequence with roughly constant N/O of $\rm \log(N/O) \approx -1.5$ for low-metallicity systems with $\rm 12 + log(O/H) < 8.0$. At high metallicities, N/O of local galaxies scales nearly linearly with O/H.
Similar scaling relations between N/O and O/H have also been found in spatially resolved regions of galaxies on scales of $\sim 1$ kpc \citep{Belfiore_no_2017,Schaefer_no_2020}.
Meanwhile, it is clear from the top panel of Figure~\ref{fig:no_com} that stars in the MW exhibit a similar scaling relation, despite that the majority of the stars are more metal-rich compared to SDSS galaxies and thus trace mainly the secondary nitrogen source. 
{At lower metallicities, field halo stars with $\rm 12+\log(O/H) < 7.7$ show slightly elevated N/O ratios, with the largest deviations near $\rm 12+\log(O/H) \approx 7$. Several factors likely contribute to this trend in the Milky Way halo.
First, below $\rm 12+\log(O/H) < 7.7$, stars from the so-called \textit{Aurora} population, which formed during the early, pre-disk phase of MW evolution, begin to appear \citep[][]{belokurov2022,Kane2025}. Up to $\sim50\%$ of Aurora's star formation may have occurred in massive clusters, potentially boosting its overall N/O ratios \citep[][]{Belokurov2023}. Second, a selection bias affects nitrogen measurements at low metallicity. As nitrogen features weaken with declining abundance and opacity, detecting them requires high S/N spectra.
\redtxt{To mitigate the S/N effect on the median trend (mainly at low O/H) of MW field stars, we did not apply any cut in the measurement uncertainties of relevant elements when plotting.}
%Our APOGEE sample prioritizes small abundance uncertainties, effectively biasing low-metallicity selection toward stars with stronger N features. 
Finally, APOGEE reports nitrogen abundances under 1D LTE assumptions. These are likely overestimated at low metallicities due to uncorrected 3D non-LTE effects \citep[see e.g.,][]{Lind2024}. Note that despite these biases, it should still be possible to carry out a relative comparison of the abundances of the MW field and the GC stars.}

%Importantly, the tail of the stellar abundance distribution at low metallicities shows a good overlap with the asymptotic distribution of the ISM abundances in local galaxies. 
It is worth noting that different populations of the MW disk stars have distinct abundance offsets. To illustrate this, we made a rough cut based on the azimuthal velocity (in galacto-centric cylindrical polar coordinates), $v_{\rm t}$, where we assumed stars with $v_{\rm t}\geq 150$ \kms come from the so-called ``thin'' disk and stars with $v_{\rm t}< 150$ \kms come from the so-called ``thick'' disk. While the thin/thick disk distinction is a clear oversimplification of the MW's complex chromo-chemo-dynamical structure, we use it here as a rough proxy to separate metal-rich stars formed at different epochs and in different Galactic environments. 
Stars in the thin disk (dashed green contours) exhibit higher N/O at fixed O/H than those in the thick disk (solid contours). This is consistent with the general view that the thin disk is significantly younger, having formed in the outer Galaxy from lower-density gas and with lower star formation efficiency.
Since the gas-phase metallicity, as quantified by oxygen, is an $\alpha$ element, the higher N/O abundances in the thin disc also reflect a lower overall $\alpha$ abundance compared to the thick disk (hence, the ``$\alpha$-poor disc''; \citealp{Minchev2013, Haywood2013, Rix2013, Bensby2014,Hayden2015,Magrini2018}).
%As a result, the thin disk is also more $\alpha$-poor (and hence is higher in N/O) compared to the thick disk \citep[see e.g.,][]{Minchev2013, Haywood2013, Rix2013, Bensby2014,Hayden2015,Magrini2018}

Compared to local galaxies and MW disk stars, \somename, by their definition, have significantly higher N/O at the same (gas-phase) O/H. 
Such a difference is particularly apparent at $\rm 12+\log(O/H)<8.0$, where N/O in local galaxies and MW disk stars roughly forms a plateau.
The stars in MW GCs, on the other hand, clearly exhibit nitrogen enhancement compared to SDSS galaxies and MW disk stars.
In the bottom panel of Figure~\ref{fig:no_com}, one can see
%those GC stars classified as the 1G stars already show a trend of increasing N/O with decreasing O/H.
that while the 1G GC stars have an overall similar distribution in terms of N/O and O/H compared to the MW field stars, the 2G GC stars show nearly ubiquitous enhancement of nitrogen with $\rm (N/O)\gtrsim (N/O)_\odot$.
The 2G GC stars also show overall lower O/H compared to the 1G GC stars. 
These lower oxygen levels in the 2G stars become readily apparent when oxygen and iron levels are compared as we show later.
%(see Figure~\ref{fig:feo_com}). 
The offset in O/H within GCs is associated with the so-called Na-O anticorrelation, where the abundance of Na increases as the abundance of O decreases. The standard explanation involves depletion of O due to the \redtxt{CNO cycle and the production of Na from the} NeNa cycle in stellar nucleosynthesis at high temperatures.
In observations, the Na-O anticorrelation has also been observed in main-sequence turn-off stars and sub-giants in GCs \citep{Gratton_nao_2001}, which rules out the possibility that this anticorrelation is driven by internal nucleosynthesis in stars at later times as the required temperature ($T>3\times 10^7$ K) is much higher than what is expected for these stars.
Therefore, the lower oxygen abundances seen in the 2G GC stars is a reflection of the abundances of the ISM they formed in, which has been enriched in Na \redtxt{via the NeNa cycle} and depleted in O via the \redtxt{CNO} cycle which operated in the hot 1G stars.
%Therefore, the reduction in oxygen in the second-generation GC stars was probably in place when they were born, resulted from NeNa cycle in the hot stars in the first generation.

The oxygen-poor \somename ($\rm 12+log(O/H)< 8.0$) appear to trace the transition from the 1G GC stars to the 2G GC stars with the decreasing oxygen abundance.
In particular, three of the metal-poor yet nitrogen-rich \somename, GHZ2 ($z=12.34$, black star) GHZ9 ($z=10.15$, gray four-pointed star) and UNCOVER-45924 ($z=4.46$, magenta seven-pointed star) overlap with the dense part of the distribution of the 2G GC stars.
Still, we caution that both GHZ9 and UNCOVER-45924 host AGN and lack high-resolution spectra for key emission lines, and thus their oxygen abundances have large uncertainties.
For \somename, the sequence of increasing N/O at decreasing O/H is also noted, for example, by \citet{Zhangyechi_2025}.
This trend appears not exact if we include more metal-rich sources at $\rm 12+log(O/H)\geq 8.0$, especially CEERS\_01019 at $z=8.68$, GS\_3073 at $z=5.55$, LyC of the Sunburst Arc at $z=2.37$, and Mrk 996 at $z\sim 0$.
Their abundances might better match those of metal-enriched GCs.
Indeed, if we focus on a single metal-enriched GC, 47 Tuc (which is the most massive MW in-situ GC), the 2G GC stars roughly reproduce the N/O and O/H of these \somename.
Thus, it is possible that the N/O-O/H anticorrelation also exists in more metal-rich \somename, but the sequence for these \somename starts at higher O/H.
It should be noted that among these \somename, there are three systems showing two abundance patterns, which are GS\_3073 at $z=5.55$, the LyC of the Sunburst Arc at $z=2.37$, and Mrk 996 at $z\sim 0$.
As described in Section~\ref{sec:method}, the two abundance patterns are derived by separating emission-line spectra into high- and low-density (pressure) components. The high-density regions consistently show elevated N/O, while the low-density components exhibit N/O ratios aligned with the local galaxy sequence. This suggests the presence of chemical stratification, where regions near super star clusters or nuclear star clusters are more nitrogen-enriched \citep{james2009,pascale_sbarc_2023,ji2024}. This also implies that the observed nitrogen enhancement in early galaxies may be confined to compact, high-gas-density regions associated with intense star formation, such as stellar clusters.

Given the above results, one question is whether there is any observational bias driving the distribution of \somename in the N/O-O/H space.
The selection of the current \somename is from surveys/programs with different selection functions and is clearly not homogeneous, {and \somename are mostly UV-bright galaxies with relatively high stellar masses as shown in Figure~\ref{fig:jades_mass}.
The intrinsic brightness of \somename probably facilitates the measurements of nebular lines but also biases their selection.
%There are also other selection biases.
}
For the metal-enriched \somename with $\rm 12+log(O/H)\geq 8.0$, for example, they might not reflect the true distribution of \somename due to the following reasons.
The LyC of the Sunburst Arc is identified because the galaxy is strongly lensed.
The galaxy, GS\_3073, hosts an unobscured AGN, which might boost the luminosities of its emission lines.
The local BCD, Mrk 996, only exhibits nitrogen enhancement when there is enough spectral resolution to separate the different density components.
%Finally, abundances of CEERS\_01019 suffer from large systematic uncertainties due to the unclear source of ionization.
Another general question is whether the nitrogen enhancement is ubiquitous at high redshift, or if we are simply biased by the bright \somename.
While this is not the main focus of this work, recent studies using stacks of high-$z$ galaxies suggest an overall enhancement in N/O at $z>4$ despite potentially large scatter in individual sources \citep{Hayes_stack_2025,Isobe_2025}.
The general enhancement of N/O would be expected if galaxies underwent cluster-dominated star formation at early times, which we further discuss in Section~\ref{sec:discuss}.

%\redtxt{[comment on selection bias]}

\subsection{Carbon abundance}
\label{subsec:carbon}

Similar to nitrogen, the enrichment of carbon has a significant contribution from the secondary process and is delayed with respect to that of oxygen \citep{Garnett1995,Nicholls_2017,Kobayashi_2020}.
This leads to a primary+secondary C/O versus O/H relation similar to the N/O versus O/H relation. As shown in the top panel of Figure~\ref{fig:co_com}, the field stars in the MW do show a fast increase in C/O at high O/H with a turnover metallicity of $\rm 12+\log(O/H)\approx 8.0$.
{At $\rm 12+\log(O/H)\lesssim 7.5$, there is again a increase in C/O of field stars. This likely reflects a combination of our signal-to-noise selection bias and the use of 1D LTE models in APOGEE abundance determinations (see above). We therefore refrain from assigning a physical interpretation to this low-metallicity upturn.}

Compared to stellar determinations of C/O, since the nebular determinations of C/O usually rely on the coverage of the strong UV carbon doublet, \ciii$\lambda \lambda 1906,1908$, or the detection of faint recombination lines of carbon, there are not many measurements in the gas phase for local galaxies.
Most of the \somename show sub-solar C/O consistent with the low-O/H tail of the MW disk stars.
One exception is the nitrogen-loud AGN host galaxy, UNCOVER-45924 at $z=4.46$, which shows a super-solar C/O at sub-solar O/H.
While the enhancement of C/O to super-solar values is also reported for a galaxy at $z=12.5$, this occurs at $\rm 12+\log(O/H) < 7$ and is potentially explained by the yield from (low-energy) supernovae of the Population III stars \citep{DEugenio_co_2024}.
For local galaxies, there is also a large spread in C/O at low O/H, which could be a combined effect of large uncertainties propagated from the UV-line measurements and the intrinsic spread in C/O caused by the bursty star-formation history in dwarf galaxies \citep{berg_2016,berg2019}.

{As shown in in the bottom panels of Figure~\ref{fig:co_com}, the distribution of the 1G GC stars closely matches the distribution of low-O/H field stars, while the distribution of the 2G GC stars exhibits slightly enhanced \redtxt{average} C/O \redtxt{(traced by the contour center)} with lowered O/H. However, compared to the field sample, the GC stars also show a wider spread in C/O, which, combined with the generally lower O/H, makes GC stars better match the abundances of \somename.} For the relatively metal-enriched \somename at $\rm 12+\log(O/H) \geq 8.0$, they still show sub-solar to solar C/O and could be compatible with the abundance pattern of the 2G stars of the massive GC 47 Tuc.
Still, the abundances of the most metal-rich \somename in our sample, Mrk 996 and possibly CEERS\_01019 (if its lines are mostly contributed by AGN-ionized regions; \citealp{isobe2023}), also overlap with the contours of MW disk stars. For UNCOVER-45924, its possibly super-solar C/O is not seen in typical GC stars.
{However, as we show later in Section~\ref{sec:discuss}, 
\redtxt{such an enhancement in C/O is observed in a particular NSC, NGC 5139.}
%in NSCs where star formation continues to occur over a longer time scale, such an enhancement in C/O is observed.
}
The overall sub-solar C/O of \somename indicates the prevalence of significantly super-solar N/C in these systems.
The anomalous N/C of \somename is also recognized in previous studies \citep[e.g.,][]{isobe2023,Isobe_2025,ji2024,schaerer_gnz9_2024,topping2024}.
%In GCs, the rise of C/O especially in the 2G stars is a combined effect of depleted oxygen and enhanced carbon.
At the equilibrium of the CNO cycle, it is expected that N/C would be significantly enhanced \citep[$\sim 2$ dex above solar,][]{Maeder_cno_2015}.
%However, the N/C ratio from, for example, the yields of AGB stars, can vary in a wide range from sub-solar to super-solar depending on the \redtxt{degree of the CNO process, which depends on the efficiency of convection in the stellar envelope and on the number of the third dredge-up episodes strongly affected by the mass loss rate adopted (\citealp{Herwig_2004}; \citealp{Karakas_2007}, although the C+N+O abundances for 2G stars are overpredicted; \citealp{Ventura_2013}).}
\realredtxt{However, the N/C ratio from, for example, the yields of AGB stars, can vary in a wide range from sub-solar to super-solar depending on the degree of the CNO processing, which is very different in different sets of models \citep[e.g.,][]{Herwig_2004,Karakas_2007,Ventura_2013}, as it depends on the efficiency of convection in the stellar envelope and on the number of the third dredge-up episodes, strongly affected by the mass loss rate adopted, although models with low mass loss rates overpredict the C+N+O abundances in 2G stars.}
%Also, the general wisdom based on observations of local galaxies and stars is that the enrichment of carbon has a smaller contribution from the AGB stars compared to nitrogen \citep{Kobayashi_2020}.
Alternatively, yields of WR stars, TDEs, and SMSs can potentially explain the elevated N/C \citep{isobe2023,watanabe2024}.

{In summary, in the C/O versus O/H plane, the low-metallicity MW field stars, GC stars, and NOEGs occupy broadly overlapping regions. While C/O alone does not single out a unique counterpart to \somename, the fact that 2G GC stars and \somename remain consistent in C/O provides a valuable sanity check. This reinforces the stronger connection already evident from their matching N/O ratios.}

\subsection{Iron abundance}
\label{subsec:iron}

As another important tracer of chemical evolution, iron is produced early by CCSNe and later by Type Ia supernovae (SNe) in large amounts.
The enrichment channel through Type Ia SNe makes Fe abundance sensitive to a timescale of $\sim 1$ Gyr after the initial burst of star formation \citep{Maoz_2012}.
\redtxt{Thus, one expects Fe to trace a longer timescale of chemical evolution compared to N.}
%Tracing such a long timescale is vital for understanding the overall chemical enrichment, as the standard chemical evolution model predicts the delay time of N and C enrichment by AGB stars should be $0.1-1$ Gyr, meaning that they should be enriched prior to or at a similar time as the mass enrichment of Fe in a single starburst.

As shown in the top panel of Figure~\ref{fig:feo_com}, the non-GC stars in the MW show a strong increase in Fe/O close to solar O/H (known as the low ``$\alpha$-knee'' if one replaces the $x$ axis with Fe/H and revert the $y$ axis, \citealp{Wallerstein_1962,Tinsley_1979}), consistent with the delay enrichment of Fe. Also, the thin disk is more oxygen-poor (or $\alpha$-poor) compared to the thick disk.
In comparison, local galaxies and \hii\ regions with gas-phase Fe/O measurements by \citet{mendezdelgado_feo_2024} show a small overlap with the MW non-GC stars roughly at $\rm 7.5 < 12+\log(O/H)<8.0$, with a tentative low-O/H tail of enhanced Fe/O. 
%(where some galaxies also overlap with the metal-poor MW halo stars).
\citet{mendezdelgado_feo_2024} argue that there is actually a low-metallicity plateau at $\rm \log(Fe/O)\approx -1.74$ indicated by the dash-dotted line in Figure~\ref{fig:feo_com}, which is also consistent with the Fe/O plateau seen in the MW field stars.
As O/H increases, the gas-phase Fe/O of local galaxies and \hii\ regions show a systematic decrease in contrast to the increasing trend in MW field stars.
The interpretation by \citet{izotov2006} and \citet{mendezdelgado_feo_2024} for this anticorrelation between Fe/O and O/H in the gas phase is the increased depletion of Fe onto dust grains at high metallicities.
From the observational studies along the MW sightlines, it has been shown that there is significant variation of Fe depletion \citep[by $1-2.5$ dex,][]{jenkins2009}, although there lack clear evidence of how the amount of depletion correlates with the metallicity.
%It should be noted that Fe/H measured by \citet{izotov2006} does increase with increasing O/H, whereas Fe/H of \hii\ regions measured by \citet{mendezdelgado_feo_2024} is nearly a constant.
%Thus, in this scenario of dust depletion, the overlapping region between galaxies and MW stars would be a coincidence, since those galaxies would have intrinsically higher Fe/O.
Besides the dust depletion, another caveat is that the derivation of the gas-phase Fe abundance mostly relies on a single Fe species, $\rm Fe^{2+}$.
%(with lower and higher ionization potentials similar to those of $\rm O^{+}$), as it produces the strong \feiii\ lines in the optical.
At low metallicities, $\rm Fe^{2+}$ is not the dominant Fe species, and thus the accuracy of the Fe abundance is subject to the accuracy of the ICF
%, the assumption of the ionizing source, and the potentially significant variation in the nebular conditions of different ionization species of Fe 
\citep[see e.g., the discussions by][]{rodriguez_2005,kojima2021,mendezdelgado_feo_2024}.
{Alternatively, pair-instability supernovae (PISNe) or hypernovae (HNe) from VMSs might produce Fe enhancement at low O/H (e.g., \citealp{kojima2021,isobe2022,watanabe2024}).}

Regarding \somename, measurements of the Fe abundance become much more difficult, although not impossible.
For the high-$z$ \somename we considered, only four of them have estimations of Fe/O, yet all with large uncertainties.
While the Fe abundance of the Sunburst Arc is derived from the optical Fe line, \feiii$\lambda 4658$ \citep{welch_sba_2024}, all other sources have different ways of estimating their Fe abundances.
For GS\_3073 at $z=5.55$, \citet{ji2024} use high-ionization Fe lines including [Fe\,{\sc iv}]$\lambda \lambda 2829,2835$ and [Fe\,{\sc vii}]$\lambda 6087$ to estimate the Fe abundance.
%, assuming the depletion of Fe at the locations of these different species are the same.
However, these Fe lines are blended with Mg\,{\sc ii}] and [Ca\,{\sc v}], respectively, in the low-resolution NIRSpec PRISM spectrum, and thus \citet{ji2024} can only make a rough estimation assuming a range of contamination.
The lower limit in Fe/O for GS\_3073 makes it compatible with the local galaxies and \hii\ regions.
For GS\_9422 at $z=5.94$, \citet{tacchella2024} use the detection of another high-ionization Fe line, [Fe\,{\sc v}]$\lambda 4227$, in the NIRSpec PRISM spectrum (and potentially in the medium-resolution NIRSpec G395M spectrum).
%, which has an ionization potential similar to that of \heii.
Notably, GS\_9422 has a gas-phase Fe/O compatible or higher than the solar value, tracing the low-metallicity tail of the distribution of local galaxies studied by \citet{izotov2006}.
Finally, for GS-z9-0 at $z=9.43$, GN-z11 at $z=10.6$, and GHZ2 at $z=12.34$, the Fe abundances reported by \citet{Nakane_gnz11_2024,Nakane_2025} are the \textit{stellar} abundances and rely on continuum fitting of the rest-frame UV spectra obtained with \jwst/NIRSpec medium-resolution ($R\sim 1000$) and low-resolution ($R\sim 100$) spectra.
This approach assumes the whole UV continuum is dominated by young stellar populations, whereas the AGN-like UV lines and extremely high densities ($n>10^9~{\rm cm}^{-3}$, i.e. BLR-like) indicate the presence of an AGN in GN-z11 \citep{maiolino_nature_gnz11_2024}.
Regardless, as shown by \citet{ji_gnz11_2024} and \citet{Nakane_gnz11_2024,Nakane_2025}, the continuum excess observed in the UV spectrum of GN-z11 would also imply an enhanced \textit{nebular} Fe abundance if one interprets the UV continuum as being dominated by an AGN accretion disk.

If we adopt the measurement of \citet{Nakane_gnz11_2024} and consider the Fe/O as the stellar abundance, GN-z11 would have a super-solar Fe/O only $\sim 430$ Myr after the Big Bang.
In fact, both GS\_9422 and GN-z11 would not have sufficient time for the enrichment by the main population of the Type Ia SNe.
Thus, one needs to invoke either small range enrichment (e.g., within a single proto star cluster) by a few early Type Ia SNe \citep[for a delay time as short as 30 Myr, see][]{Greggio_1983}, or enrichment by PISNe or HNe \citep{Nakane_2025}.
%Either of the enrichment scenarios would imply the enrichment time of Fe is comparable or even earlier than the enrichment time of N.
Either of the enrichment scenarios would imply the Fe enhancement occurred as early as the N enhancement.
%Overall, despite the different assumptions involved, the Fe/O of \somename might still be consistent with the distribution of local metal-poor galaxies.
\redtxt{Interestingly, as recently shown by \citet{Isobe_sife_2025} through stacking of JADES galaxies at $4<z<7$, supersolar Fe/O is also found at low O/H for galaxies with high specific star-formation rates and blue UV slopes. The measured Fe/O, however, cannot be simultaneously reproduced with the measured Si/O based on some models of CCSNe, PISNe, HNe, or CCSNe with the mixing-and-fallback process + Type Ia SNe inspected by \cite{Isobe_sife_2025}.}

From the bottom panels of Figure~\ref{fig:feo_com}, the MW GC stars show a plateau in Fe/O closely matching that of field stars in the 1G population and enhanced Fe/O at lowered O/H in the 2G population.
This behavior can be interpreted again as the 1G stars are born in a similar chemical environment as metal-poor field stars and yet the 2G stars are affected by the depletion of oxygen by the \redtxt{CNO cycle} through the enrichment of the 1G stars.
While the Fe/O-O/H distributions of the 2G GC stars are again more consistent with that of \somename, due to the \redtxt{uncertain} Fe abundance measurements in \somename, we cannot draw a strong conclusion.
%If we trust the seemingly decreasing trend of Fe/O with increasing O/H and adopt the oxygen depletion scenario, stars in GN-z11, for example, would have started from a super-solar metallicity of $\rm 12+\log(O/H)= 8.79^{+0.44}_{-0.23}$ to reduce the initial $\rm \log(Fe/O)$ to $-1.74$ as seen in local metal-poor \hii\ regions \citep[although not impossible based on chemical evolution models, e.g.,][]{Kobayashi_ferrara_2024}.

\subsection{Helium abundance}
\label{subsec:helium}

Last but not least, we examine the abundance of helium.
Unlike metals, helium is abundant and has a large contribution from Big Bang Nucleosynthesis (BBN).
The BBN leads to a primordial mass fraction of $Y_p\approx0.24$\,-\,0.25, and stellar nucleosynthesis and subsequent CCSNe enrichment lead to a slight increase in $Y$ in the ISM as a function of metallicity.
The abundance of helium is also connected to the abundance anomalies seen in GCs.
In GCs, significant spread in He abundances among stars are observed, with the 2G stars usually having higher $Y$ compared to the 1G stars \citep[e.g.,][]{milone_hegc_2018}.
While the average helium enhancement between the two generations is modest ($\Delta Y\sim 0.01$), the \textit{maximum} helium variation in GCs can reach $\Delta Y \gtrsim 0.05$, which depends on the initial mass and the luminosity of the GC \citep{milone_hegc_2018}.
The enhancement of helium is mainly via hydrogen burning in the core of hot stars in the first generation.
Therefore, the maximum helium variation is in principle an indicator of how complex the actual stellar populations in GCs are. 
{For example, the cluster displaying the largest He-variation reported in \citet{milone_hegc_2018}, is NGC~2808. NGC~2808 is among the most massive MW GCs \citep{baumgardt2018}, demonstrating the connection between cluster mass and He-enhancement.}
The final helium spread in observations can be explained as a combined effect of stellar-wind enrichment and accretion of pristine gas, and can potentially reflect the formation timescale of GCs \citep[i.e., higher $\Delta Y$ for GCs forming with a longer time and/or with a shorter enrichment timescale,][]{Gieles_gcmp_2025}.

In galaxies, the nebular abundance of heilum can be derived from optical transitions of \hei\ and \heii, with $\rm He^+$ usually being the dominant species.
Recently, \citet{yanagisawa_he_2024} investigated helium abundances of three \jwst-selected \somename, which are GS\_9422 at $z=5.94$, RXCJ2248-ID at $z=6.11$, and GLASS\_150008 at $z=6.23$, and they reported a tentative positive correlation between N/O and $Y$.
\citet{yanagisawa_he_2024} argued that CCSNe cannot explain the helium enhancement they found, and the helium enhancement must be related to the CNO cycle.
In the left panel of Figure~\ref{fig:heh_com}, we plot $Y$ versus metallicity for 8 \somename in comparison with the observed relations in the local Universe inferred by \citet{dopita2006} and \citet{Matsumoto2022}.
\redtxt{Our comparison to local $Y$-O/H relations is a consistency check only -- even if N/O-enhanced phases exhibit localized N enrichment and possibly small $\Delta Y$, these episodes are not expected to shift the low-$z$, galaxy-integrated $Y$-O/H locus.}
There appears to be an anticorrelation between $Y$ and O/H, \redtxt{but no clear monotonic $Y$-N/O trend}.
This is because the more oxygen enriched \somename including GN-z11, GS\_3073, Sunburst Arc, and Mrk 996 are actually more nitrogen enhanced compared to the extremely helium enhanced NOEG, RXCJ2248-ID.
In addition, we caution that the NOEG, A1703-zd6, although has low O/H, does not show any significant enhancement in $Y$.
Thus, with the current data, it is unclear whether there is any true underlying $Y$-O/H relation

%One key question is whether $Y$-O/H relation is physical.
A related question is whether there is any measurement bias or physical mechanism to enhance the apparent helium abundance.
From the perspective of measurement systematics, as already pointed out by \citet{yanagisawa_he_2024}, $Y$ would be overestimated if the helium-line emitting gas has very high electron densities.
Indeed, as shown by \citet{james2009,pascale_sbarc_2023,ji2024,Isobe_2025}, density enhancement (and variation) might be ubiquitous in \somename, and would be reasonable if their spectra are dominated by dense proto-GCs.
Specifically, for RXJ2248-ID, \citet{topping2024} report electron densities up to $10^5~{\rm cm^{-3}}$ as probed by UV lines, whereas \citet{yanagisawa_he_2024} assumed the low-density limit in their fiducial modeling.
With \pyneb and the CHIANTI atomic data set (and assuming the optically thin case or \hei$\lambda 5876$), we found the emissivity of \hei$\lambda 5876$ at $T_{\rm e}=1.5\times 10^4$ K is a factor of 1.7 higher at $n_{\rm e}=10^5~{\rm cm^{-3}}$ compared to the low-density case of $n_{\rm e}=10^2~{\rm cm^{-3}}$.
This means at $n_{\rm e}=10^5~{\rm cm^{-3}}$ and an intrinsic $Y=0.25$, the apparent helium mass fraction is $Y\approx0.36$ if one assumes $n_{\rm e}=10^2~{\rm cm^{-3}}$.
Thus, it is plausible that there is no extreme helium enhancement in \somename if helium lines are predominantly produced in dense clouds. %However, due to the typical measurement error of $\delta Y\sim 0.2$\,-\,$0.5$, it is unclear whether there is considerable helium spread in these systems or not.
\redtxt{This might also explain the seemingly low $Y$ in the most N/O-enhanced object, GS\_3073.
If the $Y$ enhancement in other \somename at low O/H is biased by underestimated gas densities, given the generally large measurement uncertainties in $Y$, and the potential intrinsic scatter due to different gas dilution, retrieving the intrinsic $Y$-O/H or $Y$-N/O trend in \somename would require more statistics.
}

%\redtxt{[discuss GC He-O?]}
Alternatively, if the observed helium lines in \somename are from low-density gas, the enhancement seems only significant at low metallicities in some \somename.
{In the left panel of Figure~\ref{fig:heh_com}, we plot individual MW GCs (small red dots) to compare with \somename, where we use the maximum helium variations in GCs from \citet{milone_hegc_2018} plus the primordial mass fraction of 0.24 to represent the maximum helium enhancement seen in individual GCs.
The O/H of GCs is averaged over their 2G stars since the helium enhancement is present in the 2G.
The enhanced helium abundances are seen both at $\rm 12+\log(O/H)<8$ and $\rm 12+\log(O/H)>8$, with the extremely helium enhanced GC, NGC 2808, reaching the location of GLASS\_150008 ($z=6.23$).}
In the middle panel of Figure~\ref{fig:heh_com}, we show the number distribution of the maximum helium spread in MW GCs from \citet{milone_hegc_2018}, which clusters at a low value of $\Delta Y=0.1$.
In the right panel of Figure~\ref{fig:heh_com}, we further plot the maximum helium spread as a function of the initial mass of the GC, showing that the extreme enhancement of $\Delta Y>0.5$ is only found at massive clusters of $M_{\rm GC;~initial}>10^6~M_\odot$.
As a result, one potential explanation for the enhanced $Y$ in some metal-poor \somename is the more efficient formation of massive GCs.
%the average mass of proto clusters we observe is lower in more metal-rich \somename.
%, driven by, for example, fragmentation of the birth clouds via enhanced cooling.
In observations of MW GCs, however, there is no clear hint of any initial mass-metallicity relation \citep[e.g., in the sample we used and presented by][]{milone_hegc_2018}.
We do note that the recovery of the initial mass of GCs might be uncertain as they lost 50\,-\,80\% of their initial masses, and the mass distributions of GCs we observed today might be biased to high-mass GCs that have lost less of their masses \citep[e.g.,][]{Kruijssen_2015,Reina-Campos_2018,Baumgardt_2019}.

Another possibility is that the enhanced $Y$ is connected to NSCs, which reserve most of their initial masses due to the deep gravitational potentials and on-average exhibit more enhanced $Y$ as shown in the right panel of Figure~\ref{fig:heh_com}, possibly due to their extended star formation histories.
The occurrence of NSCs in local early-type galaxies appears to depend on the $current$\,-\,$day$ stellar mass of the host galaxies, with an occurrence rate peaking at $M_\star\approx10^9~M_\odot$ and dropping at both lower and higher masses \citep{Neumayer_nsc_2020}.
From Table~\ref{tab:global}, the more metal-rich \somename already have $M_\star>10^{8.5}~M_\odot$ or even $M_\star\gtrsim10^{9}~M_\odot$ if we exclude the local Mrk 996, meaning they will evolve into $M_\star > 10^{9}~M_\odot$ galaxies observed today.
In such a scenario, it is plausible that the NSC growth in the more massive and metal-rich \somename is less efficient as those in lower mass \somename that eventually quenched as lower-mass early-type galaxies.
%It is worth noting that at the high-mass end the NSC formation might be disrupted by the feedback from central supermassive black holes, whose presence is clearly observed in GS\_3073 \citep{ubler2023a}.
%Thus, this scenario might imply the clusters (potentially NSCs) in massive \somename do not have enough time to accumulate large helium spread before they are disrupted.
%\redtxt{[but is He enrichment time longer than N?]}.
As a second explanation, the reduced $Y$ enhancement at high-mass and metal-rich \somename might imply less cluster-dominated star formation.
However, this scenario seems incompatible with the even stronger nitrogen enhancement in metal-rich \somename.

To summarize, in our NOEG sample, there is no clear indication of a correlation between $Y$ and N/O, although there might be enhanced $Y$ in some low-O/H \somename.
Underestimated electron densities could potentially lead to the rise of $Y$ at low O/H.
{Regardless, if the helium enhancement is real, its presence might be connected to the efficiency of forming massive GCs or NSCs in these galaxies.}

\section{Discussion}
\label{sec:discuss}

In this section, we discuss our results from two perspectives.
On one hand, we discuss the systematic uncertainties involved in current abundance measurements and how they might impact our interpretations.
On the other hand, we discuss the physical interpretations based on the similar abundance patterns of \somename and GC stars.

\begin{figure*}
    \centering
    \includegraphics[width=\textwidth]{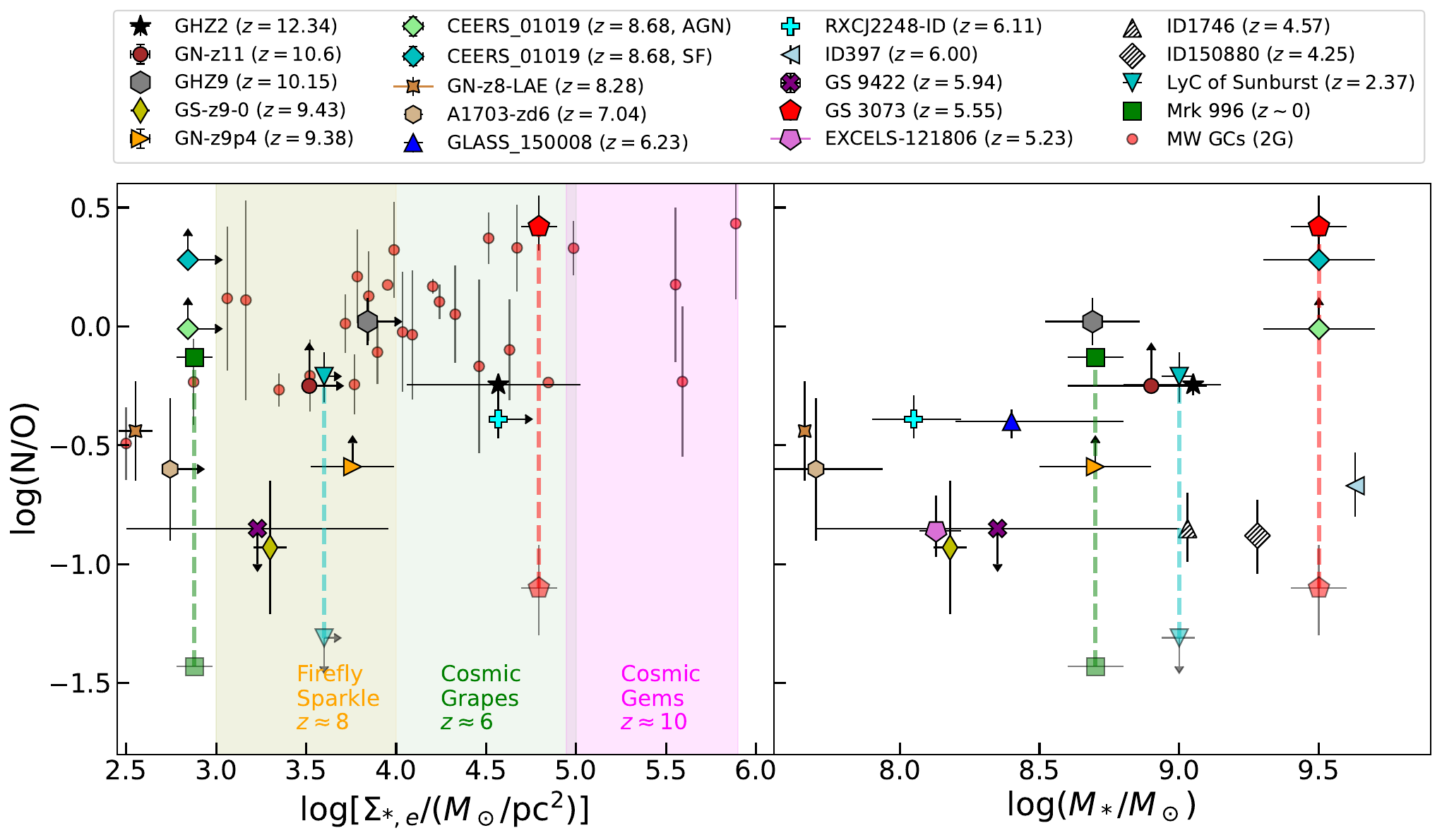}
    \includegraphics[width=\textwidth]{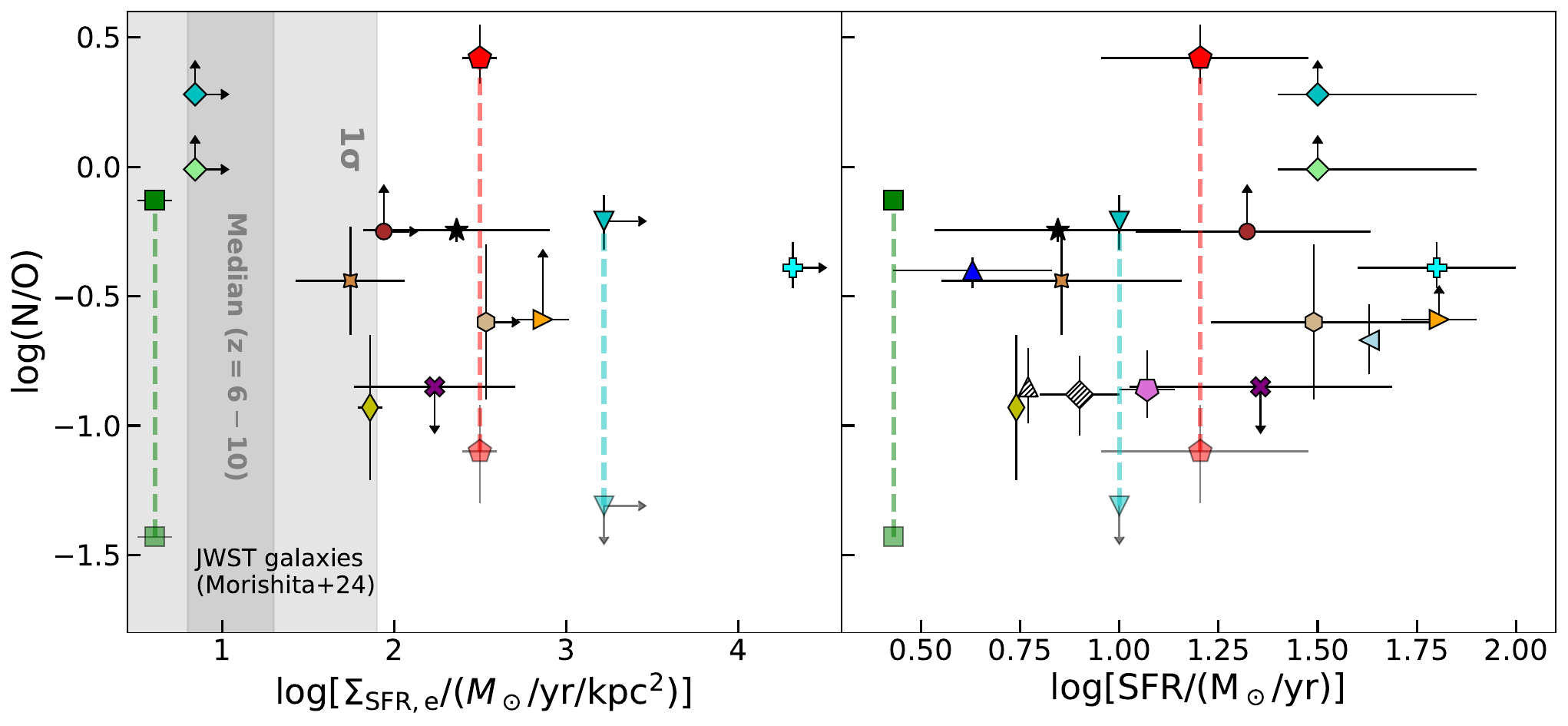}
    \caption{\textit{Top:} Stellar masses and stellar mass surface densities versus N/O for the sample of \somename.
    For systems showing two N/O values connected by dashed lines, the higher values correspond to high-density components and lower values correspond to low-density components.
    Overall, \somename have high stellar mass surface densities of $\Sigma _{\rm \star,~e}\gtrsim 10^{2.5}~{\rm M_\odot~pc^{-2}}$ consistent with progenitor environments of MW GCs and star clusters observed in lensed galaxies shown as shaded regions \citep{Vanzella_2023,Fujimoto_2024,Mowla_2024,adamo_z10cluster_2024}.
    While there is no clear correlation between N/O and stellar mass surface density, in most systems the effective radius, $R_{e}$, is not well constrained (see Table~\ref{tab:global}).
    {In comparison, we also plot the distribution of MW GCs (small red dots), where we approximate their initial stellar mass surface density with the current-day projected core density (see text). The N/O ratios for GCs are median values among 2G stars and the errobars correspond to the standard deviations.
    The GCs occupy a similar parameter space as \somename, although we note that the actual initial densities of GCs could be different due to mass segregation as GCs evolve.
    }
    In the right panel, the N/O enhancement in \somename potentially shows a positive correlation with the total stellar mass of their hosts.
    \textit{Bottom:} {SFR and SFR surface densities versus N/O for \somename. In comparison, the median SFR surface densities of \jwst galaxies (binned at median photometric redshifts of $z_{\rm photo}=6,~8,~{\rm and}~10$) compiled by \citet{Morishita_2024} along with the $1\sigma$ dispersion are shown as the gray shaded region.
    The \somename are generally actively star-forming and have SFR surface densities higher than typical high-$z$ galaxies.
    }
    %In comparison, we plotted in the right panel three shaded regions representing the stellar mass surface densities of star clusters observed in three lensed galaxies at $z=6.1,~z=8.3,~{\rm and}~ z=10.2$ \citep{Fujimoto_2024,Mowla_2024,adamo_z10cluster_2024}.
    }
    \label{fig:no_mass}
\end{figure*}

\begin{figure*}
    \centering
    \includegraphics[width=0.8\textwidth]{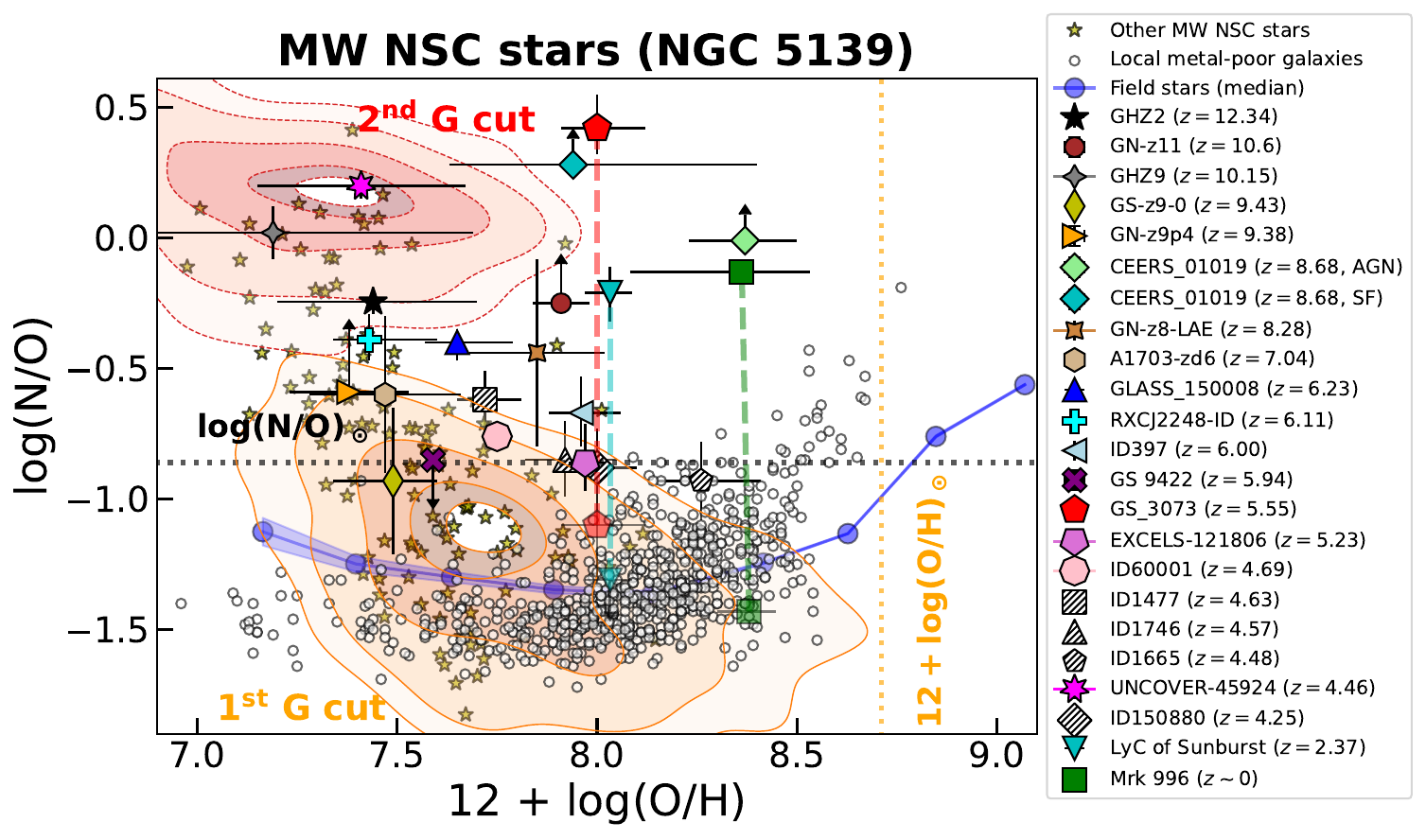}
    \includegraphics[width=0.485\textwidth]{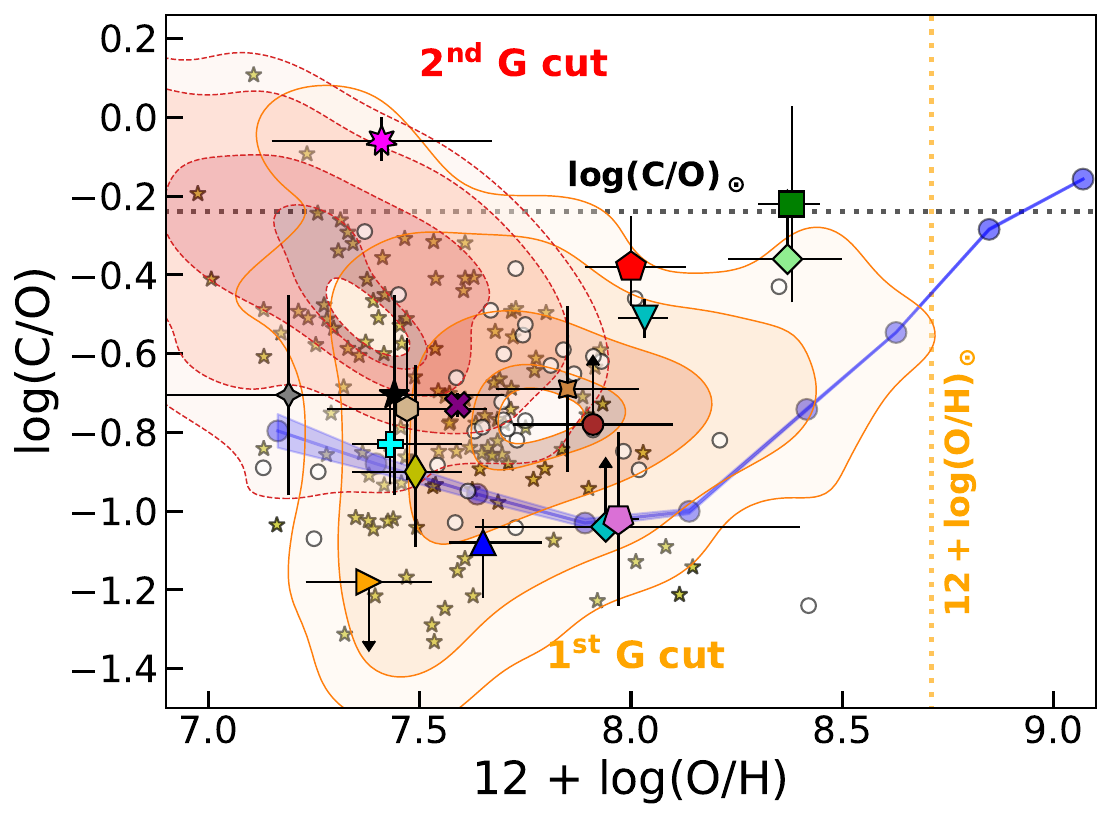}
    \includegraphics[width=0.495\textwidth]{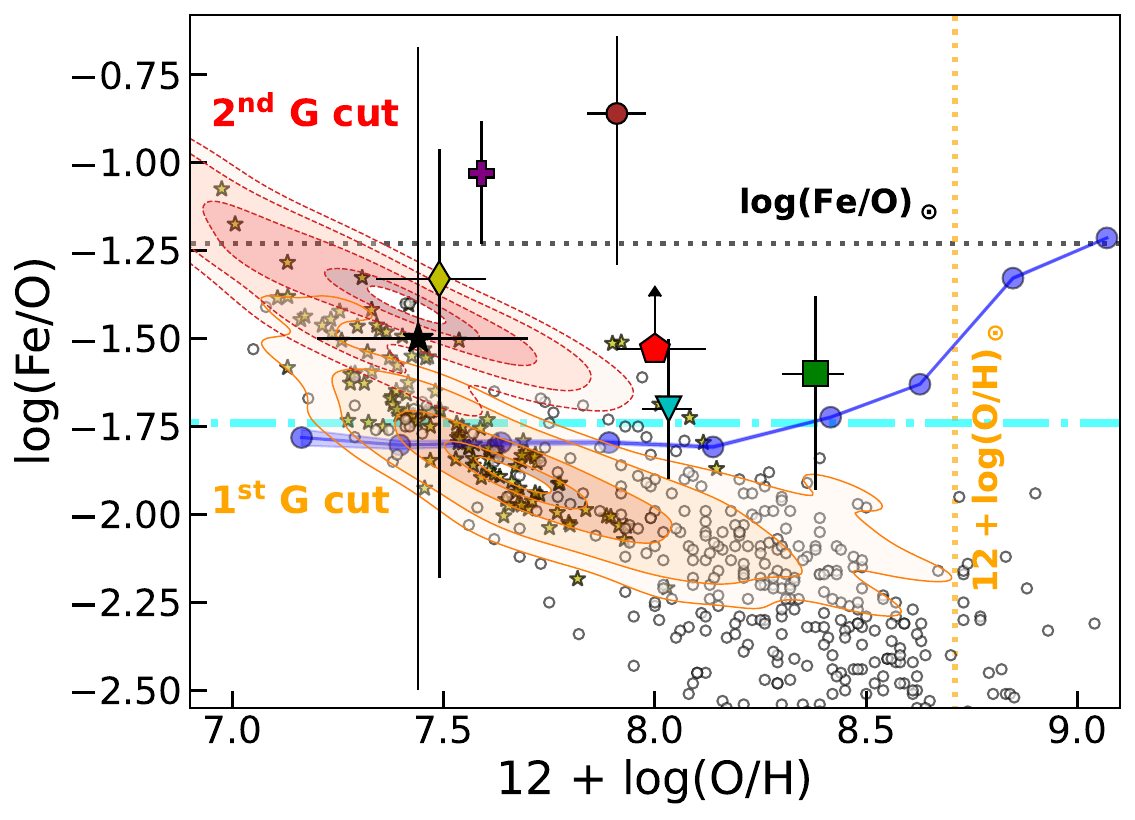}
    \caption{
    Same as Figures~\ref{fig:no_com}, \ref{fig:co_com}, and \ref{fig:feo_com}, but the MW GC stars are replaced with those likely previously associated with NSCs of dwarf galaxies.
    These include NGC 5139, NGC 6273, NGC 6656, and NGC 1851 \citep{Pfeffer_nsc_2021,McKenzie_nsc_2022}.
    %although NGC 5139 dominates the distribution.
    \redtxt{Since NGC 5139 dominates the distribution, we plot it as colored contours and plot the other NSC stars as yellow star symbols.}
    While NSCs typically have multiple stellar populations, for illustrative purposes, we apply the same cut as we did for GCs to separate the ``first generation'' from the ``second generation'', which should only be considered as a rough cut in formation time.
    %These NSCs overall show a similar distribution as GCs and overlap with the low metallicity branch of \somename, despite there is a lack of stars at higher metallicities, which could be limited by the sample statistics.
    {The NSCs show stronger enhancement in N/O and C/O at lower O/H compared to typical GCs, which potentially arise from their extended star-formation history.
    This makes NSC abundances better match the NOEG UNCOVER-45924, which has potentially enhanced N/O and C/O in its BLR.
    }
    }
    \label{fig:no_nsc}
\end{figure*}

\subsection{Validity of the abundance measurements at high redshift}

Measurements of gas-phase chemical abundances usually involve a number of assumptions.
Even when the electron temperature is constrained by auroral-to-strong line ratios such as \oiii$\lambda 4363$/\oiii$\lambda 5007$, as is the case for most of the \somename we investigated, bias can still arise due to inhomogeneous ISM conditions, such as density or temperature variations \citep[e.g.,][]{peimbert1967,peimbert2017,desired}.

As we mentioned in the last section, density variation exists in at least three \somename, with the high-density regions having $n_{\rm e}\sim 10^{5-7}~{\rm cm^{-3}}$ and the low-density regions having $n_{\rm e} \lesssim 10^4~{\rm cm^{-3}}$.
The density inhomogeneity would bias the abundance derivation if the electron density reaches the critical densities of emission lines involved in the abundance derivation.
However, since the rest-frame UV emission lines involved generally have high critical densities (e.g., $n_{\rm critical}>10^9~{\rm cm^{-3}}$ for \niii$\lambda 1750$, \niv$\lambda 1486$, \oiiip]$\lambda \lambda 1661,1666$, \ciii$\lambda 1908$, \civ$\lambda \lambda 1548,1551$), N/O derived from UV line ratios are relatively insensitive to density enhancement \redtxt{\citep{ji2024,Isobe_2025,Martinez_uvon_2025}}.
\redtxt{
An interesting question is whether ``normal'' N/O can be probed with UV lines.
There might be two limiting factors for achieving this.
First, due the enhanced line emissivity in dense gas, the observed UV lines from galaxies might preferentially sample N/O in dense gas environments (e.g., star clusters).
Second, given the sensitivity of most \jwst observations, for galaxies with no intrinsic N/O enhancement, it would be challenging to verify purely based on the upper limits of UV lines \citep{Isobe_2025,Zhu_nbias_2025}.
}

In the optical, \nii$\lambda 6583$ is the only relatively strong nitrogen transition typically used for deriving the nitrogen abundance and has a critical density of $n_{\rm critical}\sim 10^5~{\rm cm^{-3}}$. This means that a bias would arise if  $n_{\rm e} > 10^4~{\rm cm^{-3}}$, especially when using \nii$\lambda 6583$/\oii$\lambda \lambda 3726,3729$ to infer $\rm N^+/O^+$ in typical practices ($n_{\rm critical,[OII]\lambda 3727}\sim 10^4~{\rm cm^{-3}}$).
Among the 22 \somename we compiled, 14 have their N/O derived from UV semi-forbidden transitions and thus the inferred abundance ratios are insensitive to density enhancement.
For GS\_3073, due to the presence of broad emission lines in the rest-frame optical, the semi-forbidden transitions in the rest-frame UV might also have a broad component (which remains unresolved in the low-resolution NIRSpec/PRISM spectrum).
However, as shown by \citet{ji2024}, the existence of any UV broad component would only imply a stronger intrinsic nitrogen enhancement.
The UV region of UNCOVER-45924 also suffers from contamination by broad lines, and we have discussed this issue in Section~\ref{sec:method}, where we can still estimate N/O (or its lower limit) based on the density-insensitive ratio of \niii$\lambda 1750$/\oiiip]$\lambda \lambda 1661,1666$.

For the remaining 8 \somename where the abundances are derived from the optical lines including \nii\ and \oii, N/O would be overestimated if the electron density reaches $n_{\rm e}>10^{3}~{\rm cm^{-3}}$ while a low density of $n_{\rm e}<10^{3}~{\rm cm^{-3}}$ is assumed.
For 5 of such \somename reported by \citet{Stiavelli_nloud_2024}, 3 of them have electron densities derived from [S\,{\sc ii}]$\lambda \lambda 6716,6730$, which in principle probe a similar density zone as \oii$\lambda \lambda 3726,3729$.
The \somename reported by \citet{Zhangyechi_2025} and \citet{Arellano-cordova_nloud_2024} also have low-ionization zone densities.
Therefore, only two \somename, ID1477 and ID397 in our sample do not have density constraints for their optical-line based N/O.
The above results suggest little impact from density variations for most of \somename in our sample.

Next, we comment on the potential impact of temperature inhomogeneity.
Naively, emission lines from highly ionized species are in high-ionization zones that are closer to the ionizing source, implying a higher emissivity-weighted temperature compared to the lines in low-ionization zones.
Such a temperature gradient would enhance \niv\ and \civ\ relative to \niii\ and \oiiip] at fixed N/O and C/O compared to the case where the temperature is uniform, leading to overestimations of the abundance ratios if one uses the temperature of $\rm O^{2+}$ derived from \oiii$\lambda 4363$/\oiii$\lambda 5007$ \citep{ji2024}.
It is thus useful to use a higher temperature for higher ionization species by using, for example, photoionization models \citep{ji2024}.
Still, abundance ratios inferred from \niii$\lambda 1750$/\oiiip]$\lambda \lambda 1661,1666$ or \ciii$\lambda \lambda 1906,1908$/\oiiip]$\lambda \lambda 1661,1666$ should be relatively robust as $\rm N^{2+}$, $\rm C^{2+}$, and $\rm O^{2+}$ share similar ionization potentials and should be spatially close.
In addition, if high-ionization emission lines dominate the abundance derivation, one would expect C/O is also biased high as N/O, which is not commonly seen in \somename.
As an example, among the 22 \somename we compiled, 14 have their N/O derived from UV semi-forbidden transitions, whereas only {one} source, GN-z9p4, has its N/O determined through \niv\ and \oiiip] due to the non-detection of \niii\ \citep{schaerer_gnz9_2024}.

Finally, we note that there is an alternative explanation for the seemingly high N/O in \somename recently proposed by \citet{Flury_shock_2024}, who suggested prevalent ionization by slow/intermediate-velocity radiative shock in the early Universe that biases the abundance derivation usually based on assumptions of photoionization.
However, shock models rely on additional assumptions due to the intrinsically high dimensionality of their parameter space (e.g., shock velocity and magnetic field strength in addition to the typical parameters assumed for photoionization models).
Also, as shown by \citet{Flury_shock_2024}, with their shock models, nitrogen enhancement for sources such as GS\_3073 and GN-z11 is reduced, but not removed compared to local galaxies.
Detailed comparisons between the nebular properties of \somename and shocked dominated nebulae in the local Universe will be important to further test this scenario.

In summary, current observational evidence suggests that there are complex environments in \somename with varying densities and temperatures.
Still, current estimations of the abundance patterns of N/O and C/O for the majority of \somename are unlikely to be significantly biased.

%\subsection{Testing the progenitor hypothesis}

 \subsection{Environments of the abundance anomalies}

%\redtxt{[make another subsec for NSC?]}
%In this subsection, we discuss whether the presence and level of the nitrogen enhancement correlate with any other observed properties of \somename.
One avenue to investigate the origin of \somename is to identify correlations between nitrogen enhancement and other galaxy properties.
We note that due to the heterogeneous selection of \somename from samples with complex selection functions, we do not attempt to quantify the statistical occurrence rate of \somename, but rather perform a qualitative assessment of the derived properties among \somename.

\subsubsection{Compactness}

\begin{table*}
    \caption{Summary of the physical properties of star clusters observed in gravitationally lensed galaxies at high redshift.
    The $1\sigma$ relative measurement uncertainty in the total lensing magnification, $\mu$, varies among galaxies from as low as 3\% to 20\,-\,50\%, with systematic uncertainties due to lensing models potentially reaching 50\% and higher.
    However, since $M_\star$ (at fixed $M/L$) roughly scales linearly with $\mu ^{-1}$ and $R_{\rm e;~intrinsic}$ roughly scales linearly with $\mu ^{-0.5}$, the stellar mass surface density is less affected by the uncertainty in the lensing magnification.
    }
    \centering
    \begin{tabular}{l|c|c|c|c|c}
         \hline
         Name & $z$ & $N_{\rm cluster;~identified}$ & $\mu$ & $R_{\rm e;~intrinsic}$ [pc] & $\log\Sigma _{\rm \star,~ e;~intrinsic}$ [$M_\odot~{\rm pc^{-2}}$]  \\
         \hline
         Sunburst Arc$\rm ^a$& 2.37 & 15 & 15\,-\,484 & 0.9\,-\,23.7 & 3\,-\,4 \\
         Sunrise Arc$\rm ^b$& 6.0 & 6 & $\gtrsim$30\,-\,66 & 1.4\,-\,24.8 & 3\,-\,5 \\
         Cosmic Grapes$\rm ^c$& 6.07 & 15 & 22\,-\,43 & 7\,-\,62 & 3\,-\,5 \\
         Firefly Sparkle$\rm ^d$& 8.30 & 10 & 16\,-\,26 & $<4$\,-\,7 & $>3$\,-\,4 \\
         Cosmic Gems$\rm ^e$& 10.2 & 5 & 57\,-\,420 & 0.7\,-\,1 & 5\,-\,6 \\
         \hline
    \end{tabular}
    \begin{tablenotes}
        \small
        \item $\bf Notes.$ References: $\rm ^a$ \citet{vanzella_sbarc_2022}; $\rm ^b$ \citet{Vanzella_2023}; $\rm ^c$ \citet{Fujimoto_2024}; $\rm ^d$ \citet{Mowla_2024}; $\rm ^e$ \citet{adamo_z10cluster_2024}.
    \end{tablenotes}
    \label{tab:highz_cluster}
\end{table*}

%\redtxt{[discuss this first?]}
Perhaps the most common feature of \somename is the compactness of these systems, where the presence of a component with $R_e < 200$ pc is usually seen (see Table~\ref{tab:global} and Figure~\ref{fig:jades_mass}).
This makes \somename much more compact compared to other UV-bright galaxies with $R_{\rm e}\sim 1$ kpc at $z\sim 4$ and $R_{\rm e}\sim 0.3-0.6$ kpc at $z\sim 10$ \citep{Shibuya_2015,Harikane_2025}.
In the left panel of Figure~\ref{fig:no_mass}, we plot N/O as a function of the average stellar mass surface density within the effective radius for \somename in our sample.
While there is no clear correlation between N/O and $\Sigma _{\star,~e}$ (given the loose constraints on $R_e$), the stellar populations in \somename are overall densely distributed with $\Sigma _{\star,~e}\gtrsim 10^{2.5}~M_\odot~{\rm pc^{-2}}$.
Such a dense environment is again in good agreement with those inferred for the local GCs.
For example, the progenitor gas environment of MW GCs likely has a gas mass density of $\Sigma _{\rm gas}\sim 10^3~M_\odot~{\rm pc^{-2}}$ \citep{Gieles_gcmp_2025}.
{We performed another calculation to estimate the initial stellar mass surface density of MW GCs.
Specifically, we took the core densities ($\rho _{\rm c}$) and radii ($r_{\rm c}$) of MW GCs fitted by \citet{baumgardt2018}.
Simply assuming a constant density in the core, the average surface mass density of the core is $\Sigma _{\rm c}=\frac{4\pi}{3}r^3_{\rm c}\rho_{\rm c}/(\pi r_{\rm c}^2)=\frac{4}{3}\rho_{\rm c}r_{\rm c}$.
Then, if we approximate the initial densities of GCs with the current-day core densities (which is not exact due to mass segregation as GCs evolve, \citealp{Fregeau_gcms_2002}), we obtained a median surface density of $\Sigma _{\rm ini,~GC}\approx 4\times 10^3~M_\odot~{\rm pc^{-2}}$, comparable with the densities seen in \somename.
We further plotted estimated $\Sigma _{\rm ini,~GC}$ and median N/O of 2G stars for individual GCs in the left panel of Figure~\ref{fig:no_mass}, where one can see a good overlap between GCs and \somename in terms of their compact stellar populations and N/O enhancement.
}
In the early time of the evolution of the MW, star formation is likely not dominated by a smooth disk component but rather by dense star clusters in a turbulent environment \citep{belokurov2022,Belokurov2023}.
Such an early phase of cluster-dominated evolution might be common in \jwst-observed high-$z$ systems, which is evident from the dense environments seen in \somename.

Observations of highly gravitationally lensed galaxies also provide direct evidence for cluster-dominated star formation at high redshift \citep{vanzella_sbarc_2022,Vanzella_2023,Mowla_2024,Fujimoto_2024,adamo_z10cluster_2024}.
\redtxt{In Table~\ref{tab:highz_cluster}, we summarize derived properties of observed lensed galaxies at high redshift, where individual SF clumps were identified and proposed as proto-GC candidates.}
For example, \citet{Vanzella_2023} reported observations of 6 star clusters by \jwst/NIRCam in the highly magnified galaxy, the Sunrise Arc, at $z\sim 6$.
These clusters have a range of stellar mass surface densities of $\Sigma _{\star,~e} = 10^{3-5}~M_\odot~{\rm pc^{-2}}$ and contribute to 10\,-\,30\% of the recently formed stellar mass in this galaxy.
As another example, \citet{Fujimoto_2024} reported a lensed galaxy, the Cosmic Grapes, at $z=6.072$, where at least 15 star clusters with $\Sigma _{\star,~e} = 10^{3-5}~M_\odot~{\rm pc^{-2}}$ are identified.
These star clusters make up of 70\% of the stellar mass of the whole galaxy.
Notably, in the integrated \jwst/NIRSpec IFU spectrum of the Cosmic Grapes, \citet{Fujimoto_2024} found significant detection of \nii$\lambda 6583$ and \oii$\lambda \lambda 3726,3729$, and they derived a normal $\rm log(N/O)=-1.46\pm 0.04$ at $\rm 12+log(O/H)=8.11\pm 0.03$, which is in good agreement with the local scaling relation.
This result seems to suggest no N/O enhancement in these star clusters, although we note that N/O enhancement in the rest-frame UV cannot be completely ruled out, as seen in some of the \somename shown in Figure~\ref{fig:no_com}, where N/O inferred from optical lines are consistent with the local scaling relation.
At $z=8.304$, \citet{Mowla_2024} reported a gravitationally lensed galaxy, the Firefly Sparkle, where 10 star clusters are identified.
These star clusters have stellar mass surface densities of $\Sigma _{\star,~e} = 10^{3-4}~M_\odot~{\rm pc^{-2}}$, consistent with the expected densities of progenitors of the MW GCs.
Also, \citet{Mowla_2024} show that roughly 50\% of the stellar mass in the Firefly Sparkle is within the 10 star clusters, supporting the scenario of cluster-dominated star formation.
Unfortunately, there is no coverage of the optical \nii$\lambda 6583$ line in the \jwst/NIRSpec spectrum and there is no detection of any UV nitrogen lines to constrain N/O.
Finally, there is a lensed galaxy, the Cosmic Gems Arc, reported by \citet{adamo_z10cluster_2024} at $z=10.2$ with 5 extremely dense star clusters making up 15\,-\,35\% of the total stellar mass, although no spectrum is available to verify whether they exhibit any level of nitrogen enhancement.

In the {top} left panel of Figure~\ref{fig:no_mass}, we also show the ranges of the surface mass densities of the star clusters observed in the above high-redshift galaxies.
There is clearly a wide range of cluster densities in the early Universe, which covers the range of stellar densities seen in \somename.
If we exclude UNCOVER-45924, whose stellar mass might have a large systematic uncertainty due to the spectral decomposition, GS\_3073 exhibits the highest stellar mass density and the highest N/O at the same time.
The extremely dense environments could help produce the extreme nitrogen enhancement according to enrichment channels such as VMSs, EMSs, and SMSs, where the high density is a necessary condition (e.g., to allow for rapid formation via inertial flows or stellar collisions; \citealp{Gieles_gcmp_2025}).
Additionally, the dense environments could help to maintain the enriched light elements, which would also imply concentration of the abnormal abundances on cluster scales.
{Such a spatially confined enrichment is also supported by the three \somename showing two abundance patterns, which are the Mrk 996, LyC, and GS\_3073, as noted by \citet{ji2024}.
As a rough estimation, if one takes a minimum gas density of $n_{\rm H} =10^5~{\rm cm^{-3}}$ as the condition for N/O enhancement, as seen in the three \somename, the corresponding hydrogen mass density is $\rho _{\rm H}=n_{\rm H}m_{\rm H}\approx 2.5\times 10^3~M_\odot~{\rm pc^{-3}}$, comparable to the median core density of MW GCs seen today, which is $\rho _{\rm c,~med}\approx3.5\times 10^3~M_\odot~{\rm pc^{-3}}$ based on the measurements of \citet{baumgardt2018}.
The similar values of densities, although not directly related, further indicate cluster-scale enrichment seen in \somename.
}

{In addition to the stellar mass surface density, the compactness of \somename is also reflected by their star-formation rate (SFR) surface density.
We compiled measurements of total SFRs reported by the literature listed in Table~\ref{tab:global} for \somename and calculated $\rm \Sigma _{SFR,~e}=SFR/(2\pi R_e^2)$.
Most of the SFRs are estimated through SED fitting (sometimes in combination with \ha\ luminosities).
For the unobscured AGN host GS\_3073, we adopted the SFR estimated from [C\,{\sc ii}]$\lambda 158\rm \mu m$ by \citet{ubler2023a} to reduce the AGN contamination, which is much lower than the value obtained through SED fitting.
In the bottom left panel of Figure~\ref{fig:no_mass}, we plot SFR surface densities versus N/O for \somename that have SFR estimates.
Clearly, most \somename are very compact SF systems compared to typical galaxies observed at high redshift, as noted by \citet{schaerer_gnz9_2024}.
In fact, these \somename have higher SFR surface densities compared to the most intense starburst galaxies in the local Universe \citep{Kennicutt_1998,Bigiel_2008}.
Previous observational and theoretical works have shown an increase in the efficiency of forming bound clusters as the SFR surface density increases \citep[e.g.,][]{Adamo_2011,Adamo_2015,Kruijssen_2012}.
Thus, the high SFR surface densities of \somename are consistent with the picture that they are efficiently forming proto-GCs.
}

It will thus be informative for future work to compare chemical enrichment in dense star clusters and diffuse ISM at high redshift, preferably in a sample of galaxies with spatially separated star clusters.
{Notably, the compact morphologies are also frequently seen in \jwst-discovered AGN \citep[e.g.,][]{matthee2024,maiolino_agn_2024,juodzbalis_agn_2025} and the coevolution of star clusters and AGN in the gas-rich environments in the early Universe are indeed expected \citep{Neumayer_nsc_2020}.
Next, we discuss the potential connection between \somename and actively accreting black holes.
}

%Notably, galaxies with mass densities higher than $10^3~M_\odot~{\rm pc^{-2}}$ are not uncommon at $z>5$ \citep[e.g.,][]{Morishita_2024,sun_size_2024,schaerer_gnz9_2024}, implying more cluster-dominated star-forming environments and, potentially, nitrogen enhancement to be spectroscopically confirmed.
%\redtxt{[tbf; also discuss NSC]}

\subsubsection{Presence of accreting black holes}

%We start by discussing the potential connection between nitrogen enhancement and AGN activity.
%By stacking \jwst-observed galaxies at $4<z<7$ from JADES, \citet{Isobe_2025} confirm relative enhancement of nitrogen lines in the UV spectrum of Type 1 AGN.
Recently, \citet{Isobe_2025} identified a potential connection between AGN activity and N/O enhancement by finding a relative enhancement of the nitrogen line, \niii$\lambda 1750$, in a stack of Type 1 AGN in JADES at $4<z<7$.
While the nitrogen lines can be enhanced by the presence of broad lines, \citet{Isobe_2025} argue that an enhancement in N/O must also exist to explain the difference between the stacked Type 1 AGN and SF galaxies.
Among the 22 \somename we compiled, two are unambiguous Type 1 AGN (GS\_3073, UNCOVER-45924), and six have AGN signatures (GS\_9422, CEERS\_01019, GS-z9-0, GHZ9, GN-z11, and GHZ2) indicated by high-ionization lines.
%From the case of GS\_3073 and UNCOVER-45924, one can see that a density structure spanning orders of magnitude naturally arises if the galaxy hosts an obscured AGN.
Interestingly, in the pre-\jwst era, a subclass of QSOs known as ``nitrogen-loud QSOs'' has been observed, which shows enhanced nitrogen lines in the UV, implying a high N/O \citep{Baldwin_nloud_2003}.
The population of the nitrogen-loud QSOs is very rare, comprising only 1\% of QSOs from SDSS at $1.6<z<4$ \citep{Bentz_osmer_nlqso_2004,Bentz_nlqso_2004,Jiang_nlqso_2008}.
For unobscured AGN and nitrogen-loud QSOs, specifically, it has been long speculated that the nitrogen enhancement is limited to the nuclear region where a past starburst produces the current chemical imprint \citep{hamann1993, Baldwin_nloud_2003}.
The nuclear chemical imprints can be more easily seen once the dust is cleared out by feedback and the central black hole reaches the active accretion phase \citep{Hamannferland_1999}.
This points to chemical anomalies potentially dominated by NSCs, whose gravitational potential is deep enough to maintain the enriched gas.
The early formation of NSCs could also seed massive black holes \redtxt{\citep{Neumayer_nsc_2020,Isobe_2025,Partmann_2025,Vergara_bhseed_2025,Paiella_bhseed_2025}}, which might explain the ``overmassive'' black holes compared to the stellar masses of host galaxies seen in many high-$z$ AGN discovered by \jwst \citep[e.g.,][]{harikane2023,ubler2023a,maiolino_agn_2024,juodzbalis_agn_2025}.
Alternatively, some models suggest that extremely small-scale enrichment can occur in the extension of the accretion disk, where star formation can occur to produce abundance patterns with enhanced light elements for the BLR \citep[e.g.,][]{huang_agndisksf_2023}.

While the high-$z$ AGN hosts might be more N/O enhanced compared to the non-AGN hosts as shown by \citet{Isobe_2025} and from Figure~\ref{fig:no_com} focusing on GS\_3073 and UNCOVER-45924, this should not be simply interpreted as a causality, and one needs to examine other physical properties as well.
\realredtxt{Notably, the duration of a single actively accreting phase of AGN ($\rm \sim 0.1-100~Myr$; \citealp{Morganti_2017})
is comparable to the time scale of early enrichment by AGB stars ($\rm \sim 40$\,-\,300 Myr; \citealp{dantona_2023}) and those from other mechanisms (e.g., WRs, VMSs, EMSs, and SMSs; \citealp{watanabe2024,Vink_2023,Gieles_gcmp_2025,Charbonnel_2023,Nandal_ems_2024,Nandal_2025}), and one possibility is the synchronization of AGN activities and nuclear star formation \citep{hamann1993}. Still,}
a comparably large fraction of \somename in our sample do not show clear AGN signatures.
%Clearly, a comparably large fraction of \somename in our sample do not show clear AGN signatures, and the duration of a single actively accreting phase of AGN ($\rm \sim 0.1-100~Myr$; \citealp{Morganti_2017}) might be shorter compared to the typical time scale of enriching nitrogen by AGB stars ($\rm \sim 10^{2-3}~Myr$; \citealp{dantona_2023}) unless other enrichment channels with shorter time scales are involved (e.g., WRs, VMSs, EMSs, and SMSs; \citealp{watanabe2024,Vink_2023,Gieles_gcmp_2025,Charbonnel_2023,Nandal_ems_2024,Nandal_2025}) and/or the peak of the AGN accretion is modulated by star formation \citep{hamann1993}.

Since AGN activities are more frequently seen in more massive galaxies, one might wonder whether nitrogen enhancement in AGN hosts is actually a reflection of the mass dependence.
In the {top} right panel of Figure~\ref{fig:no_mass}, we plot N/O as a function of stellar masses of \somename.
Interestingly, if we ignore the low-density components of some \somename (plotted as transparent symbols), there appears to be a positive trend where more massive galaxies are more N/O enhanced.
The unobscured AGN host, GS\_3073 ($z=5.55$), is among both the most massive \somename and the most nitrogen enhanced \somename.
The other unobscured AGN, UNCOVER-45924 ($z=4.46$), is reported to be extremely massive with $\log M_\star/M_\odot=10.9\pm 0.2$ by \citet{labbe_monster_2024}.
However, recent works have suggested the stellar mass estimates for the red compact AGN hosts with flat UV slopes and strong Balmer breaks (i.e., the ``little red dots''; \citealp{matthee2024}) such as UNCOVER-45924 are significantly overestimated due to the incorrect AGN decomposition in their spectra \citep{Inayoshi_maiolino_2025,ji_qso1_2025,degraaff_lrd_2025,Naidu_lrd_2025,taylor_2025}.
Thus, we did not adopt the current stellar mass estimate for this source.
There are three sources, ID1746, ID150880, and ID397, having stellar masses of $10^{9-9.5}~M_\odot$ and yet not significantly more enhanced N/O.
However, from Figure~\ref{fig:no_com}, one can see these sources do not deviate strongly from the local scaling relations compared to many other \somename.
One might wonder whether the tentative N/O-$M_\star$ trend is driven by the mass-metallicity relation.
Compared to what is shown in Figure~\ref{fig:no_com}, it is clear that the increased N/O is \textit{not} driven by the increased O/H.
In fact, there is instead anticorrelation between N/O and O/H at $\rm 12+\log(O/H)<8.0$, which is also seen in local GC stars and is explained by the enrichment of the 1G stars.
{Instead, if we look into the dependence of N/O on the SFR in the bottom right panel of Figure~\ref{fig:no_mass}, there is a tentative positive trend, although with a large scatter similar to the N/O-$M_\star$ trend.
Regardless, both trends seem to suggest some global regulation of the N/O anomaly seen in \somename.
}
%Still, we caution that, as a ``little red dot'' (i.e., compact red objects with flat UV slopes, typically showing a Balmer break and broad emission lines if spectroscopically confirmed; \citealp{matthee2024}), the stellar mass of UNCOVER-45924 remains highly uncertain due to the potentially complex contributions from stellar populations and AGN emission to the observed spectrum \citep{labbe_monster_2024,ji_qso1_2025,degraaff_lrd_2025,Naidu_lrd_2025}.
%Regardless, without including UNCOVER-45924, there is a potential increase of N/O by 1 dex as the stellar mass increases by 1\,-\,2 dex.

{Finally, we note that the high-mass end of \somename might provide some insights into cluster-dominated star formation in early galaxies.
In the MW, the \textit{Aurora} stars formed before the emergence of the disk, which show strong evidence for a cluster-dominated mode of star formation. 
Studies by \citet{belokurov2022,Belokurov2023} have demonstrated that \textit{Aurora} stars exhibit abundance patterns characteristic of GCs and suggest a high fraction of stars formed in bound clusters. 
Importantly, \citet{Belokurov2023} estimate the total stellar mass of Aurora to be $\sim 10^9~M_\odot$, consistent with the stellar mass threshold for bursty, cluster-dominated star formation emerging from recent cosmological zoom-in simulations \citep[e.g.,][]{Dillamore_mwdisk_2024,Semenov_mwdisk_2024}. These simulations identify a critical halo mass of $\sim 10^{11}~M_\odot$ corresponding to a critical stellar mass of a few $\times 10^9~M_\odot$, below which galaxies undergo bursty star formation, sustain high ambient pressures, and efficiently form long-lived bound clusters. Once above this threshold, galaxies develop stable disks, inflow becomes smooth, pressure drops, and the bound cluster formation efficiency falls by more than an order of magnitude within $\sim 200$ Myr.}

{This transition predicts that NOEGs, galaxies whose nebular gas already shows GC-like N/O enhancements, should be found primarily below this stellar mass threshold. However, the simulations underpinning this prediction are tailored to MW-mass progenitors, and their relevance to higher-mass systems is not guaranteed. If some NOEGs turn out to have significantly larger stellar or halo masses, they may lie outside the regime where such bursty, cluster-dominated formation is expected, and the current models may not apply. While current \jwst observations show NOEGs clustering at $M_\star\lesssim10^{9.5}~M_\odot$, this could partly reflect selection effects shown in Figure~\ref{fig:jades_mass}. Broader spectroscopic coverage across stellar mass will be needed to test whether the simulation-predicted threshold truly marks the end of GC-like star formation.}

%This is because the latest cosmological zoom-in simulations have established a galaxy mass threshold of $\sim 10^{9}~M_\odot$ (set by the halo mass), below which galaxies have volatile gas cycles and bursty star formation, and can produce long-lived bound clusters \citep{Dillamore_mwdisk_2024,Semenov_mwdisk_2024}. Above the mass threshold, gravitational torques spin up a stable, cold disc, ``bursty'' inflow quiets, ambient pressure drops, and bound cluster formation efficiency collapses by an order of magnitude within $\sim 200$ Myr. This transition implies a clear observational prediction that \somename, whose nebular gas already has GC-like abundance patterns, should be common only in systems with stellar masses below a few $\times 10^9~M_\odot$, consistent with current observations. However, from Figure~\ref{fig:jades_mass}, the mass distribution of \somename can also be just reflecting that of \jwst spectroscopic sources, whose N/O might also be slightly enhanced but remain largely unconstrained \citep{isobe2023,Isobe_2025,Hayes_stack_2025}. Thus, more detailed characterization of the high-mass end of \somename is useful to confirm the simulation results.

\subsubsection{Connection to nuclear star clusters}

Given the possible mass-N/O relation in \somename, one immediate question is whether the relation is physically expected.
%Alternatively, the stellar mass-N/O distribution might be connected to the cluster formation efficiency.
%Indeed, if we connect the abundance anomalies to star cluster formations, there are known scaling relations connecting cluster formation to stellar masses.
As we mentioned in Section~\ref{subsec:helium}, the fraction of galaxies hosting NSCs at $z\sim 0$ drops with stellar masses at $M_\star>10^9~M_\odot$ \citep{Neumayer_nsc_2020}, but it is nontrivial to trace it back to their progenitors, presumably the \somename except for the most massive ones that would involve into galaxies with $M_\star>10^9~M_\odot$.
In this case, taking GS\_3073 as an example, its extreme N/O appears to contradict the expected low efficiency of NSC formation and growth, if one believes that the NSC is the key to driving the nitrogen enhancement in this scenario.
\realredtxt{However, as recently suggested by \citet{dantona_2025}, the nitrogen enhancement in GS\_3073 could represent a snapshot during the evolution of a specific NSC, NGC 5139.}

In Figure~\ref{fig:no_nsc}, we extract N/O and O/H of MW stars that are classified as NSCs of previous dwarf galaxies by \citet{Pfeffer_nsc_2021,McKenzie_nsc_2022}.
\redtxt{Since, in the APOGEE sample, 80\% of the NSC stars come from a single NSC, NGC 5139, our comparison is essentially made between this NSC and \somename. We plot the stars in NGC 5139 as colored contours and the remaining NSC stars (from NGC 6273, NGC 6656, and NGC 1851) as scatter points.}
{Although the overall distribution of \redtxt{stars in NGC 5139 (and other NSC stars)} is broadly similar to that of general GC stars shown in Figure~\ref{fig:no_com}, they exhibit subtly but systematically different abundance patterns and are more strongly concentrated towards lower metallicities, with a clear paucity of stars at the metal-rich end.}
%We do caution about the number statistics here as the data only come from four NSCs (i.e., NGC 5139, NGC 6273, NGC 6656, and NGC 1851) and the number of stars is dominated by one single NSC, NGC 5139. As a result, 
\redtxt{Since the statistics are dominated by NGC 5139}, whether the mass scaling relation of NSC should drive the N/O-$M_\star$ relation is unclear.
%There is also a local anticorrelation between the specific frequency of GCs (defined as the number of GCs divided by the stellar mass of the host galaxy) and the galaxy stellar mass at $M_\star \lesssim 10^{10}~M_\odot$ \citep[e.g.,][]{Choksi_gcscaling_2019}. \redtxt{[tbd: any physical picture? no MZR for GCs it seems.]}

{Regardless of any scaling relations associated with NSC formation, do we see direct evidence of such formation pathways in \somename? As shown in Figure~\ref{fig:no_nsc}, stars in NSCs\,—\,primarily from NGC 5139\,—\,exhibit systematically lower O/H, higher N/O, elevated C/O, and lower Fe/O relative to the general population of GC stars. These patterns have been reported in high-resolution spectroscopic studies of chemically complex clusters long suspected to be remnant NSCs \citep[e.g.,][]{Smith2000,Yong2009,Marino2011,Marino2012,Yong2015}. In these systems, carbon enhancement is found to correlate with $s$-process element enrichment, which is a signature of pollution by low-mass AGB stars. Because AGB winds are slow ($\leq25$ km s$^{-1}$) compared to energetic SN II ejecta, they can be retained within the cluster's potential well and later recycled into subsequent stellar generations. However, the characteristic lifetimes of these low-mass polluters ($M < 3~M_\odot$) are $\sim$300 Myr, requiring an extended star formation period that is only feasible in the deep potentials of NSCs.} 

{In summary, low-mass AGB wind retention is not a generic feature of globular clusters. It requires exceptional gravitational binding to maintain the prolonged star formation that typifies nuclear star clusters or massive cluster complexes embedded in dwarf galaxy nuclei. Only in these environments can slow, carbon-rich AGB winds be retained while fast, oxygen-rich supernova ejecta escape\,—\,producing the elevated C/O that distinguishes NSC-type clusters from ordinary GCs. This elevated C/O is particularly noteworthy given its apparent similarity to that observed in the unobscured AGN host UNCOVER-45924. The chemical abundance pattern in UNCOVER-45924 thus raises the intriguing possibility that it harbors not only an accreting supermassive black hole but also an emerging or fully formed NSC. Confirming this connection will require follow-up observations at higher spectral resolution to reduce systematic uncertainties in the derived abundances.}

\section{Conclusions}
\label{sec:conclude}

In this work, we investigated the connection between a sample of 22 ``N/O-enhanced'' galaxies {(NOEGs)} with enhanced nitrogen-to-oxygen ratios (i.e., super-solar N/O at sub-solar O/H) in their interstellar medium (ISM) at $0 \lesssim z\lesssim 12$ with field stars and globular cluster (GC) stars in the Milky Way (MW) through their chemical abundance patterns, including O/H, N/O, C/O, Fe/O, and He/H.
These abundances trace different chemical enrichment mechanisms and time scales.
We assume that the chemical abundances measured in the atmospheres of evolved MW stars capture the chemical composition of the ISM of their host galaxies at the time of formation.
%For old stars in the MW, it is expected that their stellar abundances would correspond to the abundances in the gas of high-redshift galaxies as their progenitor environments.
We summarize our conclusions below.

\begin{itemize}
    \item The \somename occupy a unique parameter space in the N/O versus O/H diagram, where MW disk stars are rarely populated but MW stars in GCs including accreted nuclear star clusters (NSCs) candidates are commonly found (Figure~\ref{fig:no_com}).
    At low metallicities ($\rm 12+log(O/H)<8.0$), currently discovered \somename show a tentative anticorrelation between N/O and O/H also present in GCs, which, for GC stars, is usually interpreted as evidence of oxygen depletion through \redtxt{the CNO cycle} as GCs evolve and self-enrich the gas in the clusters.
    While there are also more metal rich \somename at $\rm 12+log(O/H)>8.0$, they might still overlap with more metal-enriched GCs in the abundance space that are not representative in the MW GCs.
    \item In terms of the carbon abundance, \somename generally show sub-solar C/O consistent with either the low-metallicity MW field stars or GC stars (Figure~\ref{fig:co_com}).
    Only one NOEG, UNCOVER-45924 at $z=4.46$, shows potentially super-solar C/O, {which is more consistent with the abundance pattern of NSCs compared to GCs, although further observations are needed to verify this connection (Figure~\ref{fig:no_nsc}). The presence of an accreting supermassive black hole in this NOEG might point towards a coevoluation scenario for the NSC and the black hole.}
    \item In terms of the iron abundance, which is a typical tracer of a longer enrichment time scale, \somename exhibit a tentative anticorrelation between Fe/O and O/H similar to the trend observed in the ISM of local metal-poor galaxies and \hii\ regions (Figure~\ref{fig:feo_com}).
    This anticorrelation was previously interpreted as increased dust depletion with increased metallicity.
    While there is potential enhancement of Fe/O for \somename at low O/H reaching solar or super-solar Fe/O similar to the second generation GC stars, the measurements of Fe/O in these \somename are highly uncertain.
    Besides the depletion of oxygen via \redtxt{the CNO cycle}, enhancement of Fe/O at low metallicities can also be caused by atypical supernova enrichment such as through the pair-instability supernovae or the hypernovae.
    %It is vital to have Fe abundance measurements in general high-redshift galaxies in addition to \somename to better understand the Fe/O evolution and whether it is indeed related to the N/O enhancement.
    \item The helium enhancement is found in both GC stars and potentially a few metal-poor \somename (Figure~\ref{fig:heh_com}).
    %For the abundance of helium, which shows little spread on average among GC stars but can have significant enhancement if taking the maximum value, enhancement is potentially present in a few metal-poor \somename.
    Unlike previous work, we found no clear indication of any correlation between the helium enhancement and the N/O enhancement.
    However, there appears to be a tentative anticorrelation between the helium enhancement and the oxygen abundance in \somename.
    The enhancement of helium can either be interpreted as a bias induced by underestimated gas densities, suggested by the complicated density variations seen in some \somename, or a true enhancement caused by different initial masses of the proto-GCs involved.
    The latter interpretation implies more massive proto-GCs or proto-NSCs found in the low metallicity environments with a longer enrichment time, which require further direct observational evidence from star clusters at high redshift to verify.
    \item Observations of \somename suggest a wide range of environments, which, however, share a common feature of high stellar densities consistent or higher than the expected progenitor environment of MW GCs and comparable to those directly observed in a few star clusters in high-redshift lensed galaxies (Figure~\ref{fig:no_mass}).
    The dense environments would allow for the formation and growth of very-to-extremely massive stars ($M_\star \sim 10^2-10^4~M_\odot$), which is one of the proposed channels for producing the abundance anomalies in GCs and \somename.
    In addition, NOEGs predominantly show high star-formation rate surface densities, which might imply efficient formation of bound clusters.
    There is also evidence of accreting supermassive black holes in a few \somename (and, vice-versa, evidence for nitrogen enrichment in galaxies hosting accreting supermassive black holes found by \jwst\ at high-$z$ shown by previous work), indicating a possible connection from \somename and formation of black holes seeds (via collapses of very-to-extremely massive stars or runaway merging), as suggested by various models. Yet, the connection between the black hole accretion and nitrogen enhancement might also be caused by a common correlation with the stellar mass.
    There is a tentative N/O-stellar mass relation in \somename, where more massive \somename are more nitrogen enhanced, although verifications from a larger sample with quantified selection functions will be needed to verify the correlation.
\end{itemize}

As a concluding remark, the chemical similarities between local GCs and high-redshift \somename produce key evidence that connects the two populations.
The seemingly high occurrence rates of \somename in the early Universe is qualitatively consistent with the picture of bursty, cluster-dominated star formation recently proposed for the early phase of galaxy evolution.
Once galaxies transform from cluster-dominated star formation to disk-dominated star formation at later epochs, the chemical anomalies including the N/O enhancement in the ISM and later formed field stars should be erased and the former \somename would converge to the local scaling relations.
\redtxt{While the above scenario is promising, we do note that there are alternative theories for massive cluster formation from, rather than prior to, the disk \citep[e.g.,][]{Krumholz_2010,Kruijssen_2012,clarke_2019}.
}
Future deep observations of a statistical sample of high-redshift galaxies with determinations of chemical abundances in spatially separated star clusters and inter-cluster regions will be important to further understand this connection.

\begin{comment}

\section{TODO}

\begin{itemize}
    \item abundance corrections? (0.15 dex? 0.25 dex? uncertain?)
    %\item Carbon abundance
    \item Ne abundance?
    %\item He diagram (primordial + 30\%)
    %\item Global properties; stellar mass?
    \item NSC vs. GC
    \item Ubiquity of chemical stratification (maybe remove)?
    \item halo mass threshold --> how ubiquitous for cluster dominance?
\end{itemize}
    
\end{comment}

\section*{Acknowledgements}
\realredtxt{We thank the anonymous referee, whose thoughtful comments improved the clarity of the this work.}
We thank Wyn Evans and Hui Li for helpful discussions.
We thank Charlotte Simmonds for providing the photometric/spectroscopic catalogues of JADES galaxies.
XJ and RM acknowledge ERC Advanced Grant 695671 “QUENCH” and support by the Science and Technology Facilities Council (STFC) and by the UKRI Frontier Research grant RISEandFALL.
VB acknowledges support from the Leverhulme Research Project Grant RPG-2021-205: ``The Faint Universe Made Visible with Machine Learning''.
RM acknowledges funding from a research professorship from the Royal Society.
YI is supported by JSPS KAKENHI Grant No. 24KJ0202.
AK was supported by the NASA ATP grant 80NSSC20K0512 and the National Science Foundation grant AST-2408267.
HÜ acknowledges funding by the European Union (ERC APEX, 101164796). Views and opinions expressed are however those of the authors only and do not necessarily reflect those of the European Union or the European Research Council Executive Agency. Neither the European Union nor the granting authority can be held responsible for them.

\redtxt{This work is based on observations made with the NASA/ESA/CSA James Webb Space Telescope. The data are available at the Mikulski Archive for Space Telescopes (MAST) at the Space Telescope Science Institute, which is operated by the Association of Universities for Research in Astronomy, Inc., under NASA contract NAS 5-03127 for JWST.}

\section*{Data Availability}

All \jwst\ data used in this paper are available through the MAST portal.
The APOGEE data are available through \url{https://www.sdss4.org/dr17/data_access/}.
All analysis results of this paper will be shared on reasonable request to the corresponding author.

\bibliographystyle{mnras} % style aa.bst
\bibliography{ref} % your references ref.bib

\begin{appendix}

\section{Derivation of N/O and C/O for UNCOVER-45924}
\label{appendix:monster}

\begin{figure*}
    \centering
    \includegraphics[width=\textwidth]{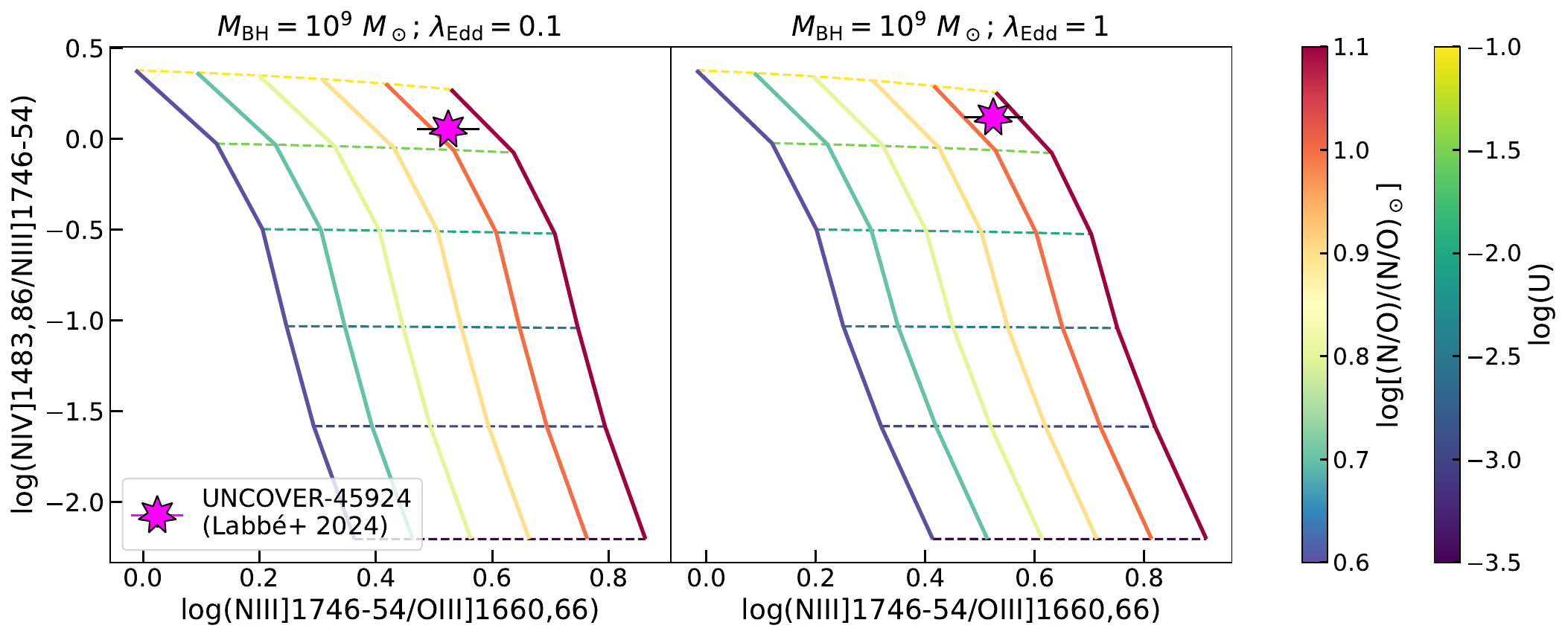}
    \caption{Comparison between observed line ratios of UNCOVER-45924 measured by \citet{labbe_monster_2024} and those predicted by \cloudy photoionization models, where all emission lines involved are assumed to come from a BLR cloud with $n_{\rm H}=10^9~{\rm cm^{-3}}$ and $N_{\rm H}=10^{23}~{\rm cm^{-2}}$.
    The two panels show two sets of models with different Eddington ratios, $\lambda _{\rm Edd}$.
    For each set of models, values of N/O and $U$ are varied independently and $\rm 12+\log(O/H)$ is fixed to 7.41 (see Section~\ref{sec:method}).
    As shown by the models, the $x$ axis, \niii/\oiiip], mainly traces N/O;
    the $y$ axis, \niv/\niii, mainly traces $U$.
    The best-fit model for UNCOVER-45924 has an N/O roughly 1 dex above solar.
    }
    \label{fig:cloudy_monster}
\end{figure*}

\begin{figure*}
    \centering
    \includegraphics[width=\textwidth]{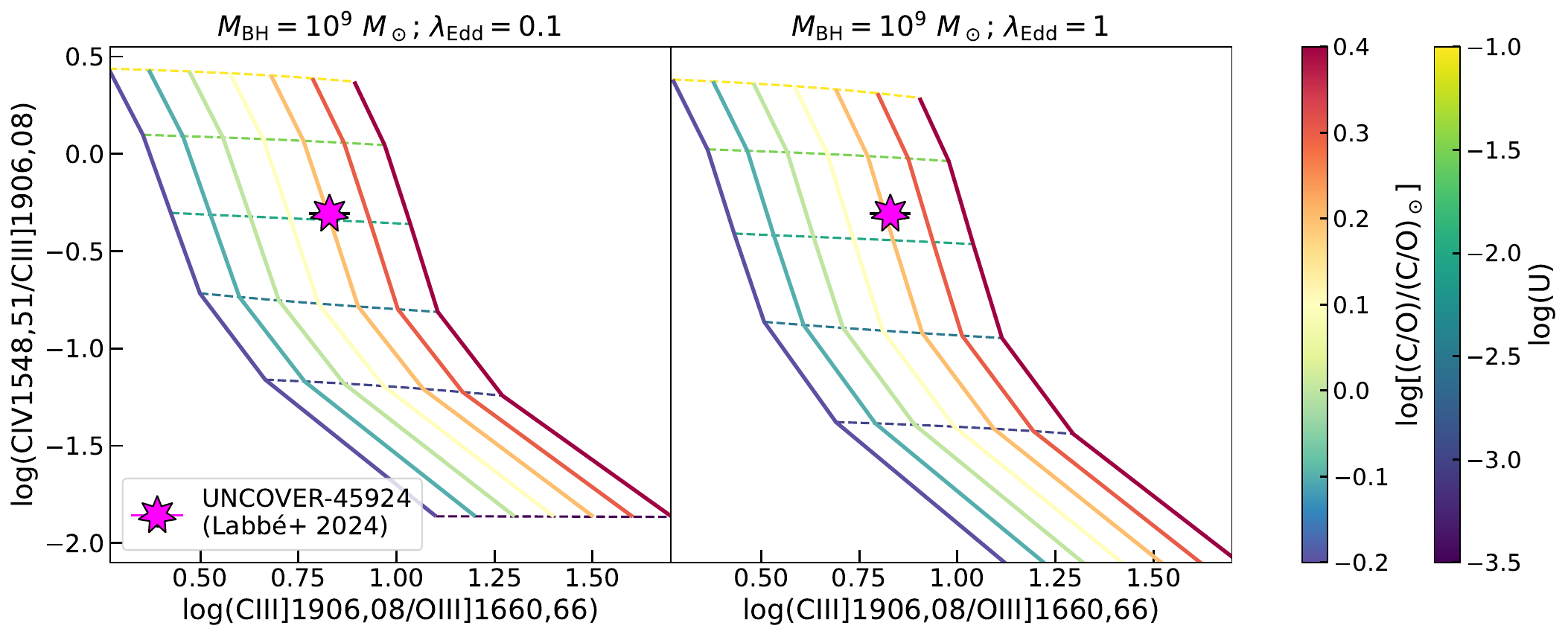}
    \caption{Comparison between observed line ratios of UNCOVER-45924 measured by \citet{labbe_monster_2024} and those predicted by \cloudy photoionization models, where all emission lines involved are assumed to come from a BLR cloud with $n_{\rm H}=10^9~{\rm cm^{-3}}$ and $N_{\rm H}=10^{23}~{\rm cm^{-2}}$.
    The two panels show two sets of models with different Eddington ratios, $\lambda _{\rm Edd}$.
    For each set of models, values of C/O and $U$ are varied independently and $\rm 12+\log(O/H)$ is fixed to 7.41 (see Section~\ref{sec:method}).
    As shown by the models, the $x$ axis, \niii/\oiiip], mainly traces N/O;
    the $y$ axis, \civ/\ciii, mainly traces $U$.
    The best-fit model for UNCOVER-45924 has an C/O roughly 0.2 dex above solar.
    }
    \label{fig:cloudy_monster_c}
\end{figure*}

\begin{table}
        \centering
        \caption{Input parameters for \textsc{Cloudy} photoionization models.}
        \label{tab:models}
        \begin{tabular}{l c}
            \hline
            \hline
            Parameter & Values \\
            \hline
            $\rm 12+\log(O/H) $ & 7.41\\
            \hline
            $\rm [N/O]$ & 0.6, 0.7, 0.8, 0.9, 1.0, 1.1\\
            \hline
            $\rm [C/O]$ & $-0.2$, $-0.1$, 0, 0.1, 0.2, 0.3, 0.4\\
            \hline
            $\log U$& $-3.5$, $-3$, $-2.5$, $-2$, $-1.5$, $-1$ \\
            \hline
            $\log (n_{\rm H}/{\rm cm^{-3}})$& 9 \\
            \hline
            $\log (N_{\rm H}/{\rm cm^{-2}})$ & 23 \\
            \hline
            AGN SED & $M_{\rm BH}=10^9~M_\odot$, $\lambda _{\rm Edd}=0.1,~1$\\ 
             & \citep{pezzulli_2017} \\
            \hline
            Dust & No dust\\
            \hline
            Atomic data & CHIANTI (v7, \citealp{chianti0};\\
             & \citealp{chianti_v7})\\
            \hline
        \end{tabular}
\end{table}

In this appendix, we describe the derivation of N/O and C/O for UNCOVER-45924 \citep{labbe_monster_2024} based on \cloudy \citep[c17.03,][]{ferland2017} photoionization models.
While we follow the argument of \citet{labbe_monster_2024} to use BLR models to describe the relevant emission lines, if these lines actually come from an NLR with lower densities (i.e., $n_{\rm e}<10^{8}~{\rm cm^{-3}}$), the inferred N/O is actually higher due to the density dependence of \niii$\lambda 1750$/\oiiip]$\lambda \lambda 1661,1666$.
We summarize our fiducial model set in Table~\ref{tab:models}.
The choice of $n_{\rm H}=10^{9}~{\rm cm^{-3}}$ is to obtain a conservative estimate of N/O.
We set $N_{\rm H}=10^{23}~{\rm cm^{-2}}$ as a typical value for BLR emission-line modeling \citep[e.g.,][]{ferland2009}, and we note that higher column densities would not impact the results significantly as high ionization lines are mainly produced close to the illuminated face of the clouds \citep{ji2024}.
We allow the ionization parameter to vary in a wide range reaching $\log(U)=-1$.
We take the input AGN SED from the analytical model of \citet{pezzulli_2017}, where we set a black hole mass of $10^9~M_\odot$ and two Eddington ratios ($\lambda _{\rm Edd}\equiv L_{\rm bol}/L_{\rm Edd}$) of 0.1 and 1 possible for UNCOVER-45924 \citep{labbe_monster_2024}.
Finally, we assume the BLR is dust free.

Figures~\ref{fig:cloudy_monster} and \ref{fig:cloudy_monster_c} show comparisons between line ratios measured by \citet{labbe_monster_2024} corrected for dust extinction (see Section~\ref{sec:method}, although the corrections are modest due to the lines involved have similar wavelengths) and our model grids.
In each diagram, the $y$ axis is designed to trace the ionization parameter and the $x$ axis is designed to trace mainly N/O or C/O at fixed O/H.
From these two figures, UNCOVER-45924 exhibits a very high ionization parameter of $\log(U)=-2$\,-\,$-1.25$.
Ratios of \niv/\niii\ and \civ/\ciii\ give $U$ values differ by 0.5 dex, which might be caused by a spatial stratification of different ionization species, an additional unresolved narrow component in \niv, or the radiative transfer of \civ.
To estimate N/O and C/O, we interpolate models in the logarithm space of $U$ and N/O (or C/O) with different $\lambda _{\rm Edd}$ using the \textsc{Python} function \textsc{griddata} with a cubic spline function.
We then search for the most likely model points for UNCOVER-45924 and take the average N/O (or C/O) of models with different $\lambda _{\rm Edd}$, resulting in $\rm \log(N/O)=0.20^{+0.06}_{-0.05}$ and $\rm \log(C/O)=-0.06^{+0.06}_{-0.05}$ with a systematic uncertainty of 0.10 dex from the different values of ionization parameters.

\end{appendix}

%   \bibliographystyle{mnras} % style aa.bst
%   \bibliography{ref} % your references ref.bib

%\appendix

%\section{Bump detection over different visits of observations}
%\section{Detection of the small blue bump in Jackknifed Prism spectra}
%\label{appendix:a}

% Don't change these lines
\bsp	% typesetting comment
\label{lastpage}
\end{document}